\documentclass[options]{JHEP3}

\usepackage{epsf,epsfig}
\usepackage{amssymb}
\usepackage{graphicx}
\usepackage{amsmath}
\usepackage{amsfonts}
 
\usepackage{graphicx}

\newcommand{\insertplot}[5]{\begin{figure}
 \hfill\hbox to 0.05in{\vbox to #5in{\vfill
 \inputplot{#1}{#4}{#5}}\hfill}
 \hfill\vspace{-.1in}
 \caption{#2}\label{#3}
 \end{figure}}
 \newcommand{\inputplot}[3]{
 \special{ps: plotfile #1}
}
\newcounter{fig}

\newcommand{\beq}{\begin{equation}}
\newcommand{\eeq}{\end{equation}}
\newcommand{\beqs}{\begin{eqnarray}}
\newcommand{\eeqs}{\end{eqnarray}}

\newcommand{\be}{\begin{equation}}
\newcommand{\ee}{\end{equation}}
\newcommand{\bea}{\begin{eqnarray}}
\newcommand{\eea}{\end{eqnarray}}

\numberwithin{equation}{section}

\abstract{ 
We present numerical evidence for the existence of several types of static 
black hole solutions with a nonspherical event horizon topology
in $d\geq 6$ spacetime dimensions. 
These   asymptotically flat configurations are found for a specific metric ansatz
and can be viewed as higher dimensional
  counterparts of the $d=5$ 
  static black rings, dirings and black Saturn.
Similar to that case, they are supported against collapse by conical singularities.
The issue of rotating generalizations of these solutions is also considered.  }

\keywords{ black holes, numerical solutions}\preprint{ }

\title{ New generalized nonspherical black hole solutions  } 
  
 \author{
 {\large Burkhard Kleihaus}$^{\dagger}$, {\large Jutta Kunz}$^{\dagger}$, {\large Eugen Radu}$^{\star \diamond }$  
  and 
 {\large Maria J. Rodriguez}$^{\ddagger}$
\\ 
\\
$^{\dagger}${\small Institut f\"ur Physik, Universit\"at Oldenburg, Postfach 2503
D-26111 Oldenburg, Germany} 
\\
$^{\ddagger}$ {\small Max-Planck-Institut f\"ur Gravitationsphysik, Albert-Einstein-Institut, 14476 Golm,
 Germany }
   \\
$^{\star}${\small  Department of Computer Science,
National University of Ireland Maynooth,
Maynooth,
Ireland} \\
$^{\diamond}${\small School of Theoretical Physics -- DIAS, 10 Burlington
Road, Dublin 4, Ireland }
 }
 
\begin{document}

\section{ Introduction}

In recent years the interest in the properties of gravity in more than $d = 4$ dimensions has increased
significantly. This interest was enhanced by the development of string theory, which requires
a ten-dimensional spacetime, to be consistent from a quantum point of view.
An unexpected result in this area was Emparan and Reall's discovery  
of the black ring in $d=5$ spacetime dimensions \cite{Emparan:2001wn,Emparan:2001wk}. 
This asymptotically flat solution of the Einstein equations  has a horizon with topology $S^2\times S^1$,
while the Myers-Perry black hole \cite{Myers:1986un} has a horizon topology $S^3$.
This made clear that a number of well known results in $d=4$ gravity
do not have a simple extension  to higher dimensions. 
For example, the $d=5$ gravity allows for  multi-black hole configurations regular outside and on the horizon.
In this case, at least one of the
constituents possesses a nonspherical topology of the horizon, the simplest examples being the
black Saturn \cite{Elvang:2007rd} 
(a black ring with a central black hole), a diring \cite{Iguchi:2007is,Evslin:2007fv}
(two concentric coplanar black rings) and bicyling black rings \cite{Elvang:2007hs} (two black rings in orthogonal planes).

However, while one can construct an encyclopedia of general relativity exact solutions in four and five dimensions, the
situation for $d>5$ is more patchy (see \textit{e.g.} \cite{Rodriguez:2010zw}).  
For most of the cases, the known solutions are very special, with
a large amount of symmetry.
Moreover, it becomes clear that as the dimension increases, the phase structure
of the solutions
becomes increasingly intricate and diverse.
The main obstacle stopping the progress in this field seems to be the absence of
closed form solutions (apart from the Myers-Perry black holes), which were very useful in $d=5$.
No general framework seems to exist for $d>5$,
and the issue of constructing black objects
with a nonspherical horizon topology 
was considered by using various approximations
or numerical methods.
Most of the results in this area have been found by using the method of matched asymptotic expansions \cite{Emparan:2007wm,Emparan:2009vd}.
The central assumption  is that some black objects, in certain ultra-spinning regimes,
can be approximated by very thin
black strings or branes curved into a given shape.
However, this method has limitations; black holes whith no black membrane behavior (e.g. at high spins) 
would not be captured by this approach \cite{Astefanesei:2010bm}.

Although it would clearly be 
preferable to have analytic solutions\footnote{However, one should
not exclude the possibility that most of these solutions will
remain analytically intractable within a nonperturbative approach.},
 some of the $d>5$ black holes with a nonspherical horizon topology  
 can be constructed numerically, 
 within a nonperturbative approach, 
 as solutions of partial differential
equations with suitable boundary conditions.

The main purpose of this paper is to present a general  
framework for a special class of  static configurations
with a symmetry group $R_t\times U(1)\times SO(d-3)$
and to present numerical evidence for the existence of such solutions with nonspherical horizon topology.
For $d=5$, this framework reduces to that used in  \cite{Emparan:2001wk}
to construct generalized Weyl solutions.
However, for higher values of the spacetime dimension, the solutions can be found only numerically.
We argue that the basic properties 
of the $d=5$ case still hold for $d>5$ configurations
with a symmetry group $R_t\times U(1)\times SO(d-3)$, in particular the rod structure of the solutions.
The simplest example of a
$d>5$ black object with a nonspherical horizon obtained within this approach
was studied in Ref. \cite{Kleihaus:2009wh} and has a horizon topology 
 $S^2\times S^{d-4}$.
 In this work, on the one hand, we extend these results and discuss the basic 
 features of two new types of configurations
representing composite black objects with 
$\left(S^2\times S^{d-4}\right) \times S^{d-2}$ 
-- a generalized black Saturn -- and also 
$\left(S^2\times S^{d-4}\right) \times \left(S^2\times S^{d-4}\right)$ 
horizons --a generalized diring. On the other hand, within a slightly more 
general metric ansatz, we consider rotating solutions in either 
the $S^2$ or the $S^{d-4}$ spheres.

This paper is organized as follows:
in the next  Section we present a systematic discussion of this approach together with its limitations,
while in Section III
we present our numerical results.
All solutions are found within a nonperturbative approach, 
by directly solving Einstein equations which for our ansatz reduce to a set of 
four nonlinear partial differential equations.

Since all these solutions are plagued by conical singularities 
which seem to be unavoidable in the absence of rotation,
the issue of spinning solutions is addressed in Section IV.
The results reported there are only partial,
and so far we could not construct spinning balanced 
solutions. However, we expect that they will be useful for further work in this direction.

Also, we have found that all new static solutions in this work
have similar qualitative properties as their five dimensional
counterparts.
Therefore in the Appendix A we present  the basic properties
of the corresponding $d=5$ solutions, which are
known in closed form.
Appendix B  introduces  
a new coordinate system which 
simplifies the numerical calculations and leads to high accuracy 
($e.g.$ it
has allowed to recover numerically the
 spinning balanced black ring starting with the static solution).

\section{The general formalism}

\subsection{The field equations and a metric ansatz}
We consider the Einstein action
\begin{eqnarray} 
\label{action-grav} 
I=\frac{1}{16 \pi G_d}\int_{\cal M}~d^d x \sqrt{-g} R
-\frac{1}{8\pi G_d}\int_{\partial {\cal M}} d^{d-1}x\sqrt{-h}K,
\end{eqnarray}
in a $d-$dimensional spacetime, with $d\geq 5$.
The last term in  (\ref{action-grav}) is the Hawking-Gibbons surface term \cite{GibbonsHawking1},
which
is required in order to have a well-defined variational principle. 
$K$ is the trace 
of the extrinsic curvature for the boundary $\partial\mathcal{M}$ and $h$ is the induced 
metric of the boundary.  
Also, $G_d$ is Newton's constant in $d-$dimensions; for simplicity, we shall set $G_d=1$ in this work.

The upshot for our computations is that 
the line element of the static solutions of interest can be cast in the following form
(where $0\leq \psi\leq 2\pi$, $0\leq \theta \leq \pi/2$ and $d\Omega^2_{d-4}$ the unit metric on $S^{d-4}$,
while  $0\leq \rho<\infty,$ $-\infty< z<\infty$)
\begin{eqnarray}
\label{metric-canonical} 
ds^2=-e^{2U_0(\rho,z)} dt^2+e^{2\nu(\rho,z)} (d\rho^2+dz^2)+e^{2U_1(\rho,z)} d\psi^2+e^{2U_2(\rho,z)} d\Omega_{d-4}^2.
 \end{eqnarray} 
The solutions constructed within this ansatz are static and axisymmetric,
 with a symmetry group $R_t\times U(1)\times SO(d-3)$ (where $R_t$ denotes the time translation). 
While in principle it is possible to choose any kind of boundary conditions, we will only concentrate on black hole solutions 
which asymptote to flat spacetime. 
Moreover, the coordinates in (\ref{metric-canonical}) have a rectangular boundary and thus are suitable for numerical calculations.

A suitable combination 
of the Einstein equations,
 $G_t^t=0,~G_\rho^\rho+G_z^z=0$, $G_{\psi}^{\psi}=0$
 and  $G_{\varphi}^{\varphi}=0$ (with $\varphi$ an angle on $\Omega_{d-4}$),
yields the following set of equations for the functions $U_0,~U_1,~U_2$ 
 \begin{eqnarray}
\label{eqU0}
\nonumber
&&\nabla^2 U_0+(\nabla U_0)^2+(\nabla U_0)\cdot( \nabla U_1)+(d-4)(\nabla U_0)\cdot( \nabla U_2)=0,
\\
\label{eqU2}
&&\nabla^2 U_1+(\nabla U_1)^2+(\nabla U_0)\cdot( \nabla U_1)+(d-4)(\nabla U_1)\cdot( \nabla U_2)=0,
\\
\label{eqUx} 
\nonumber
&&\nabla^2 U_2+(d-4)(\nabla U_2)^2+(\nabla U_0)\cdot( \nabla U_2)+(\nabla U_1)\cdot( \nabla U_2)-(d-5)e^{2\nu-2U_2}=0,
\end{eqnarray}%
and 
 \begin{eqnarray}
\label{eqnu}
&&\nabla^2 \nu-(\nabla U_0)\cdot( \nabla U_1)-(d-4)(\nabla U_0)\cdot( \nabla U_2)
 -(d-4)(\nabla U_1)\cdot( \nabla U_1)
\\
\nonumber
&&{~~~~~~~~~~~~~~~~~~~~~~~~~~~~~~~~~~~~~~~~~~~~~~}+\frac{1}{2}(d-4)(d-5)\left( e^{2\nu-2U_2}-(\nabla U_2)^2 \right)=0,
\end{eqnarray}%
for the metric function $\nu$,
where we define 
\begin{eqnarray}
\label{rel}
(\nabla U) \cdot (\nabla V)=\partial_\rho U \partial_\rho V+ \partial_z U \partial_z V,~~~
\nabla^2 U=\partial_\rho^2U+\partial_z^2 U.
\end{eqnarray}
The remaining Einstein equations $G_z^\rho=0,~G_\rho^\rho-G_z^z=0$
yield two constraints. Following \cite{Wiseman:2002zc}, we note that
setting $G^t_t=G^{\varphi}_{\varphi} =G^\rho_\rho+G^z_z=0$
in $\nabla_\mu G^{\mu \rho}=0$ and $\nabla_\mu G^{\mu z}=0$, we obtain the Cauchy-Riemann relations
\begin{eqnarray}
2\partial_z\left(\sqrt{-g} G^\rho_z \right) +
  \partial_\rho\left(\sqrt{-g}(G^\rho_\rho-G^z_z) \right)= 0,~
 2\partial_\rho\left(\sqrt{-g} G^\rho_z \right)
-\partial_z\left(  \sqrt{-g}(G^\rho_\rho-G^z_z) \right)= 0.
\end{eqnarray}
Thus, the weighted constraints satisfy Laplace equations,
and the constraints are fulfilled
when one of them is satisfied on the boundary 
and the other at a single point
\cite{Wiseman:2002zc}. 
As we shall see, this is the case for all configurations discussed in the next Section.

  Although the Einstein equations take a simple form in terms of  ($U_i,\nu$), 
for the purposes of this paper it is more convenient to work with a set a functions
$f_i$ defined as follows\footnote{Some divergencies are avoided in this way. For example, $f_i\to 0$
would correspond to $U_i\to -\infty$ which is clearly not suitable for a numerical approach.}
\begin{eqnarray}
\label{fi}
e^{2\nu(\rho,z)}=f_1(\rho,z),~~e^{2U_2(\rho,z)} =f_2(\rho,z),
~~e^{2U_3(\rho,z)} =f_3(\rho,z),~~e^{2U_1(\rho,z)} =f_0(\rho,z).
\end{eqnarray} 
This leads to a line element
\begin{eqnarray}
\label{metric} 
ds^2=-f_0(\rho,z)dt^2+f_1(r,z)(d\rho^2+dz^2)+f_2(\rho,z)d\psi^2+f_3(\rho,z)d\Omega_{d-4}^2,
 \end{eqnarray}
which was used in our numerical  study of the $d>5$ solutions.

One might be concerned 
that (\ref{metric})  is too restrictive to leave room for new interesting black hole solutions. 
In higher dimensions, a priori this is not the case\footnote{
 Black holes have to be of positive Yamabe type
\cite{Galloway:2005mf} and, if stationary, they have to be axisymmetric \cite{Hollands:2006rj}.},
and, in the next Section, we shall present numerical evidence for the existence of 
nontrivial solutions which share the basic properties of some $d=5$
configurations with a nonspherical horizon topology.

Other more general metric proposals 
which may describe higher dimensional
black hole solutions with a nonspherical horizon topology
have been presented in \cite{Harmark:2009dh}. 
However, due to their complexity, they will be out of our present scope.
 
\subsection{Known solutions}

\subsubsection{Minkowski space-time}

In  $d\geq 5$ dimensions,
the flat spacetime metric can be written in the form (where $0\leq r <\infty$, $0\leq \psi\leq 2\pi$, $0\leq \theta \leq \pi/2$)
\begin{eqnarray} 
\label{m1}
ds^2=-dt^2+dr^2+r^2(d\theta^2+\sin^2\theta d\psi^2+\cos^2\theta d\Omega_{d-4}^2 ),
\end{eqnarray}
thus with the metric on $S^{d-2}$ written in terms of a warped product of $S^2$ and $S^{d-4}$.
 Then, for all dimensions, the coordinate transformation
 \begin{eqnarray} 
\label{ct1}
r=\sqrt{2}(\rho^2+z^2)^{1/4},~~~\theta=\frac{1}{2}\arctan (\rho /z),
\end{eqnarray}
leads to the equivalent form of (\ref{m1})
 \begin{eqnarray} 
\label{m2}
ds^2=- dt^2+\frac{1}{2\sqrt{\rho^2+z^2}} \,(d\rho^2+dz^2)+(\sqrt{\rho^2+z^2}+z)\, d\psi^2+(\sqrt{\rho^2+z^2}-z)\, d\Omega_{d-4}^2,~~{~~~~}
\end{eqnarray}
where $0\leq \rho<\infty,$ $-\infty< z<\infty$.
 
An interesting observation here concerns the 
value of the determinant $\Delta$ for the non-conformal part of the metric\footnote{The choice of this 
determinant has been proven to be crucial in recent progress
 on finding new classes of solutions  \cite{Emparan:2001wk}
 and also for the metric proposals in \cite{Harmark:2009dh}.}  
($i.e.$ 
the line element (\ref{metric})
without the $(\rho,z)$-part).
One can see that even for the simplest case of a Minkowski space-time within the parametrization (\ref{metric}),
 $\Delta=-\rho^2$  for $d=4,5$ only. 

\subsubsection{Schwarschild-Tangherlini black hole}\label{sec:Schw}

The simplest  example of a $d>5$ nontrivial solution that can be studied
within this approach corresponds to the Schwarzschild-Tangherlini black hole.
Usually this metric is written in the form
\begin{eqnarray}\label{STmetric}
ds^2&=& -f(r)\,dt^2 + f(r)^{-1}\,dr^2+r^2( d\theta^2
+\sin^2\theta\, d\psi^2 +\cos^2\theta\, d\Omega^2_{d-4}),
\end{eqnarray}
with $f(r)=1-\mu/r^{d-3}$.
This $d-$dimensional static black hole solution has an isometry group $R_t \times SO(d-1)$. 
By a change of coordinates one can bring the metric to the desired conformal form (\ref{metric}).
The change of coordinates is
\begin{eqnarray}\label{eq:coordchange}
\rho= \frac{\alpha}{2}  \sin 2\theta\ \sinh G(r),\qquad
z=  \frac{\alpha}{2} \cos 2\theta \cosh G(r),
\end{eqnarray}
which yields
\begin{eqnarray}
d\rho^2+dz^2= \frac{\alpha^2}{2}(\cosh 2 G(r)-\cos 4\theta) \left(\frac{G'(r)^2}{4} dr^2+d\theta^2\right)\,. 
\end{eqnarray}
By simply integrating $G(r)$ one finds
\begin{eqnarray}
G(r)=2\int \sqrt{\frac{g_{rr}}{g_{\theta\theta}}}\,dr=2 \int \frac{1}{\sqrt{f(r)\,r^2}}\, 
dr=\log[2^{\frac{4}{d-3}}\left(r^{\frac{d-3}{2}}+\sqrt{r^{d-3}-\mu}\right)^\frac{4}{d -3}]+k,~~{~~}
\end{eqnarray}
where $k=-{\frac{2}{d-3}}\,\log(4 \mu)$ is the integration constant. And,
finally we fix $\alpha=2^{\frac{d-7}{d-3}}\mu^{\frac{2}{d-3}}$ to match asymptotically flat space.

The transformation (\ref{eq:coordchange}) simplifies drastically in $d=5$
\begin{eqnarray}\label{eq:coordchangefive}
\rho= \frac{1}{2}  \sin 2\theta\ \left(1-\frac{\mu}{r^2}\right)^{1/2}r^2,\qquad
z=  \frac{1}{2} \cos 2\theta \left(1-\frac{\mu}{2r^2}\right)r^2\,,
\end{eqnarray}
matching the findings of \cite{Emparan:2001wk}, and in $d=7$ where
\begin{eqnarray}\label{eq:coordchangeseven}
\rho= \frac{1}{2}  \sin 2\theta \left(1-\frac{\mu}{ r^4}\right)^{1/2}r^2,\qquad
z=  \frac{1}{2} \cos 2\theta\,r^2\,. 
\end{eqnarray}
A straightforward but cumbersome computation leads to the following expression for the
Schwarzschild-Tangherlini black hole in the $(\rho,z)$ coordinates 
%
\begin{eqnarray}
ds^2&=&-\left(\frac{v^{(d-3)/2}-1}{v^{(d-3)/2}+1}\right)^2\,dt^2+ \frac{\mu(v^{(d-3)/2}+1)^{4/(d-3)}}
 {4v\left(z^2\frac{(v^2-1)^2}{(v^2+1)^2}+\rho^2\frac{(v^2+1)^2}{(v^2-1)^2}\right)}\,(d\rho^2+dz^2)
\label{metricST} 
 \\
  \nonumber
&&+\frac{(v^{(d-3)/2}+1)^{\frac{4}{d-3}}}{2v(v^2+1)}(\mu(v^2+1)+2zv)\,d\psi^2
+\frac{(v^{(d-3)/2}+1)^{\frac{4}{d-3}}}{2v(v^2+1)}(\mu(v^2+1)-2zv)\,d\Omega_{d-4}^2,
 \end{eqnarray}
where
\begin{eqnarray}
\nonumber
&&v=\frac{1}{\mu}\left(\rho^2+z^2+{\cal P}+\sqrt{2}
\sqrt{(\rho^2+z^2)^2+\mu^2(\rho^2-z^2)+(\rho^2+z^2){\cal P}}\right )^{1/2},
\\
\nonumber
&&~~{\rm and~~}
{\cal P}=\sqrt{\mu^4+2\mu^2(\rho^2-z^2)+(\rho^2+z^2)^2 }. 
\end{eqnarray}
Moreover, one can also show that 
for $d=5$ these expressions reduce to those in \cite{Emparan:2001wk}.

\subsection{The rod structure of black hole solutions}
 
\subsubsection{Five dimensional structure}
 For $d=5$, the coordinates in (\ref{m1}) are the usual Weyl coordinates, 
 while the sphere $\Omega_{d-4}$ reduces to a single angular coordinate $\varphi$, 
 with $0\leq \varphi \leq 2 \pi$. 

In this case, it is most convenient to choose
the three functions $U_i$ as to satisfy the condition
\begin{eqnarray}
\label{cons}
\sum_i U_i=\log \rho.
\label{eq:det}
\end{eqnarray}
This is compatible with the vacuum Einstein equations
(\ref{eqU2}), which for the choice (\ref{cons}) imply also
\begin{eqnarray}
\label{eq-U}
 \frac{\partial^2 U_i}{\partial \rho^2} + \frac{1}{\rho}\frac{\partial
 U_i}{\partial \rho} + \frac{\partial^2 U_i}{\partial z^2} = 0.
\end{eqnarray}
One can see that (\ref{eq-U}) is just Laplace's equation in a (fictious) three-dimensional flat space with
metric 
$
 ds^2 = d\rho^2 + \rho^2 d\Theta^2 + dz^2,
$
whose solutions are well-known.

 From the other components of the Einstein
equations $G_{\rho}^\rho-G_{z}^z=0$ and $G_{\rho}^z=0$,
we obtain the equations which determine the function $\nu(\rho,z)$
for a given solution of the  equation (\ref{eq-U})
\begin{eqnarray}
\label{eq-nu1}
\nu'=-\frac{1}{2\rho}+\frac{\rho}{2}
\left( 
U_1'^2+U_2'^2+U_3'^2-\dot U_1^2-\dot U_2'^2-\dot U_3^2
\right),
~~
\dot \nu=\rho(\dot U_1'+\dot U_2'+\dot U_3'),
\end{eqnarray}
where a  prime denotes the derivative with respect to 
$\rho$ and a dot denotes the derivative with respect to $z$.
Solutions with the ansatz (\ref{metric-canonical}) and with $U_1,U_2,U_3$ and $\nu$
satisfying the equations (\ref{eq-U}), (\ref{eq-nu1})
are usually called generalized Weyl solutions \cite{Emparan:2001wk}.

This approach has proven very fruitful, a variety of 
physically interesting configurations being discussed in the literature.
They can be uniquely characterized 
by the boundary conditions on the $z-$axis, known as the {\it rod-structure} \cite{Emparan:2001wk,Harmark:2004rm,Hollands:2007aj}.
One finds that the physically relevant solutions for $U_i$  
can
also be thought of as Newtonian potentials produced by thin rods 
of zero thickness with linear mass density $1/2$, placed
on the axis of symmetry
in the auxiliary three-dimensional flat space. 
Then the constraint (\ref{cons}) states that these sources must add up
to give an infinite rod.

In this approach, the $z-$axis is divided into $N$ intervals (called rods of the 
solution), $[-\infty, z_1]$,
$[z_1,z_2]$,$\dots$, $[z_{N-1},\infty]$.
As proven in \cite{Harmark:2004rm},  in order to avoid curvature naked 
singularities at $\rho=0$, it is a 
necessary condition that only one of the functions
$f_0(0,z)$, $f_2(0,z)$, $f_3(0,z)$ becomes zero for a given rod,
except for isolated points between the intervals.

 For the static case discussed here, a horizon corresponds to 
 a timelike rod where $f_0(0,z)=0$ while $\lim_{\rho \to 0}f_0(\rho,z)/\rho^2>0$.
 There are also spacelike rods corresponding to compact directions  specified by the
 conditions $f_{a}(0,z)=0$, $\lim_{\rho\to 0}f_{a}(\rho,z)/\rho^2>0$, with $a=2,3$.
A semi-infinite spacelike rod corresponds to an axis of rotation, the associated coordinate being
a rotation angle. 
Demanding regularity of the solutions at $\rho=0$
imposes a periodicity $2\pi$ for both $\psi$ and $\varphi$.
(However, when several  $\psi$- or $\varphi$-rods are present, 
it may be impossible to satisfy simultaneously all the periodicity
conditions, see $e.g.$ the examples in Appendix A).

One of the main advantages of this approach is that
the topology of the horizon is automatically imposed by the rod structure. 
This provides a simple way to construct a variety of solutions with 
nontrivial topology of the horizon (including multi-black objects). 
Since (\ref{eq-U}) is linear, one can superpose different solutions 
for the same potential $U_i$.
The nonlinear nature of the Einstein gravity manifests itself through the equation (\ref{eqnu}) 
for the metric
function $\nu$.

\subsubsection{Higher dimensional structure}
The central point in this approach\footnote{Some aspects of the proposal in this work
can be found also in Ref.
\cite{Kudoh:2006xd},
which considers the numerical construction of the five-dimensional black 
rings with two independent angular momenta.} is that the rod structure, as defined above for the $d=5$ case,
can be used also for $d>5$ solutions constructed
within the ansatz (\ref{metric}). 
This fixes the boundary conditions  along the
$z-$axis for the functions $f_i$ and thus the topology of the horizon.

However, note that the interpretation of a rod
as corresponding to a zero thickness  source  with linear mass density $1/2$, placed
on the axis of symmetry
in an auxiliary three-dimensional flat space  is no longer valid for $d>5$.
Also, the relation (\ref{cons}) fails to be satisfied in this case, as one can see already for the simplest case of a 
Minkowski space-time.

A crucial observation here is that, supposing the existence of a power series expansion in $\rho$, 
the Einstein equations imply\footnote{Such an expansion is also required by the regularity
of the Kretschmann scalar at $\rho=0$ (but it does not guarantee it automatically).}
 the following form
of the metric functions $f_i$ close to 
the $z$-axis, valid for any $d\geq 5$
\begin{eqnarray}
\label{rods}
 f_i(\rho,z)=f_{i0}(z)+\rho^2f_{i2}(z)+O(\rho^3),
\end{eqnarray}
where $f_{ik}(z)$ are 
solutions of a complicated set of nonlinear second order ordinary differential equations which we shall not present here. 
Then, similar to the $d=5$ case,
we suppose that the $z-$axis is divided into $N$ intervals--the rods of the solution.
Except for the isolated
points between the rods, one assumes that
 only one of the functions
$f_0(0,z)$, $f_2(0,z)$, $f_3(0,z)$, becomes zero for a given rod, while 
the remaining functions stay finite at $\rho = 0$ in general.
(In fact, if more than one of these functions is going to zero
  for a given $z$ inside a rod,
one can prove following the arguments in \cite{Harmark:2004rm} that
we have a curvature singularity at that point.)
Again, one imposes the condition that the 
$N$ intervals must add up
to give an infinite rod.

A finite timelike rod corresponds to an event horizon, where\footnote{$f_{ik}(z)$ here 
should not be confused with those in (\ref{nrod1}).} 
\begin{eqnarray}
\label{nrod2}
&&f_0(\rho,z)=\rho^2f_{02}(z)+\rho^4f_{04}(z)+\dots,~~f_1(\rho,z)=f_{10}(z)+\rho^2f_{12}(z)+\dots,
\\
\nonumber
&&f_2(\rho,z)= f_{20}(z)+\rho^2f_{22}(z)+\dots,~~f_3(\rho,z)=f_{30}(z)+\rho^2f_{32}(z)+\dots.
\end{eqnarray}
 with $\lim_{\rho\to 0}\rho^2 f_1/f_0=c_3$. As we shall see in the next Subsection, this 
 fixes the Hawking temperature of the solutions. 

For a rod in the $\psi$-direction,
one finds the following expansion of the metric functions as $\rho\to 0$:
\begin{eqnarray}
\label{nrod1}
&&f_0(\rho,z)=f_{00}(z)+\rho^2f_{02}(z)+\dots,~~f_1(\rho,z)=f_{10}(z)+\rho^2f_{12}(z)+\dots,
\\
\nonumber
&&f_2(\rho,z)= \rho^2f_{22}(z)+\rho^4f_{24}(z)+\dots,~~f_3(\rho,z)=f_{30}(z)+\rho^2f_{32}(z)+\dots.
\end{eqnarray}
%
  
 {\small \hspace*{1.5cm}{BH: $S^{d-2}$ horizon} 
 \hspace{4.3cm} 
 {GBR: $S^2 \times S^{d-4}$ horizon$^{~~~~~~~~~~}$ } 
 }
\begin{figure}[ht]
\hbox to\linewidth{\hss%
	\resizebox{6cm}{4cm}{\includegraphics{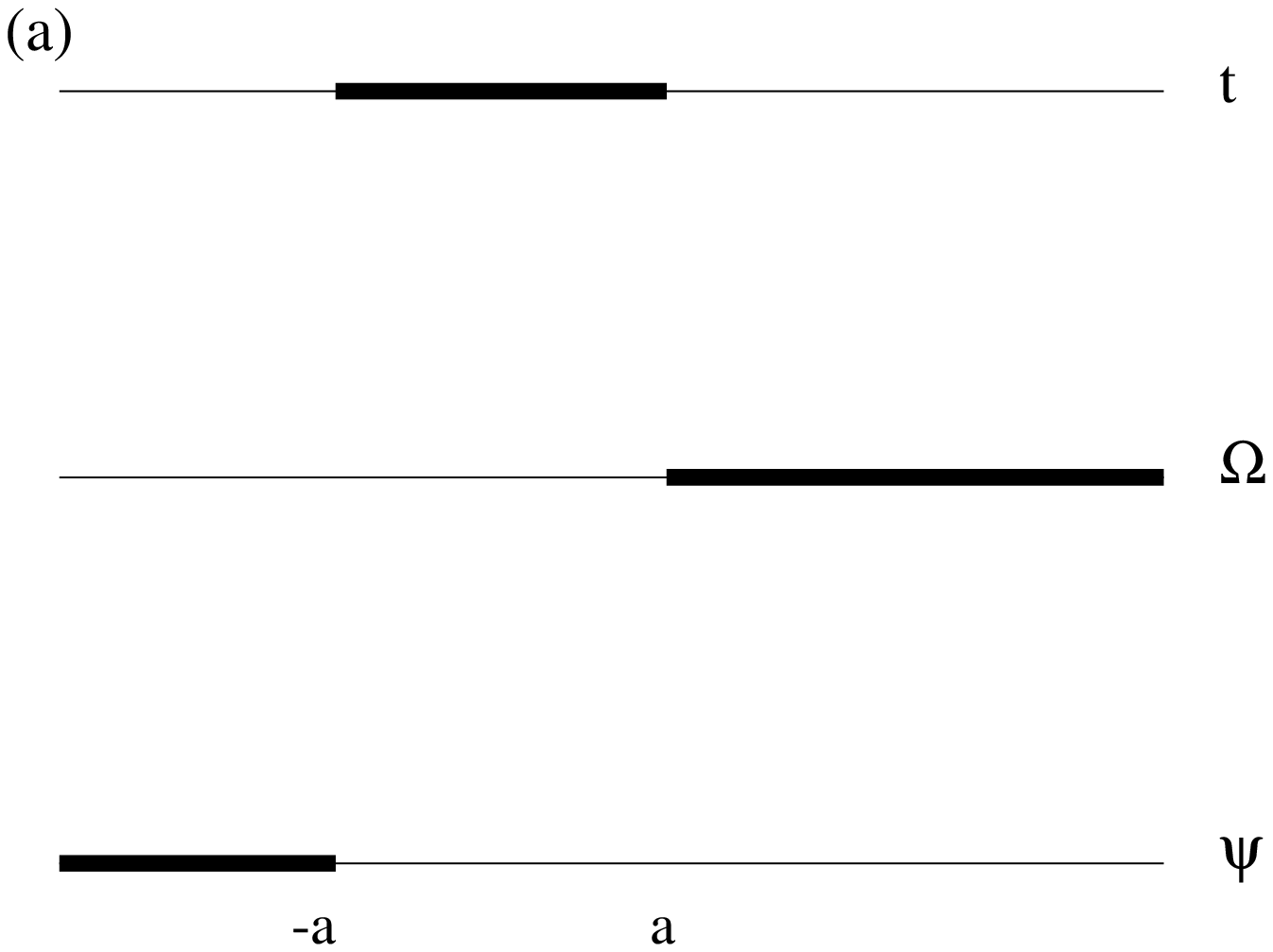}}
\hspace{15mm}%
        \resizebox{6cm}{4cm}{\includegraphics{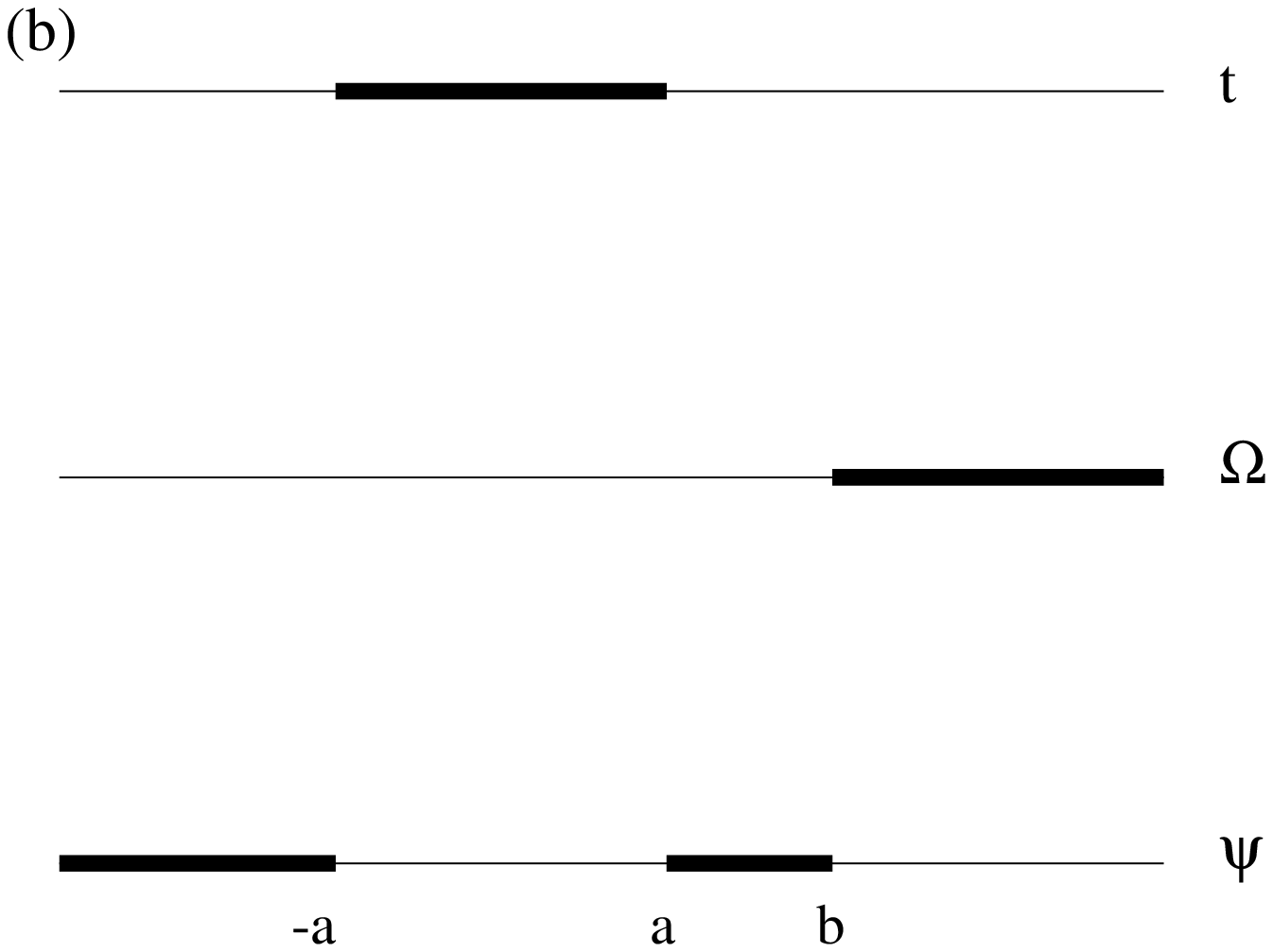}}	
\hss}
\label{rod-sch-ring}
 \end{figure}
\vspace*{0.8cm}
 {\small \hspace*{1.cm}{ GBS: $(S^2 \times S^{d-4})\times S^{d-2}$ horizon} \hspace{1.6cm} {GBD: $(S^2 \times S^{d-4})\times (S^2 \times S^{d-4})$ horizon} }
\begin{figure}[ht]
\hbox to\linewidth{\hss%
	\resizebox{6cm}{4cm}{\includegraphics{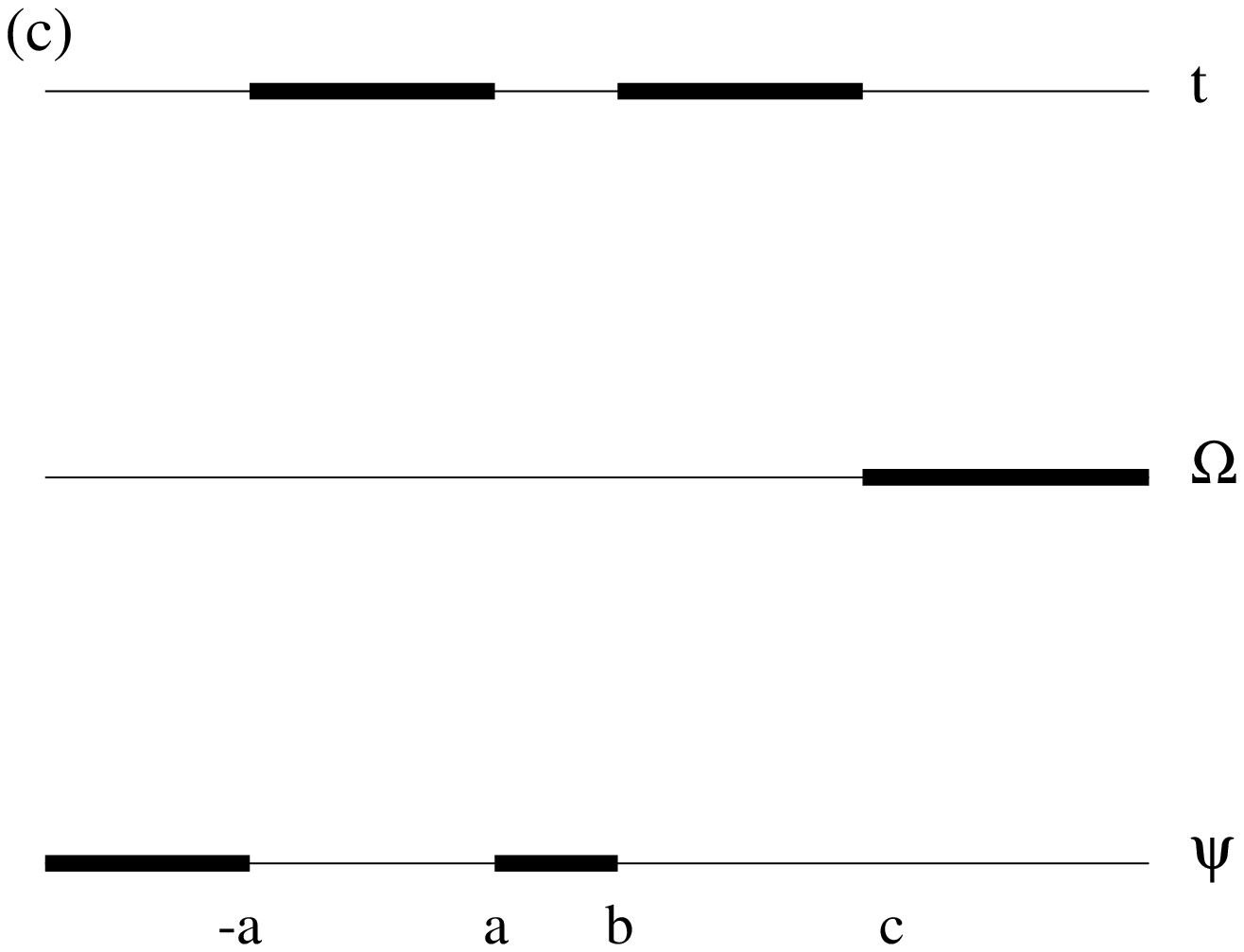}}
\hspace{15mm}%
        \resizebox{6cm}{4cm}{\includegraphics{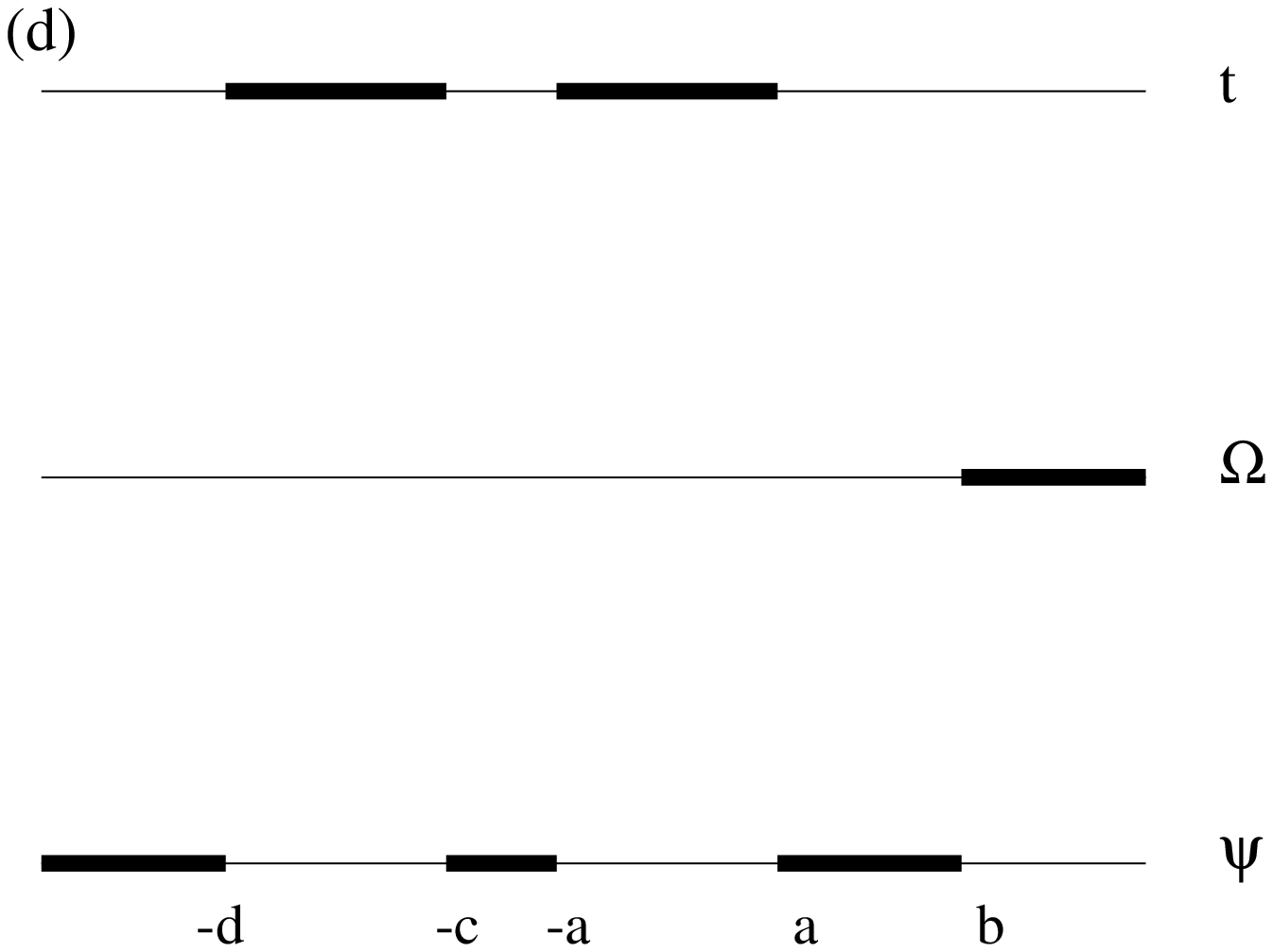}}	
\hss}
 \label{rod-string-saturn}
\end{figure}
\\
\\
{\small {\bf Figure 1.}
Rod structure of several static   black objects in $d\ge 5$ spacetime dimensions. 
These include the Schwarzschild-Tangherlini black hole (BH), the generalized black ring (GBR), 
the generalized black Saturn (GBS) and the generalized black di-ring (GBD). 
In the diagrams the thin lines represent the $z-$axis and the thick lines denote the rods.
}
\vspace{0.7cm}
\\
The important feature here is that
the constraint equation $G_\rho^z=0$ implies  $f_{10}(z)/f_{22}(z)=c_1>0$, $i.e.$
a well-defined periodicity for the  coordinate $\psi$.
The value of $c_1$ is not fixed apriori and follows from the details of the solutions.

For $d=5$, a similar result is found when interchanging $f_2$ and $f_3$, 
$i.e.$ for a rod in the $\varphi$-direction.  
The periodicity of $\varphi$ there is arbitrary, being again fixed by the constraint equation $G_\rho^z=0$, 
$i.e.$ $\lim_{\rho\to 0}\rho^2 f_1/f_3=c_2>0$. 
However, the picture is very different\footnote{Formally, this is a consequence of the presence of the $(d-5)$ coefficients in the eqs. (\ref{eqU2}),  (\ref{eqnu}).}
 for $d>5$, in which case 
 we find
\begin{eqnarray}
\label{nrodOmega}
&&f_0(\rho,z)=f_{00}(z)+\rho^2f_{02}(z)+\dots,~~f_1(\rho,z)=f_{10}(z)+\rho^2f_{12}(z)+\dots,
\\
\nonumber
&&f_2(\rho,z)= f_{20}(z)+\rho^2f_{22}(z)+\rho^4f_{24}(z)+\dots,~~f_3(\rho,z)= \rho^2f_{12}(z)+\dots,
\end{eqnarray}
$i.e.$  the Einstein equations impose the following requirement for an $\Omega-$rod:
$\lim_{\rho\to 0}\rho^2 f_1/f_3=1$,
which is an important new feature.
As we shall argue in the Section 2.5, this condition prevents 
us to construct $d>5$ static black rings (or multi-Schwarzschild-Tangherlini black holes) within the ansatz (\ref{metric}).

Thus, depending on the physical situation we consider,
 the boundary conditions along the $z-$axis are fixed by the above relations.
The obvious boundary conditions for large $\rho,|z|$ are that $f_i$ approach the 
Minkowski background
functions (which are read from \ref{m2}):
 \begin{eqnarray}
\label{Mink}
 f_0(\rho,z)=1,~~
 f_1(\rho,z)=\frac{1}{2\sqrt{\rho^2+z^2}},~
f_2(\rho,z)= \sqrt{\rho^2+z^2}+z,~
f_3(\rho,z)= \sqrt{\rho^2+z^2}-z.~~~~{~~}
\end{eqnarray} 
This is in fact the simplest solution of the Einstein equations in $d\geq 5$ dimensions, as we have seen already in the Section 2.2.1.
There the function $f_2$ vanishes for 
$\rho=0, z<0$, and $f_3$ for $\rho=0, z>0$,
which, in the language of the Weyl formalism, 
 corresponds to two semi-infinite  rods $ [-\infty,0]$ and $ [0,\infty]$.

Similar
to the  $d=5$ case, the topology of an event horizon is fixed by the boundary conditions satisfied by $f_2$ and $f_3$
at the ends of the corresponding (finite) timelike rod.
For example, if either end of this rod continues with rods
of different directions ($\psi$ and $\Omega$), then the event horizon
has an $S^{d-2}$ topology and (for a single horizon) the solution 
is a Schwarzschild-Tangherlini black hole (see Figure 1a).
A black object with $S^2\times S^{d-4}$ topology of the horizon 
is a 'generalized black ring' and
has the metric function $f_2$  vanishing at both ends of 
the  finite timelike rod associated with the horizon, see Figure 1b.
(For $d=5$, this corresponds to the static black ring in \cite{Emparan:2001wk}.)
One can consider as well a 'generalized black Saturn' combining both types of black objects above, with two different horizons
(Figure 1c). 
The rod structure for a
 solution consisting of two black holes, both of them with $S^2\times S^{d-4}$ topology of the horizon (thus a 'generalized 
 black diring')
is shown in Figure 1d.
The basic properties of these objects are discussed in the next Section.

It is tempting to conjecture that, similar to the $d=5$ case \cite{Hollands:2007aj},
a $d>5$  solution within the ansatz (\ref{metric}), is uniquely specified by its rod structure. 

\subsubsection{Physical quantities}

The ADM mass $M$ of the solutions\footnote{The discussion here follows the
general formalism in \cite{Herdeiro:2009vd}, which contains also several applications for
$d=4,5$ exact solutions.} can be read from the asymptotic
expression for the metric function $g_{tt}$
\begin{eqnarray}
\label{gtt}
-g_{tt}=f_0\sim 1-\frac{16 \pi M}{(d-2)V_{d-2}(\rho^2+z^2)^{(d-3)/4}}+\dots~.
\end{eqnarray}%
Supposing we have an event horizon at $\rho=0$  for some $z_1\leq z\leq z_2$,
the horizon metric is given by\footnote{If there are several horizons, then one should write
such an expansion for each of them.}
\begin{eqnarray}
d\sigma^2=f_1(0,z)dz^2+f_2(0,z)d\psi^2+f_3(0,z)d\Omega_{d-4}^2.
\end{eqnarray}
Two quantities associated with the event horizon are
the event horizon area $A_H$ and
the Hawking temperature. For the metric ansatz (\ref{metric}) these are given by
 \begin{eqnarray}
\label{AHTH}
A_H=\Delta \psi V_{d-4}\int_{z_1}^{z_2}dz \sqrt{f_1(0,z)f_2(0,z)f_3^{d-4}(0,z)}, ~~~
T_H=\frac{1}{2\pi}\lim_{\rho\to 0} \sqrt{\frac{f_{0}(\rho,z)}{\rho^2 f_{1}(\rho,z)}},
\end{eqnarray} 
where $V_{d-4}$ is the area of the unit sphere $S^{d-4}$ and $\Delta \psi$ the periodicity of the 
angular coordinate $\psi$ on the horizon.

A solution with $n$ different event horizons satisfies  the Smarr law
\begin{eqnarray}
\label{smarr}
M=\frac{d-2}{4(d-3)}\sum_{k=1}^n T_H^{(k)} A_H^{(k)}~.
\end{eqnarray}

Considering now the case of a space-like $\psi-$rod for some $z_3\leq z\leq z_4$,
one starts by writing the line element on this $(d-2)$-dimensional surface ${\Sigma}$
\begin{eqnarray}
d\sigma^2=f_1(0,z)dz^2+ f_3(0,z)d\Omega_{d-4}^2- f_0(0,z)dt^2.
\end{eqnarray}
The first quantity of interest is the proper
length of the rod
 \begin{eqnarray}
\label{L}
 L=\int_{z_3}^{z_4}dz\sqrt{f_1(0,z)},
\end{eqnarray}
which,   for a finite rod, differs from the coordinate distance $\Delta z=z_4-z_3$ (although it is proportional to it).

All solutions in this work possess a conical singularity  for (at least) a region of the $z$-axis.
 To define a
conical singularity for a rotational axis with angle $\psi$, one computes
the proper circumference $C$ around the axis and its proper radius $R$ and
defines
\begin{eqnarray}
\alpha&=&\frac{dC}{dR}\bigg|_{R=0}=\lim_{\rho\rightarrow 0}\frac{\sqrt{%
g_{\psi\psi}}\Delta\psi}{\int_0^{\rho}\sqrt{g_{\rho\rho}}d\rho}%
=\lim_{\rho\rightarrow 0}\frac{\partial_{\rho}\sqrt{g_{\psi\psi}}%
\Delta\psi}{\sqrt{g_{\rho\rho}}},
\end{eqnarray}
where $\Delta\psi$ is the period of $\psi$. 
The asymptotic flatness implies $\Delta\psi=2 \pi$. Then the presence of a conical
singularity is now expressed\footnote{Note that, in some sence, 
fixing $\delta$ is the analogue of computing the Hawking temperature on the Euclidean section.} by means of:
\begin{eqnarray}
\label{delta}
\delta&=&2\pi-\alpha=2\pi\left(1-\lim_{\rho\rightarrow 0} \sqrt{\frac{f_{2}(\rho,z)}{\rho^2f_{1}(\rho,z)}}\right),
\end{eqnarray}
such that $\delta>0$ corresponds to a conical deficit,
while $\delta<0$ corresponds to a conical excess.
A conical deficit  can be interpreted as a `string' stretched along  on a certain segment of the $z-$axis, 
while a conical  
excess is a `strut' pushing  apart the rods connected to that segment (in fact, the `struts' and `strings'
 are $(d-3)$-dimensional surfaces, $i.e.$ higher dimensional
 analogues of the $d=4$ cosmic strings). 
Also,  a constant rescaling  of
$\psi$  can be used to eliminate possible conical singularities on a
given segment, but in general, once this is fixed, there will remain
conical singularities at other $\psi$-segments.  
For all solutions in this work, we have prefered to set the conical singularity on a finite $\psi$-rod
such that our solutions are asymptotically flat, meaning that $\Delta\psi=2 \pi$.
Moreover, in the presence of a conical singularity,
the manifold ${\cal M}$ naturally factorizes as ${\cal M}=\mathcal{C}_\alpha \times \Sigma$, where $\mathcal{C}_\alpha$ is the two-dimensional conical surface $\rho-\psi$
and $\Sigma$ is the remaining $(d-2)$-dimensional  surface, which may be seen as the world-volume of the conical defect \cite{Fursaev:1995ef}.

For practical reasons, we have found it convenient to introduce the quantity
\begin{eqnarray}
\label{rel-delta}
\bar \delta=\frac{\delta/(2\pi)}{1-\delta/(2\pi)},
\end{eqnarray}
which has a finite range ($\bar \delta \to -1$ for $\delta \to \pm \infty$)  and measures the 
'relative angular deficit/excess'.

  As argued in \cite{Herdeiro:2009vd,Herdeiro:2010aq},
the asymptotically flat black objects with conical singularites still admit a thermodynamical description.
  Moreover, when working with the appropriate set of thermodynamical variables, the Bekenstein-Hawking law still holds for
such solutions. The mass-energy which enters the first law of thermodynamics 
does not, however, coincide with the ADM mass; it differs from the latter by the energy associated with the conical singularity, as seen by an asymptotic, static observer.

The Ricci scalar of ${\cal M} $, ${}^{(\alpha)}R$, can be represented near $\Sigma$ in the following local form
\cite{Fursaev:1995ef}:
\begin{eqnarray}
\label{R-tot}
{}^{(\alpha)}R=R+2(2\pi -\alpha) \delta_{\Sigma},
\end{eqnarray}
where 
$R$ is the curvature computed in the standard way on the
smooth domain ${\cal M}/\Sigma$ of ${\cal M}$.
Here, 
$\delta_{\Sigma}$ is the Dirac $\delta$-function, with $\int_{{\cal M} } f \delta_\Sigma=\int_{\Sigma} f$.
A direct integration of (\ref{R-tot}) leads to \cite{Regge}
\begin{eqnarray}
\label{new}
\int_{{\cal M} }{}^{(\alpha)}R=\int_{{\cal M} /\Sigma}R+2(2\pi-\alpha) Area,
\end{eqnarray}
where $Area$ is the area of $\Sigma$, $i.e.$ the space-time area of the conical singularity's world-volume. 
For the metric ansatz in this work, the expression of the extensive parameter $Area$ is
\begin{eqnarray}
\label{area}
Area=  \beta V_{d-4} \int_{z_3}^{z_4}dz\sqrt{f_0(0,z)f_1(0,z) f_3^{d-4}(0,z)},
\end{eqnarray}
$\beta=1/T_H$ being the periodicity of the Euclidean time. 

In the presence of conical singularities,
  the solutions cannot  be viewed as vacuum   solutions and 
there is a matter source (the struts) which supports the conical singularities.
The   stress energy tensor  associated 
with the singularity can be computed by using the Einstein equations $G_{ij}=8\pi T_{ij}$.
The results in \cite{Fursaev:1995ef} also show that the singular part of the Ricci tensor has components only in the $\rho-\psi$ plane,
such that $R_i^j=0$ for the remaining components.
It follows that the only non-vanishing components of $T_{i}^j$
are 
\begin{eqnarray}
\label{tik}
T_i^j=-\delta_{i}^j\frac{1}{8\pi }(2\pi-\alpha)\delta_{\Sigma},~~~{\rm with}~~(i,j)\neq (\rho,\varphi).
\end{eqnarray}
A direct consequence of this result is that the conical deficit/excess as defined by (\ref{delta}),
 $\delta=2\pi -\alpha$,
corresponds to 
the pressure exerted by the strut. This is found  by integrating the $T_z^z$-component over $\mathcal{C}_\alpha$
\begin{eqnarray}
\label{res1}
P=\int_{\mathcal{C}_\alpha}  T_z^z =-\frac{\delta}{8\pi } .
\end{eqnarray}
Moreover, as seen from (\ref{tik}),
the energy density $\mu$ of the matter source supporting the conical singularity
is also $\mu=-{\delta}/{8\pi }$.
Thus $\delta<0$ (the case of the solutions in this work)
corresponds to a negative energy density source.

Another quantity of interest is the total energy associated with the strut as seen by a static
observer placed at infinity. 
This is found  by integrating the $T_t^t$-component over a $t=t_0={\rm constant}$ hyper-surface, 
\begin{eqnarray}
\label{Eint}
E_{int}=-\int_{t=t_0} T_t^t = \frac{\delta}{8\pi } \frac{Area}{\beta}=-P{\cal A},
\end{eqnarray}
 where we have defined
\begin{eqnarray}
\label{def-A}
{\cal A}\equiv \frac{Area} {\beta}.
\end{eqnarray}

Thermodynamics of a system with a
conical singularity in the bulk can be approached\footnote{In fact, 
it is possible to define the thermodynamic quantities also with the conical singularity stretching towards the boundary,
see $e.g.$ \cite{Astefanesei:2009mc}.} by using the path-integral formulation of quantum
gravity \cite{GibbonsHawking1}. 
The first step here is to evaluate the total tree level Euclidean action of the system.
The new feature introduced by the conical singularity is to add an extra contribution to
this quantity which can be 
evaluated by using the relation (\ref{new}).
  Then the  total action is 
 \begin{eqnarray}
 \label{tot-action}
I=I_0-\frac{\delta}{8\pi}{\cal A}\beta, \label{action}
\end{eqnarray}
 where $I_0$ is the usual tree level action \cite{GibbonsHawking1}
found when neglecting the conical singularity. 

As argued in \cite{Herdeiro:2009vd}, the first law of thermodynamics for static vacuum solutions with a  conical singularity reads
  \begin{eqnarray}
\label{firstlaw} 
d{\cal M}=T_H dS + {\cal T} d {\cal A},
  \end{eqnarray} 
  where ${\cal A}$ (as defined by (\ref{def-A})) is the extensive parameter which takes into account 
  the presence of the conical singularity and ${\cal T}=P=- {\delta}/{8\pi}$
 is the associated ``tension".  
 
  In a canonical ensemble, one keeps the Hawking temperature $T_H$ and the  extensive parameter ${\cal A}$ fixed.
The free energy $F$ is 
 \begin{eqnarray}
\label{F} 
 F[T_H,{\cal A}]=T_H I={\cal M}-T_H S.
  \end{eqnarray}
Then
the entropy $S$, mass ${\cal M}$ and tension ${\cal T}$ of the physical system are given by
\begin{eqnarray}
\label{r1} 
 S=-\frac{\partial F}{\partial T_H}\bigg|_{{\cal A}},~~
~{\cal M}=F+T_HS,~~{\cal T}= \frac{\partial F}{\partial {\cal A}}\bigg|_{T_H}.
 \end{eqnarray}  
 This approach has been applied in \cite{Herdeiro:2009vd}
 for several $d=4,5$ static
 solutions with conical singularities in the bulk which are known 
 in closed form, the effects of rotation being considered in \cite{Herdeiro:2010aq}.
The results there show that $S=A_H/4$ in all cases, as expected.

In principle, one can use the same approach to discuss the thermodynamics of the solutions in this work.
The only complication appears for multi-black objects.
For thermal equilibrium, the individual black holes should have the same Hawking temperature.
Moreover, if there are several different finite $\psi$-rods, then the corresponding tension parameters $\delta_i$
should be equal.
 
\subsection{Numerical procedure}
All new solutions in this work are found within a nonperturbative approach
by solving numerically a set of four nonlinear partial
differential equations.
These equations  have dependence on two variables and are subject to suitable boundary conditions
which follow  from the required rod  structure and asymptotic flatness.

In this scheme, the input parameters provided to the solver are the positions of the rods 
and the value $d$ of the spacetime dimension.
The numerical integration eventually converges and provides an output
consisting of the   functions $f_i$ and their first and second derivatives
with respect to $\rho$ and $z$.
The relevant physical data ($e.g.$ the ADM mass, the Hawking temperature(s) etc.)
are computed from this numerical output.

In practice, we 
have found it convenient to take
\begin{eqnarray}
\label{ans1}
f_i=f_i^{0}F_i ,
\end{eqnarray}
where $f_i^{0}$ are some background functions\footnote{Although sometimes it 
is not stated explicitly, the use of a suitable set of background
functions is a common feature of 
numerical studies in general
relativity. For example, for axisymmetric configurations, this factorizes the trivial
angular dependence of the metric functions.
}, given by the metric functions of a $d=5$ 
 solution with some rod structure. 
The numerical computation is performed working with the functions $F_i$.
The advantage of this approach is that, since $F_i>0$ everywhere, the functions $f_i$ will automatically satisfy the
desired rod structure also for the $d>5$ solutions.
Moreover, this choice `absorbes' the divergencies of the functions $f_2$ and $f_3$ as $\rho\to \infty$,  $z\to \pm \infty$ 
originating in the imposed asymptotic behaviour (and also the divergencies of $f_1$
at the end of the finite rods\footnote{Note that for all $d=5$
solutions the function $f_1(0,z)$ behaves as $1/|z-z_i|$ as
$z\to z_i$, with $z_i$ the end point of a finite rod.
This behaviour is recovered for all $d>5$ solutions in this work.}). 

The equations satisfied by $F_i$ can easily be derived from those satisfied by $f_i$.
As for the boundary conditions, the relations (\ref{nrod1}), (\ref{nrod2})
imply 
\begin{eqnarray}
\label{bc-psi}
\nonumber
&&\partial_\rho F_i|_{\rho=0}=0, 
\end{eqnarray}
for~a $\psi$-rod or an event horizon, and
\begin{eqnarray}
\label{bc-Omega}
&&\partial_\rho F_0|_{\rho=0}=\partial_\rho F_1|_{\rho=0}=
\partial_\rho F_2|_{\rho=0}=0,~~F_1|_{\rho=0}=F_3|_{\rho=0} ,
\end{eqnarray}
for an  $\Omega$-rod.
 As $\rho\to \infty$ or $z\to \pm \infty$ one imposes the obvious condition $F_i=1$.

The constraint equation $G_\rho^z=0$ implies 
 $F_2/F_1=const.$ on a $\psi$-rod.
Now, to be consistent with the assumption of asymptotic flatness, one finds $const.=1$
for a $\psi-$rod extending to infinity. The value of this ratio for a finite rod 
results from the numerical results. 

In the numerical calculations, one starts with the corresponding $d=5$ solution, ($i.e.$ $F_i=1$),
and slowly increases the parameter $d$.
This leads to nontrivial deformations of $F_i$.
The physical values of $d$ are integers.
For all types of black objects in this work, we have studied   solutions in $d=6$ dimensions in a systematic way. 
A number of $d=7$ solutions were also obtained but we did not fully investigate  this case  
except for single black objects.
We think that solutions with $d>7$ are also very likely to exist;
however, we could not construct them within the approach in this work
and their study may require a different numerical method\footnote{We believe this is not a  worrisome
aspect. For example, a similar situation
was found in the past for nonuniform black strings in Kaluza-Klein theory, which at the beginning could be  constructed numerically 
for 
$d=6$ only \cite{Wiseman:2002zc}. However, subsequent work has managed to extend these solutions to $d=5$
\cite{Kleihaus:2006ee} and $7\leq d\leq 11$ \cite{Sorkin:2006wp}.
Kaluza-Klein caged black holes were so far constructed within a nonperturbative approach for $d=5,6$ only \cite{Headrick:2009pv}.}.

In practice, one introduces also new compactified coordinates 
$x=\rho/(1+\rho)$, $u=\arctan(z)$, with $0\leq x \leq 1$, $-\pi/2\leq u \leq \pi/2$. 
The equations for $F_i$ are then discretized on a non-equidistant grid in
$x$ and $u$. 
Typical grids used have sizes around $80 \times 150$,
covering the full integration region. 
(See \cite{schoen} and \cite{kk}
for further details and examples for the numerical procedure.) 
All numerical calculations  
are performed by using the program FIDISOL/CADSOL \cite{schoen},
which uses a  Newton-Raphson method\footnote{Thus providing a good
initial guess solution
is essential
for the convergence of the numerical process.}.
This software provides also an error estimate for each unknown function.
For $d=6$, 
the typical  numerical error 
for the functions is estimated to be lower than $10^{-3}$.
This error increases to several percent for most of the $d=7$ solutions. 

Perhaps the most problematic aspect of this approach is that the derivatives of the functions
$F_i$ diverge at the end points of the rods (although the functions $F_i$ are smooth everywhere).
However, this appears already for the $d=5$ exact solutions\footnote{For a $d=5$ black ring, 
the derivatives of the metric functions
are finite everywhere when writing the solution in a ring-coordinate system.
Although one can devise a ring-coordinate system (or more complicated versions adapted to the
solutions in this work) for $d>5$ as well, we could not use
it in practice within a numerical scheme. The main problem seems to be that the spacelike 
infinity is approached for a single point in that case.
}  and 
is an unavoidable feature of the Weyl-type coordinate system (\ref{metric}).
To assure that these divergencies are coordinate
artifacts,  we have verified that the Kretschmann scalar stays finite everywhere, in particular at $\rho=0$
(here we ignore  the $\delta$-Dirac terms in the expression of Riemann tensor in the presence 
of a conical singularity \cite{Fursaev:1995ef}). 
However, this non-smooth behaviour makes the numerics more involved as compared to
other problems which were solved with similar methods and the same software.
In particular, this approach requires a careful construction of the mesh 
in the region close to the end points of the rods. 
One should also remark that the presence of a conical singularity for the solutions in this work has a rather neutral
effect on the numerics, since the solver does not notice it directly.

Furthermore, as a test of the software, we have recovered numerically the $d=6,7,8$ exact Schwarzschild-Tangherlini black 
hole starting with the $d=5$ solution, which, for the ansatz  (\ref{metric}) possesses
already the basic features of other more complex black objects.
Another accuracy test of our solutions was provided by the Smarr relation (\ref{smarr}).

A further numerical test is presented in Appendix B, where 
the $d=5$ balanced black ring is recovered numerically.
The background functions in this case
correspond to those of the static exact solution.

\newpage
 \setlength{\unitlength}{1cm}
%
\subsection{The issue of $d>5$ black ring solutions}
One may expect that the approach proposed in this work 
can be used to construct
a static black ring as well.
But to be able to do so 
the  metric has to have an unbroken $SO(d-3)$ symmetry group. The approximate solutions 
for \textit{thin} black rings \cite{Emparan:2007wm}
 do indeed preserve this symmetry and are cohomogenity-2. Therefore, we expect
their static metrics in the $(\rho,z)$ coordinates to be within our ansatz.

The rod structure of these solutions in the generalized approach we are considering would  
consist  of 
a semi-infinite space-like $\Omega-$rod $[-\infty,-a]$
(with $f_3(0,z)=0$ there),
a finite time-like rod $[-a, a]$ (the horizon, $f_0(0,z)=0$), a second (finite)
$\Omega$-rod $[a,b]$ and a  
semi-infinite space-like $\psi$-rod $[b,\infty]$ (with a vanishing $f_2(0,z)$). 
Thus, it is obvious that the horizon topology of such an object  would be 
$S^{d-3}\times S^1$.

 In principle, one may consider as well multi-black objects with several $\Omega-$rods,
the simplest case corresponding to a two Schwarzschild-Tangherlini black hole system, both with a $S^{d-2}$ topology
of the horizon. 
Indeed, such solutions are known in closed form in four \cite{BW,IsraelKhan}, and five dimensions \cite{Tan:2003jz}.


On general grounds, we know that
higher dimensional black rings are possible and
 Ref.~\cite{Emparan:2007wm} found an approximate construction of such solutions
based on the matched asymptotic expansion method.
We note that the solutions in 
Ref.~\cite{Emparan:2007wm} describe {\it balanced} spinning black rings,
but unbalanced and, in particular, static solutions should also exist. 
The analysis in Ref.~\cite{Emparan:2007wm} shows, 
that when the equilibrium condition is not
satisfied, naked singularities arise in the plane of the ring.
For higher-dimensional rings, the singularities are thus
necessarily stronger than in five dimensions, where conical
singularities suffice to balance the rings.

This observation is in correspondence with our analysis in (\ref{nrodOmega}), 
which shows that
for $d>5$ one cannot assign to $\Omega_{d-4}$ 
 any higher dimensional counterpart of a conical defect since  $\lim_{\rho\to 0}\rho^2 f_3/f_1=1$ in that case. 
One may conclude that the
static black rings do not admit at $\rho=0$ a power series expansion of the form (\ref{rods}),
since the static limit of a higher 
 dimensional back ring
will possess a stronger singularity than a conical one.

Clearly, for the envisaged numerical construction of such higher-dimensional
static black rings
the presence of curvature singularities provides a strong obstacle.
\footnote{
Moreover, a $d-$dimensional static 
black ring might not be described within the ansatz (\ref{metric})
because the sphere $\Omega_{d-4}$ could be deformed in that case 
($i.e.$ 
the metric functions would have a dependence on (at least) one coordinate on that sphere).}
Therefore we conclude that we cannot study $d-$dimensional static black rings 
within the metric (\ref{metric}) in a straightforward manner
and a different approach seems to be required.
\newpage
\begin{figure}[ht]
\hbox to\linewidth{\hss%
	\resizebox{8cm}{6cm}{\includegraphics{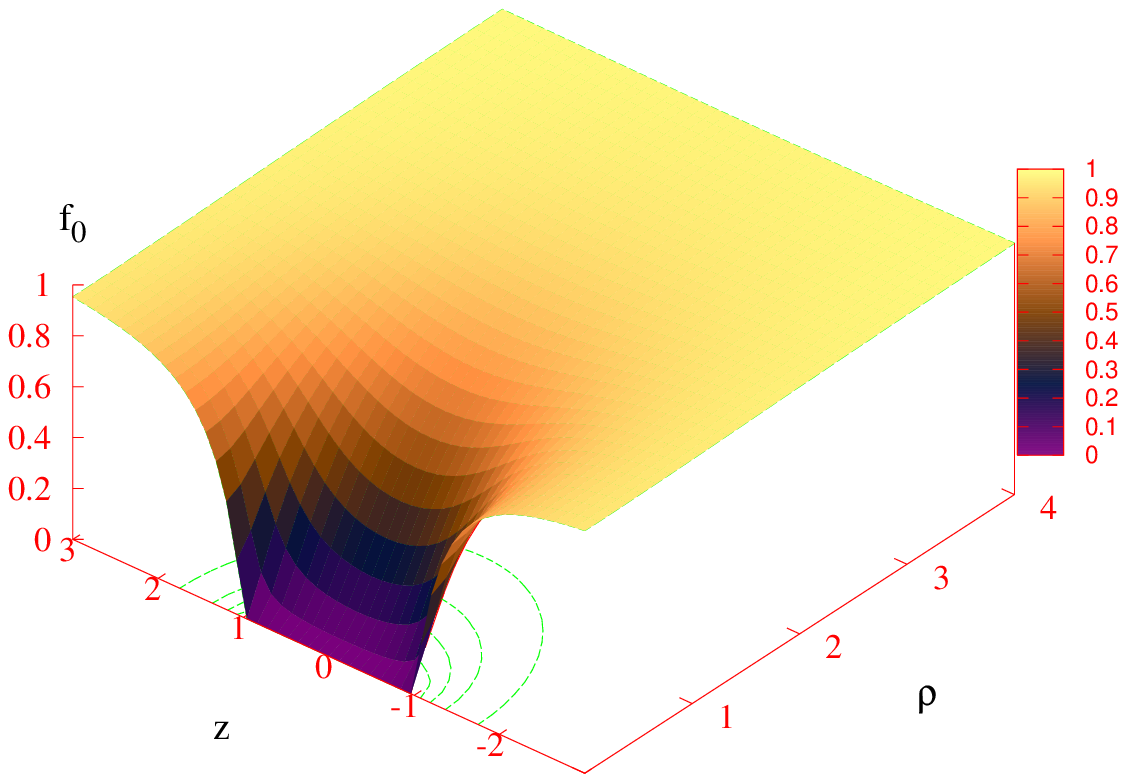}}
\hspace{15mm}%
        \resizebox{8cm}{6cm}{\includegraphics{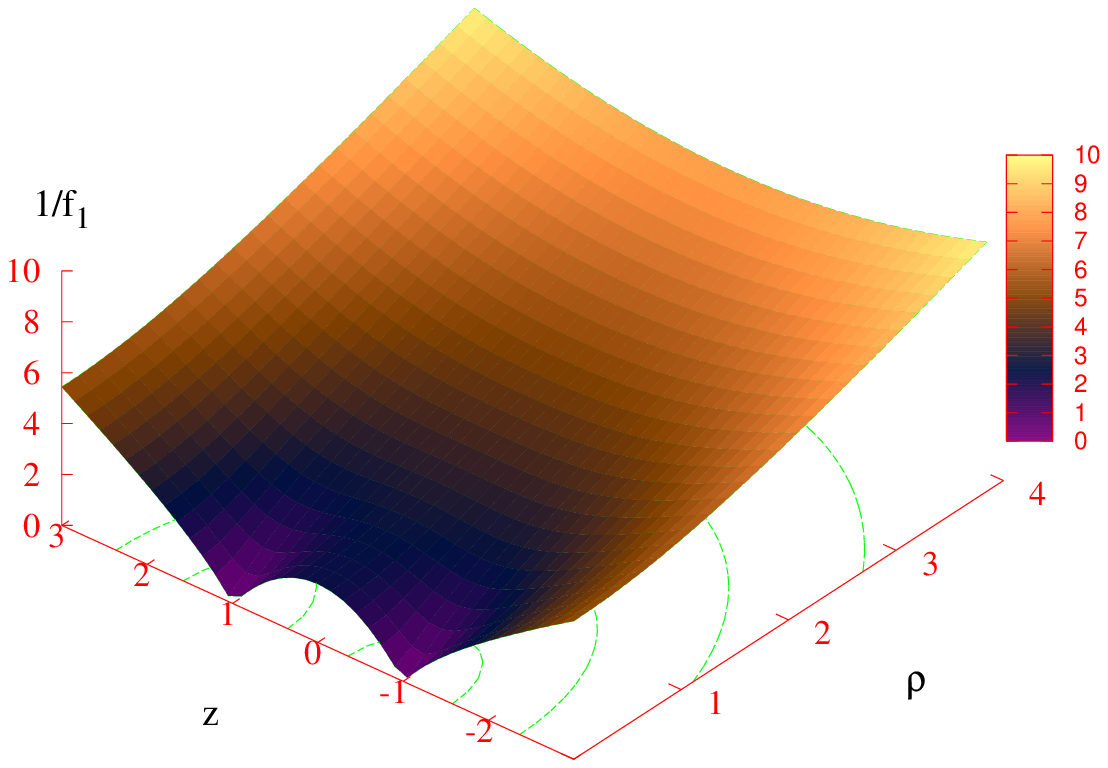}}	
\hss}
 \end{figure}
\vspace*{-1.1cm}
 {\small \hspace*{3.cm}{\it  } }
\begin{figure}[ht]
\hbox to\linewidth{\hss%
	\resizebox{8cm}{6cm}{\includegraphics{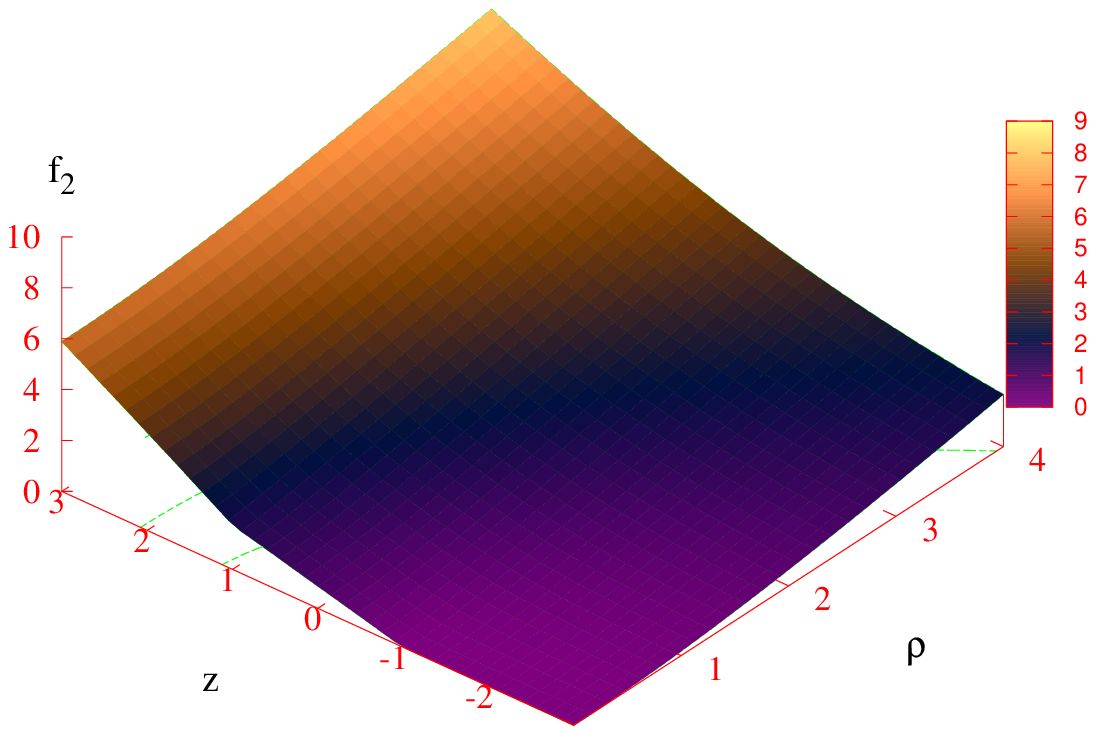}}
\hspace{15mm}%
        \resizebox{8cm}{6cm}{\includegraphics{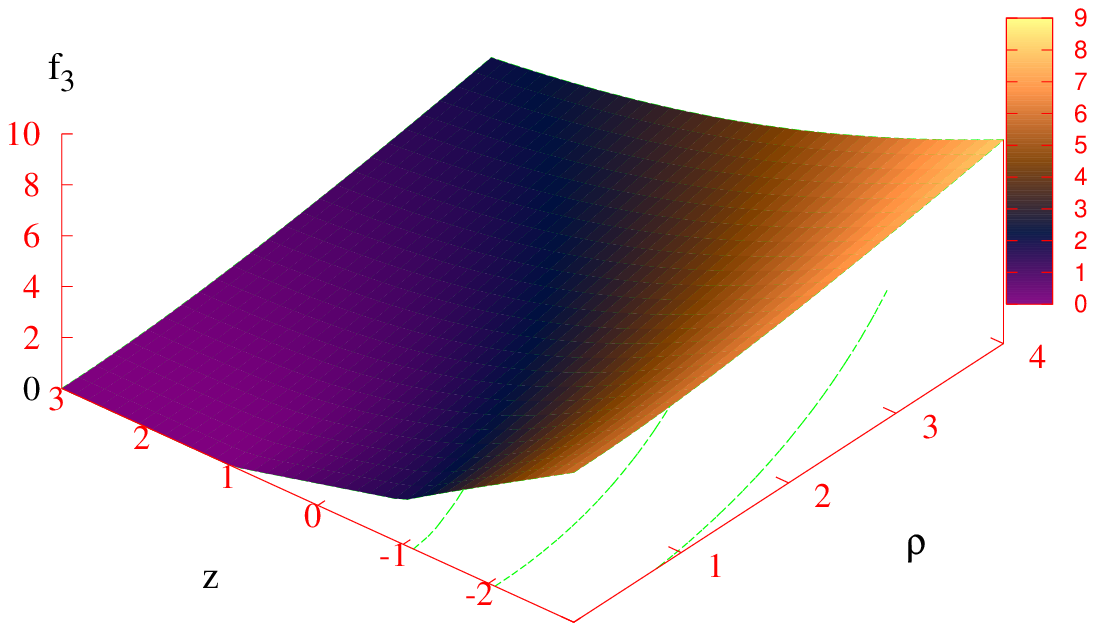}}	
\hss}
\end{figure}
\\
{\small {\bf Figure 2.}
  The profiles of the metric functions $f_i$ for a  
$d=8$ Schwarzschild-Tangherlini solution with $a=1$.
} 
\\ 
\\

\section{General results for static black holes}

\subsection{Uni-horizon black holes}

\subsubsection{Schwarzschild-Tangherlini black hole: $S^{d-2}$ horizon}\label{sec:ST}
 The simplest example of a $d\geq 5$ black object that can be studied
within this approach corresponds to the Schwarzschild-Tangherlini black hole.
In Section \ref{sec:Schw} we have shown its  form within the metric ansatz  (\ref{metric}).

The rod 
structure of this black hole consists of  (see Figure 1)
 \begin{itemize}
\item A semi-infinite space-like rod $[-\infty,-a]$
in the $\partial / \partial \psi$ direction (with $f_2(0,z)=0$ there),
\item A finite time-like rod $[-a,a]$
in the $\partial / \partial t$ direction (the horizon, $f_0(0,z)=0$),
\item A semi-infinite space-like rod $[a,\infty]$
in the $\Omega$-direction (with a vanishing $f_3(0,z)$).
\end{itemize}
Thus the topology of the horizon is $S^{d-2}$ as required.
Requiring the absence of a conical singularity imposes a 
periodicity $2 \pi$ for the coordinate $\psi$.
The only parameter here is $a>0$, which fixes the ADM mass of solutions (the only global
charge for a Schwarzschild-Tangherlini black hole).

To get the feeling for the type of functions we will be dealing with later in the numerical computations,
we show in Figure 2 the form of the metric components, namely the metric functions $f_i$, exhibiting 
this rod structure for a typical $d=8$ Schwarzschild-Tangherlini solution.

In principle, most of the physically relevant 
properties of the Schwarzschild-Tangherlini  black hole 
can also be rederived within the metric ansatz (\ref{metric}).
However, the required computation is much more difficult for that coordinate system.

\subsubsection{Generalized black ring: $S^2\times S^{d-4}$ horizon}\label{sec:GBR}

Very likely, these are the simplest $d>5$ black objects with
a nonspherical topology of the horizon which can be constructed
within a nonperturbative approach.

The rod structure in this case 
consists of  (see Figure 1b)
 \begin{itemize}
\item A semi-infinite space-like rod $[-\infty,-a]$
in the $\partial / \partial \psi$ direction (with $f_2(0,z)=0$ there),
\item A finite time-like rod $[-a,a]$
in the $\partial / \partial t$ direction (the horizon, $f_0(0,z)=0$),
\item A a second (finite)
$\psi$-rod $[a,b]$ (with $f_2(0,z)=0$),
 \item A semi-infinite space-like rod $[b,\infty]$
in the $\Omega$-direction (with a vanishing $f_3(0,z)$).
\end{itemize}
The problem has two length scales $a$ and $b$, roughly corresponding to the event horizon radius and the
radius of the round $\Omega_{d-4}$-sphere.
These two input parameters fix the ADM mass $M$ and the tension $\delta$ which are the global charges here.

The basic properties of these  solutions  were discussed in \cite{Kleihaus:2009wh}
for $d=6,7$. For $d=5$, they correspond to the static 
black ring found in \cite{Emparan:2001wk} (its basic properties are reviewed in Appendix A.1).
The $d>5$ configurations share the basic properties
of the  five dimensional counterparts. This is why we propose  to call them
'generalized black rings'.

Unfortunately, all static solutions with a $S^2\times S^{d-4}$ horizon topology 
  have a conical excess $\delta$ on the finite $\psi-$rod.
In terms of the dimensionless ratio $a/b$,
 the  generalized black 
rings  smoothly interpolate between two limits. 
 First, as $a/b \to 1$, one
finds $\bar \delta \to  -1 $ 
($i.e.$ the conical excess $\delta \to -\infty$) and the Schwarzschild-Tangherlini metric is approached (the
finite $\psi$-rod vanishes). As the second $\psi-$rod extends to infinity ($a/b\to 0$), the 
radius   on
 the  horizon  of the round $S^{d-4}$-sphere increases and asymptotically it
becomes a $(d-4)$-plane, while  $\bar \delta\to 0$. 
After a suitable rescaling, one finds the 
four dimensional Schwarzschild black hole uplifted to $d$ dimensions  ($i.e.$ a black $(d-4)$-brane).
%
\newpage
\begin{figure}[ht]
\hbox to\linewidth{\hss%
	\resizebox{8cm}{6cm}{\includegraphics{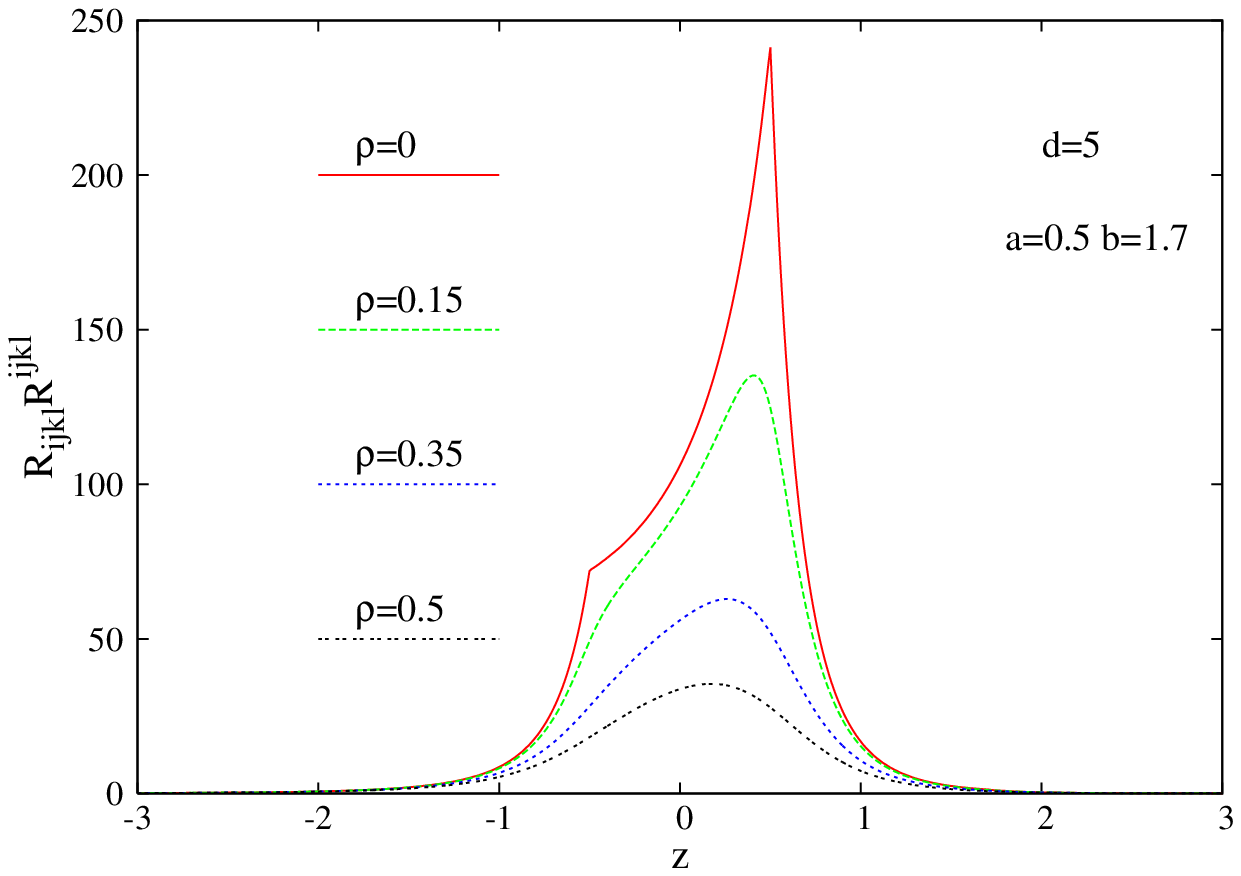}}
\hspace{15mm}%
        \resizebox{8cm}{6cm}{\includegraphics{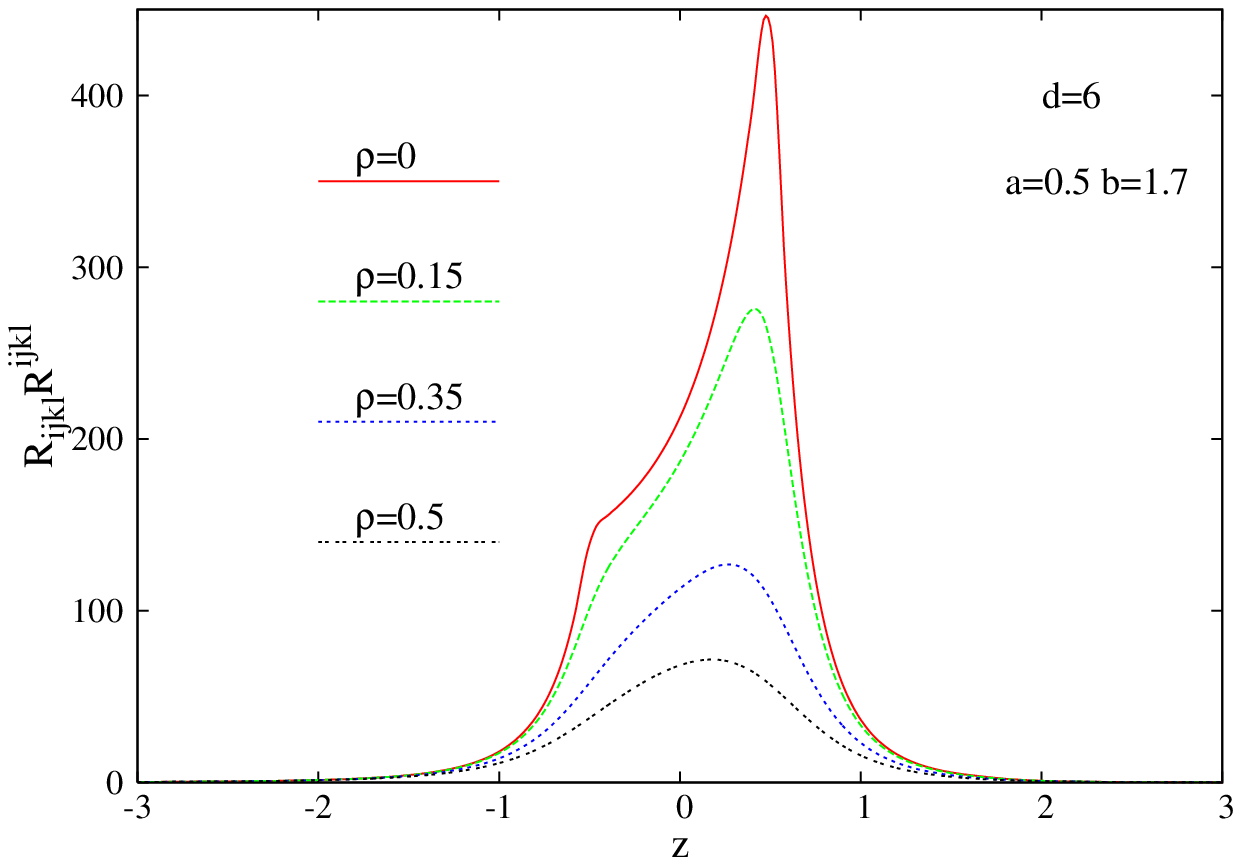}}	
\hss}
 \end{figure}
  \vspace*{0.3cm}
{\small {\bf Figure 3.}
  The Kretschmann scalar $R_{ijkl}R^{ijkl}$ is plotted as a function of $z$ 
for several values of $\rho$ for $d=5,6$ black objects with $S^2\times S^{d-4}$ 
 event horizon topology.  
} 
\\

\begin{figure}[ht]
\hbox to\linewidth{\hss%
	\resizebox{8cm}{6cm}{\includegraphics{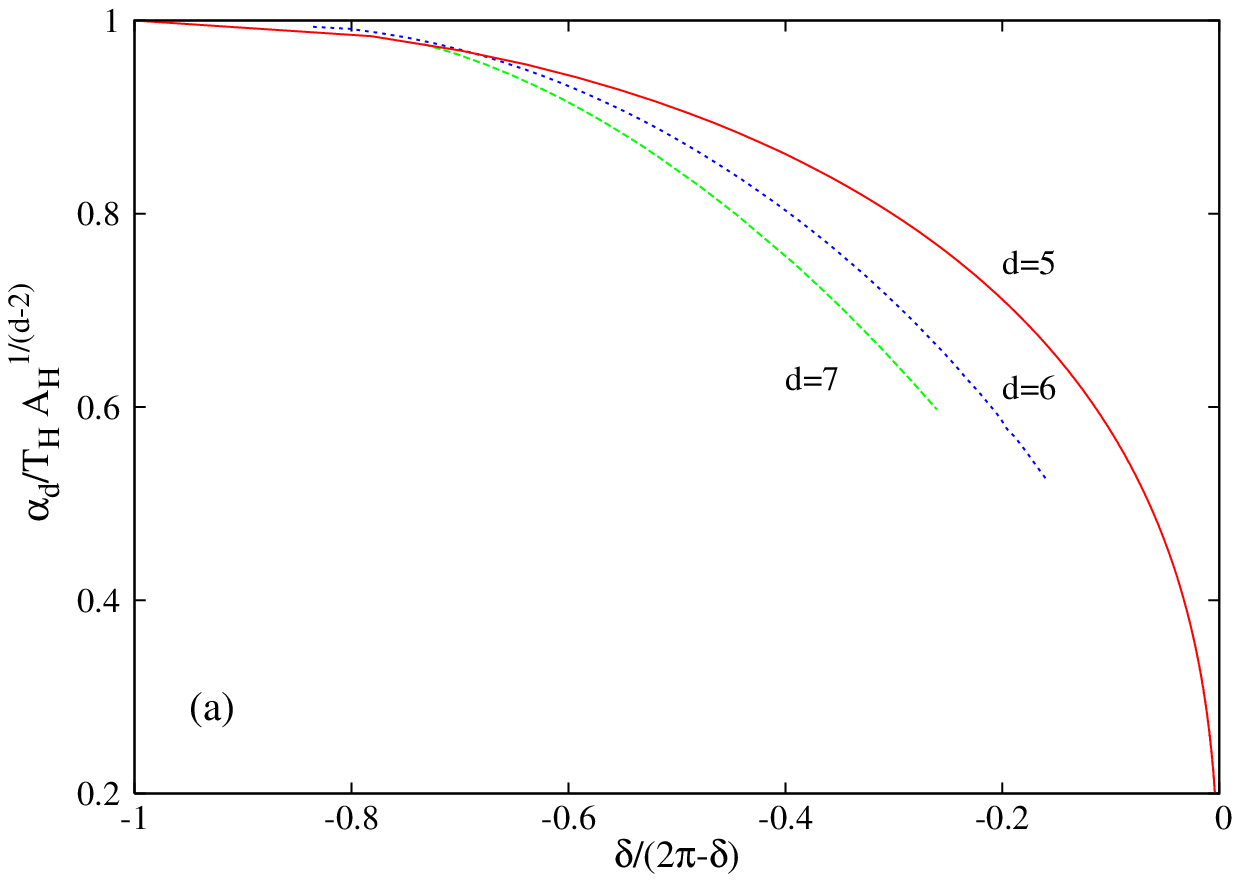}}
\hspace{15mm}%
        \resizebox{8cm}{6cm}{\includegraphics{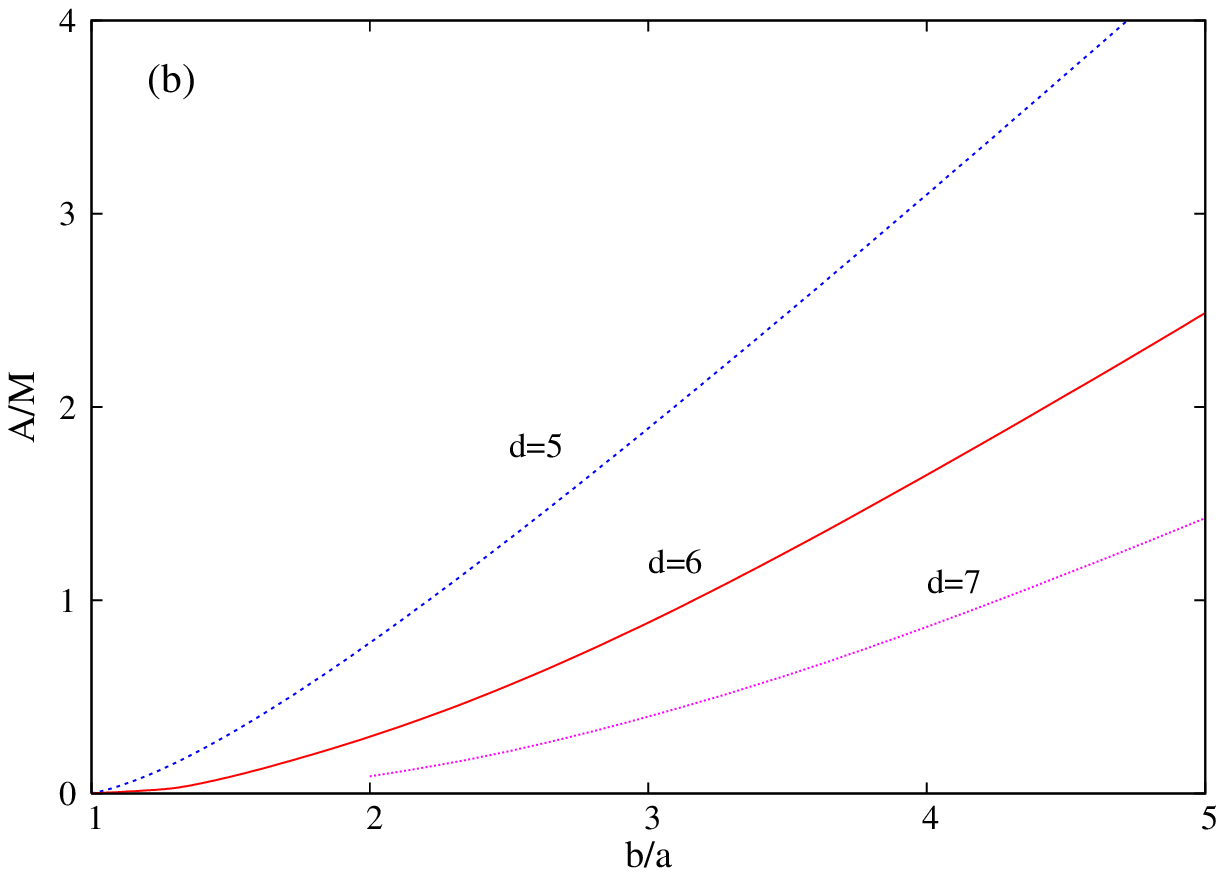}}	
\hss}
 \end{figure}
  \vspace*{0.3cm}
{\small {\bf Figure 4.} Features of generalized black ring solutions.
 { (a):} The scale free ratio $ T_H^{-1} A_H^{1/(d-2)}$
is shown as a function of the relative angular excess $\bar \delta=\delta/(2\pi-\delta)$. 
The  parameter  $\alpha_d=(d-3)2^{(5-2d)/(d-2)}\pi^{(3-d)/2(d-2)}\Gamma((d-1)/2)^{1/(2-d)}$
 has been chosen such that the point $(1,-1)$ on the plot corresponds to the
Schwarzschild-Tangherlini black hole.
{ (b):}  The scale free ratio ${\cal A/M}$ is shown as a function of the ratio 
between the two length scales $b/a$.
 }

The Kretschmann scalar $R_{ijkl}R^{ijkl}$ of typical $d=5,6$ solutions is shown in Figure 3.
There one can see the nonsingular character of this type of configurations 
(here we do not consider the $\delta$-Dirac terms in the expression of Riemann tensor 
in the presence of a conical singularity \cite{Fursaev:1995ef}).

In Figure 4 we show the scale free ratio $ T_H^{-1} A_H^{1/(d-2)}$ as a function of the 
relative angular excess $\bar \delta=\delta/(2\pi-\delta)$ for $d=5,6,7$ solutions.
The dimensionless ratio  ${\cal A/M}$ is also shown there as a function of the ratio $b/a$
of the two length scales.
One can see that, as expected, the pattern for $d=5$ repeats for the higher dimensional 
configurations.
Further details on these solutions including typical profiles of the functions 
$f_i,F_i$ are given in Ref. \cite{Kleihaus:2009wh}.

\subsection{Multi-horizon black holes}
The general ansatz proposed in Section 2 offers the possibility 
to construct multi-black hole solutions as well.
As discussed above, a limitation of this approach is that all configurations would have a
single $\Omega$-rod extending to infinity.
(For example, this prevents us from constructing $d>5$ multi-Schwarzschild-Tangherlini 
solutions.)
However, the number of rods in the $\psi$ or $t$-directions are not constrained
(although the numerical accuracy decreases with the number of rods).

In what follows, we present numerical evidence for the 
simplest asymptotically flat multi black objects, 
corresponding to 'generalized black Saturns' and 'generalized black dirings' (although more complex configurations are very likely to
exist).

\subsubsection{Generalized black Saturn: $(S^2\times S^{d-4}) \times S^{d-2}$ horizon}\label{sec:GBS}
The  $d=5$ static black Saturn
 describes a multi black hole configuration, with a black ring with horizon topology $S^2\times S^1$ around a 
Schwarzschild-Tangherlini black hole.
This exact solution is found as a static limit of a rotating solution originally presented in \cite{Elvang:2007rd}
(see Appendix A.2 for a review of its basic properties).
 
We are interested in the generalization of this type of configuration to $d>5$ within the metric ansatz (\ref{metric}).
There we have again two black objects, with
a topology of the horizon $S^2\times S^{d-4}$ for the generalized black ring and $ S^{d-2}$  
for the central black hole.
(Note that the horizon of the central black hole is not a round sphere).

This type of solution is found by imposing the following rod structure:
\begin{itemize}
\item A semi-infinite space-like rod $[-\infty,-a]$
in the $\partial / \partial \psi$ direction (with $f_2(0,z)=0$),
\item A finite time-like rod $[-a,a]$
in the $\partial / \partial t$ direction ($f_0(0,z)=0$),
\item A finite space-like rod $[a,b]$
in the $\partial / \partial \psi$ direction  (where $f_2(0,z)=0$ again),
\item A second finite time-like rod $[b,c]$
in the $\partial / \partial t$ direction, ($f_0(0,z)=0$),
\item A semi-infinite space-like rod $[c,\infty]$
in the $\Omega$-direction (with a vanishing $f_3(0,z)$).
\end{itemize}
 This rod structure is illustrated in Figure 1c. 

The profiles of a tyical $d=7$ solution are shown in Figure 5.
Thus the problem has three input parameters $a,b,c$ fixing the mass of the individual components 
and the distance between them.
In practice, one can always fix one of the parameters $a,b,c$ ($i.e.$ the length scale of the problem) 
and vary the other two.

\newpage
\setlength{\unitlength}{1cm}
\begin{picture}(15,20.85)
\put(-1,0){\epsfig{file=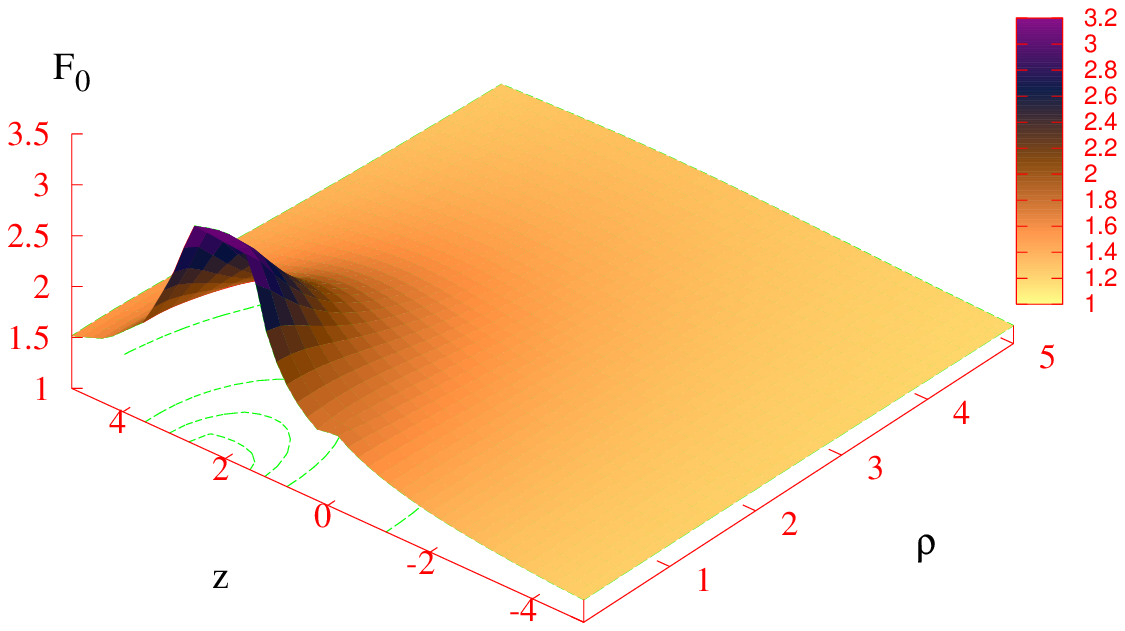,width=7.5cm}}
\put(7,0){\epsfig{file=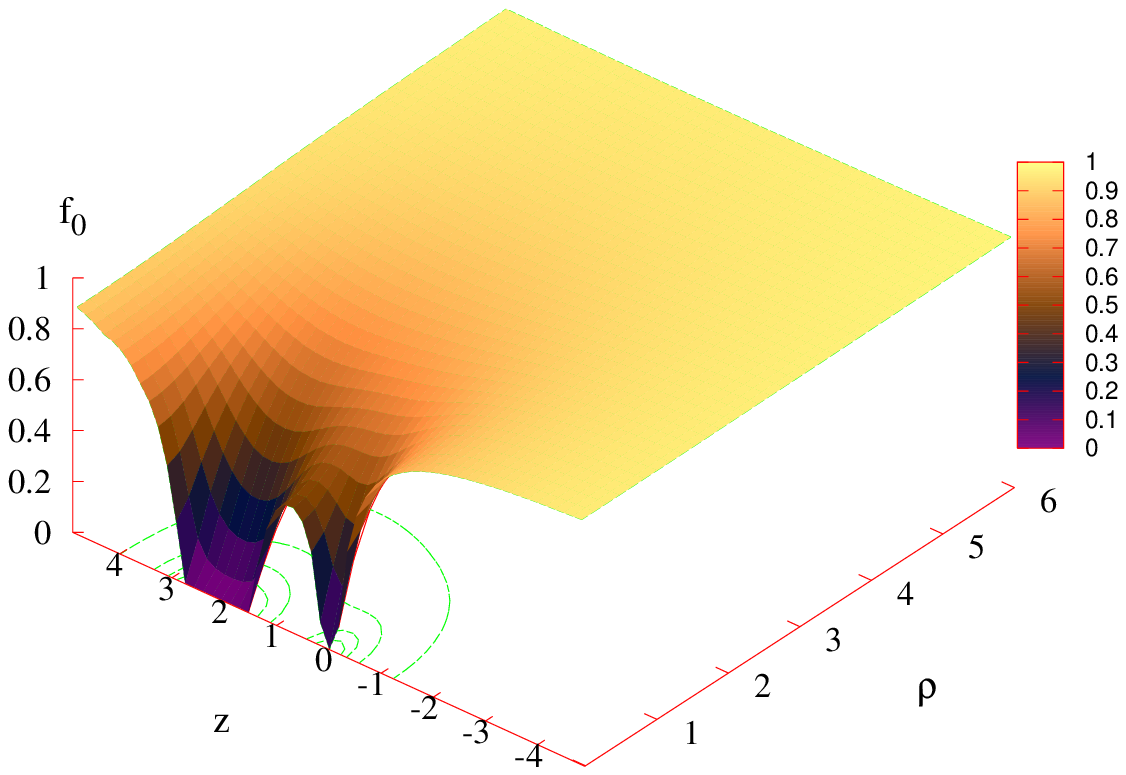,width=7.5cm}}
\put(-1,6){\epsfig{file=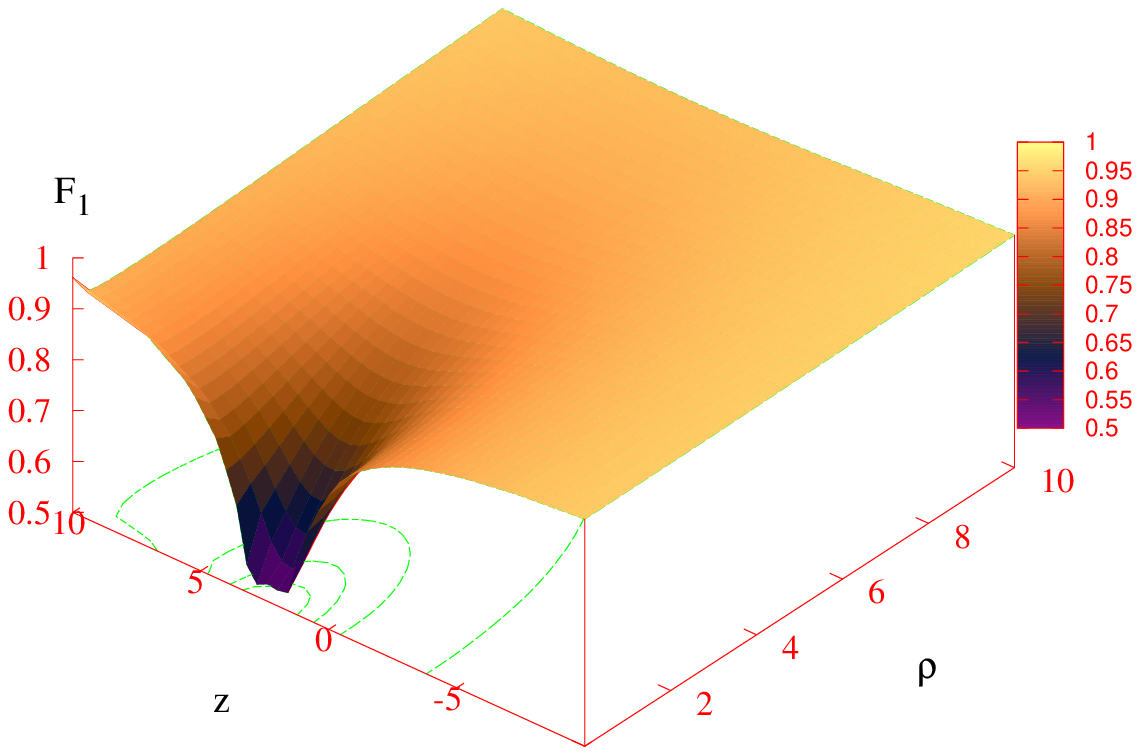,width=7.5cm}}
\put(7,6){\epsfig{file=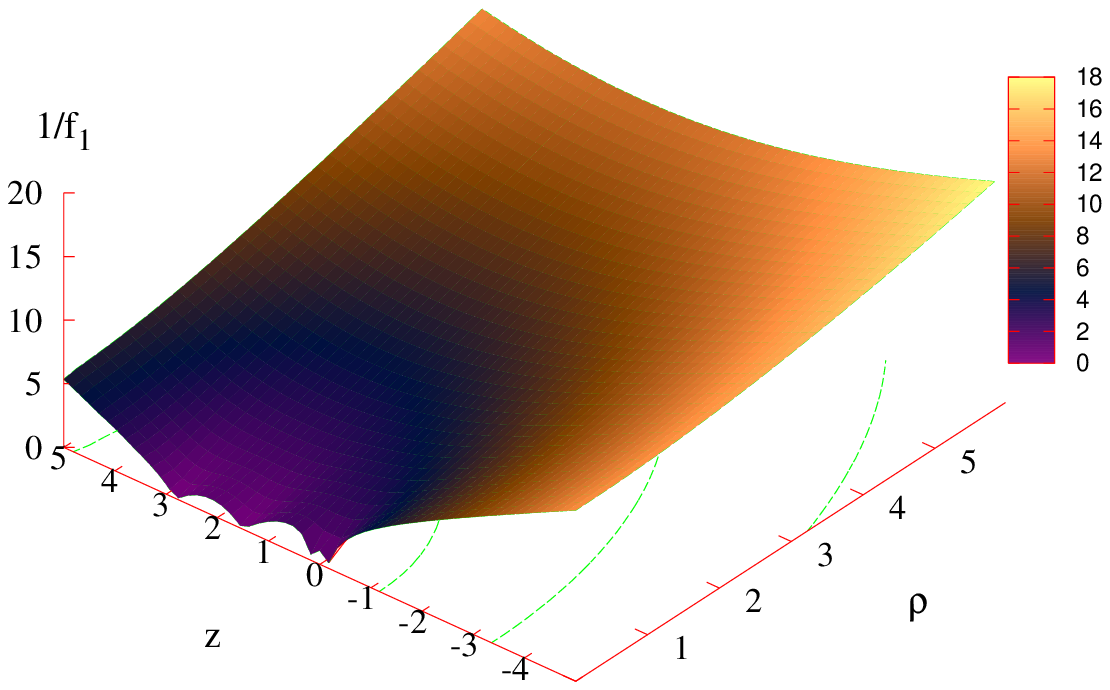,width=7.5cm}}
\put(-1,12){\epsfig{file=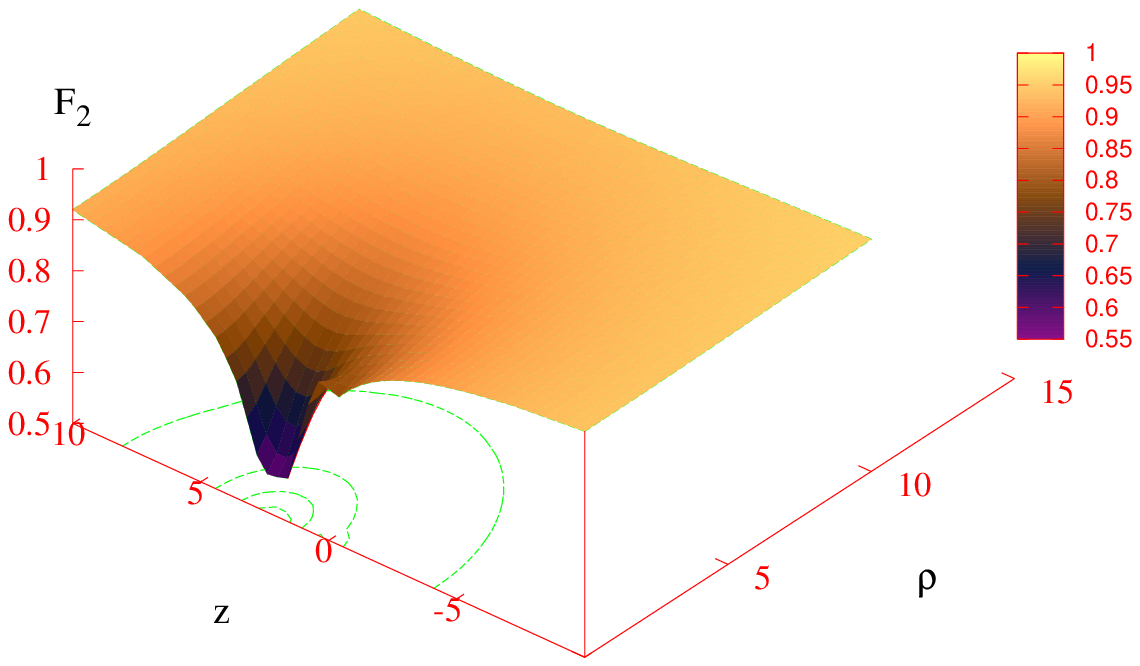,width=7.5cm}}
\put(7,12){\epsfig{file=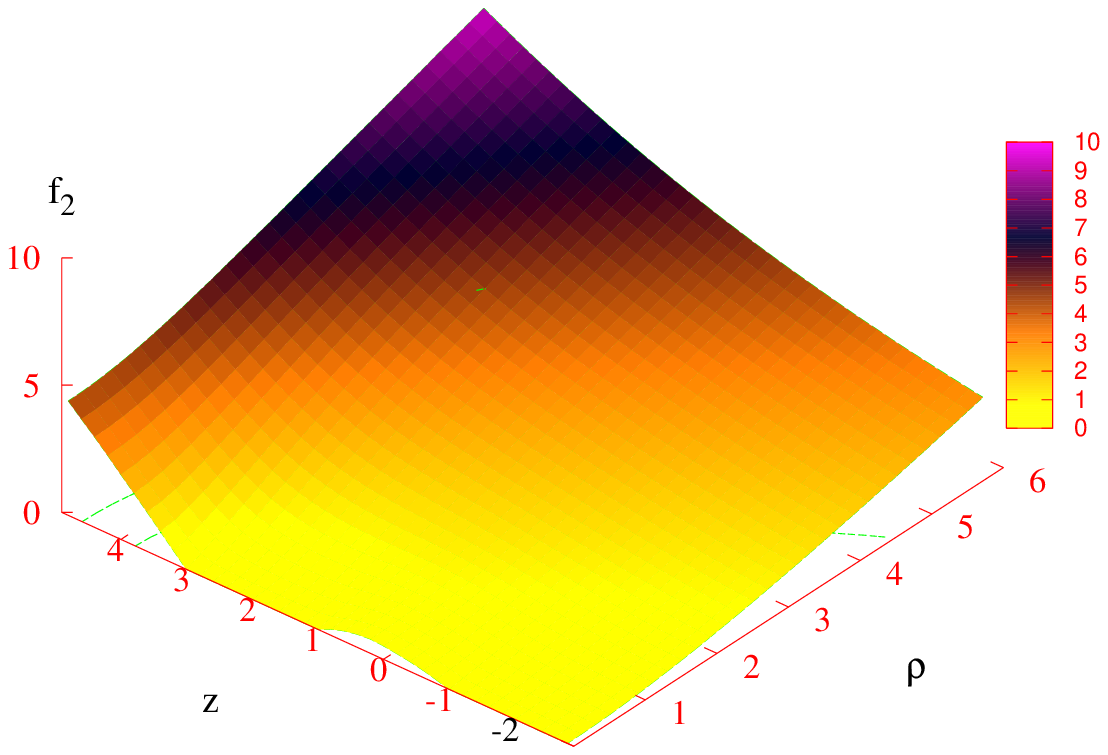,width=7.5cm}}
\put(-1,18){\epsfig{file=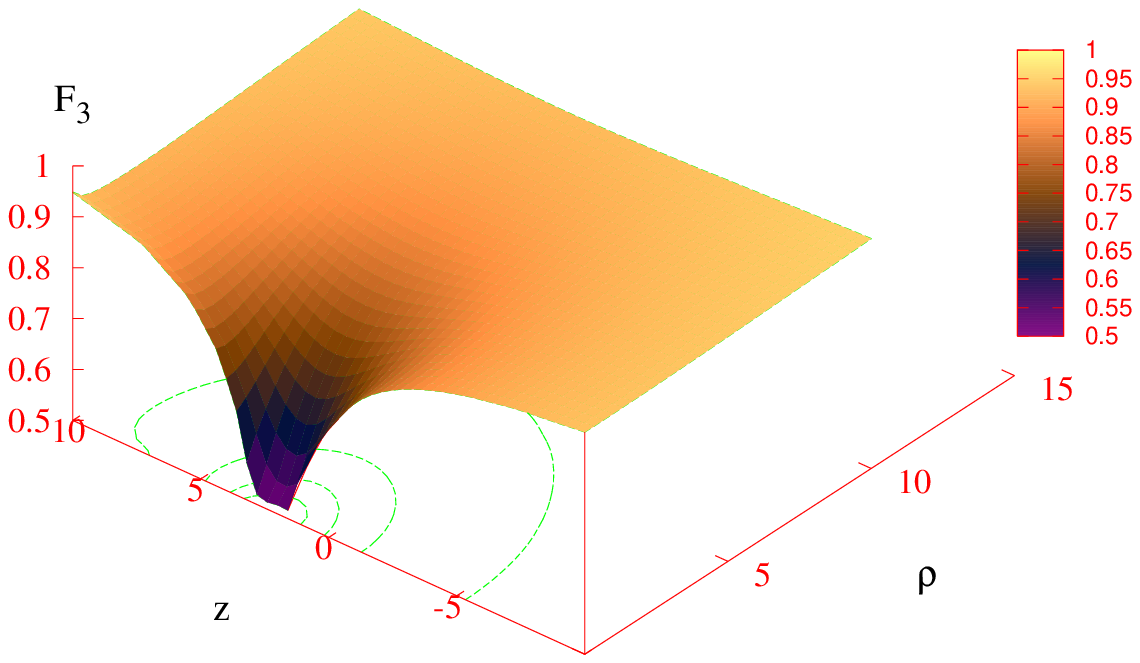,width=7.5cm}}
\put(7,18){\epsfig{file=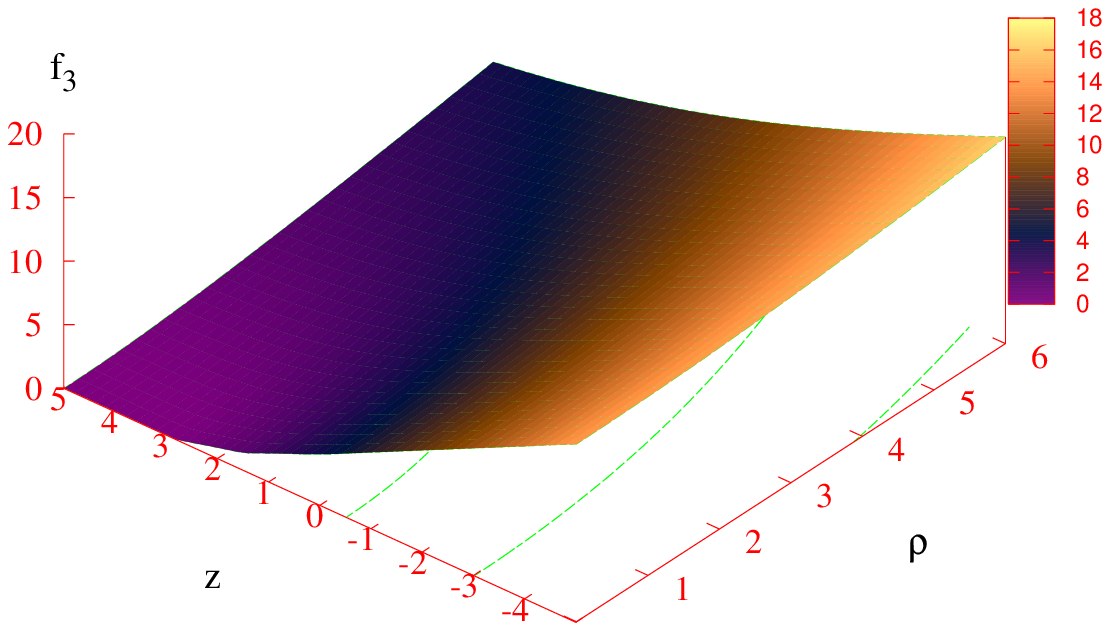,width=7.5cm}}
\end{picture} 
\\
{\small {\bf Figure 5.}
 The profiles of the functions $F_i$ used in the numerical calculations and of the metric functions $f_i$ 
are shown for a typical
$d=7$ black Saturn-type solution with $a=0.14$, $b=1.5$, $c=2.8$.
} 

 \newpage
\setlength{\unitlength}{1.cm}
\begin{picture}(15,18)
\put(-1,1){\epsfig{file=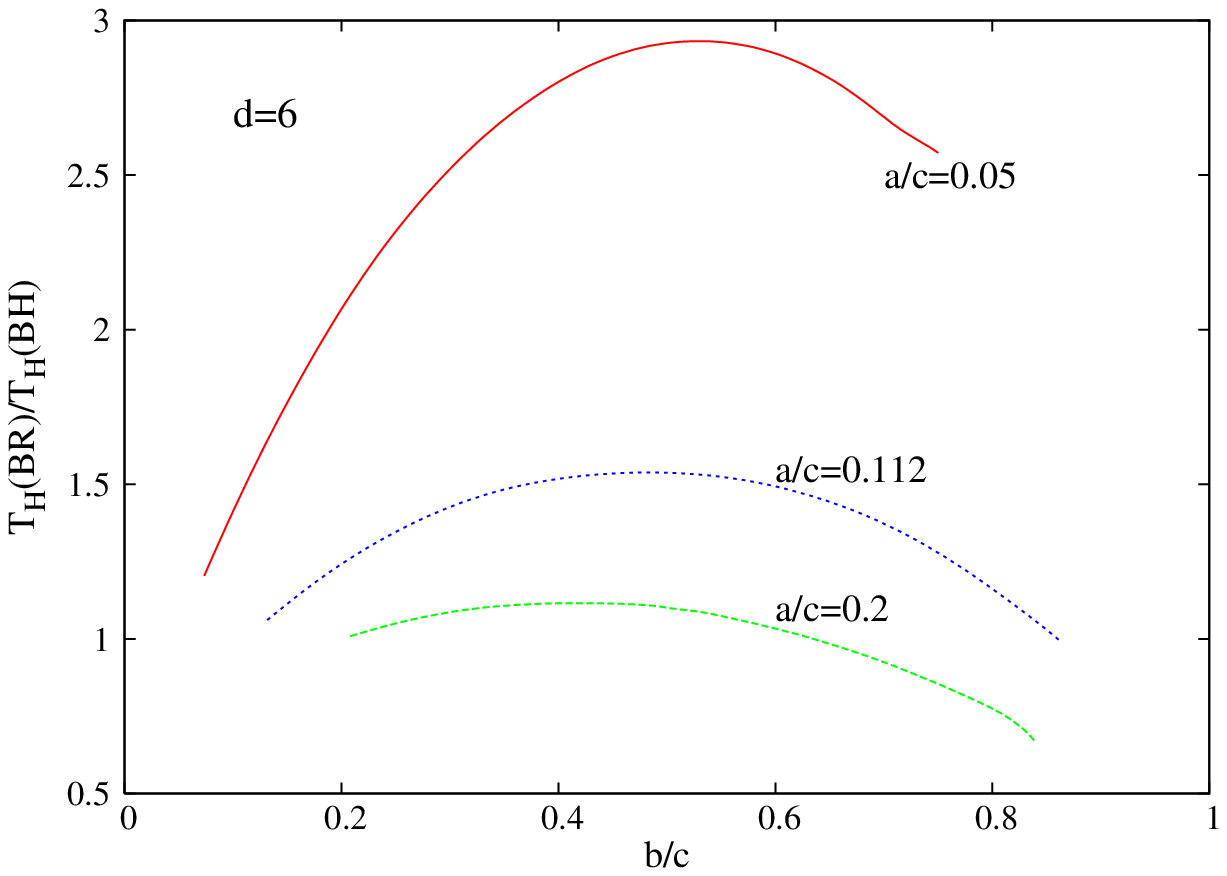,width=7.5cm}}
\put(7,1){\epsfig{file=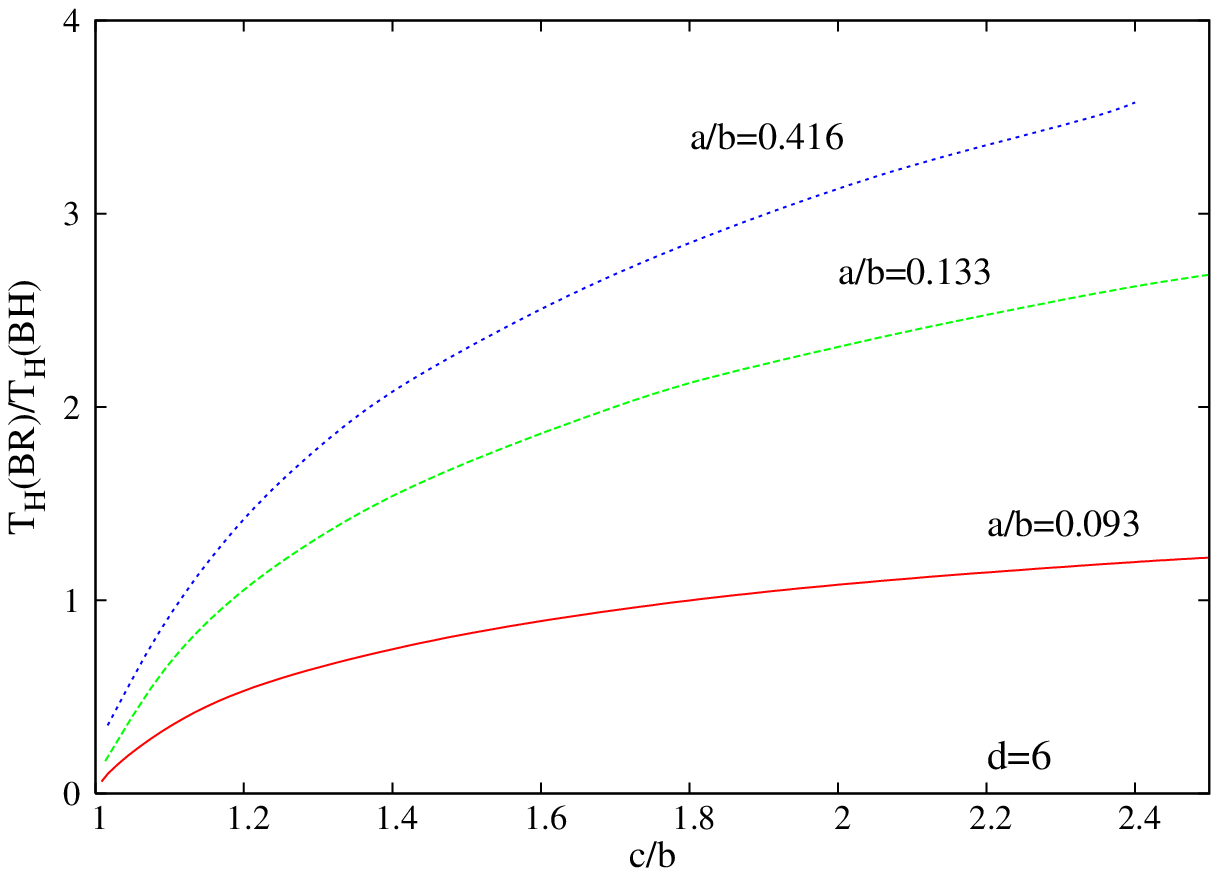,width=7.5cm}}
\put(-1,7){\epsfig{file=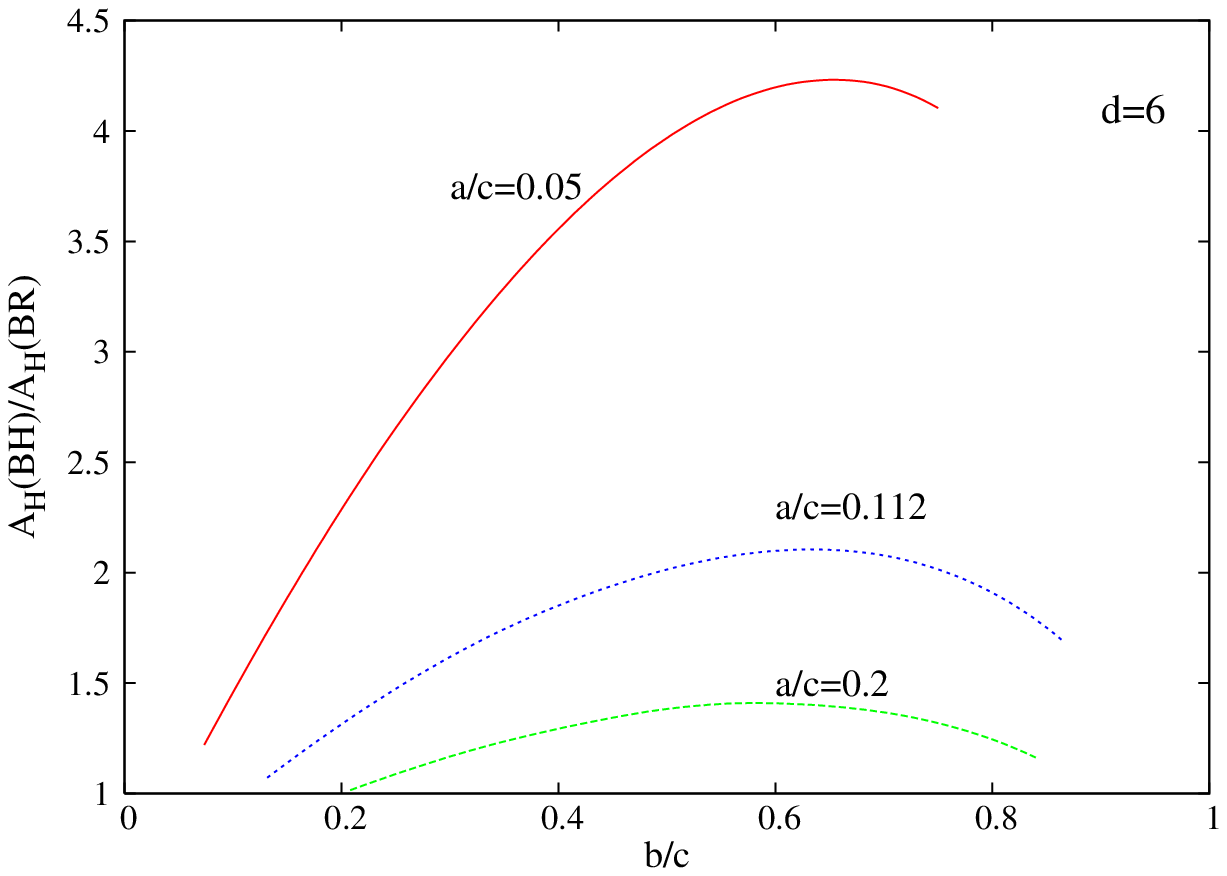,width=7.5cm}}
\put(7,7){\epsfig{file=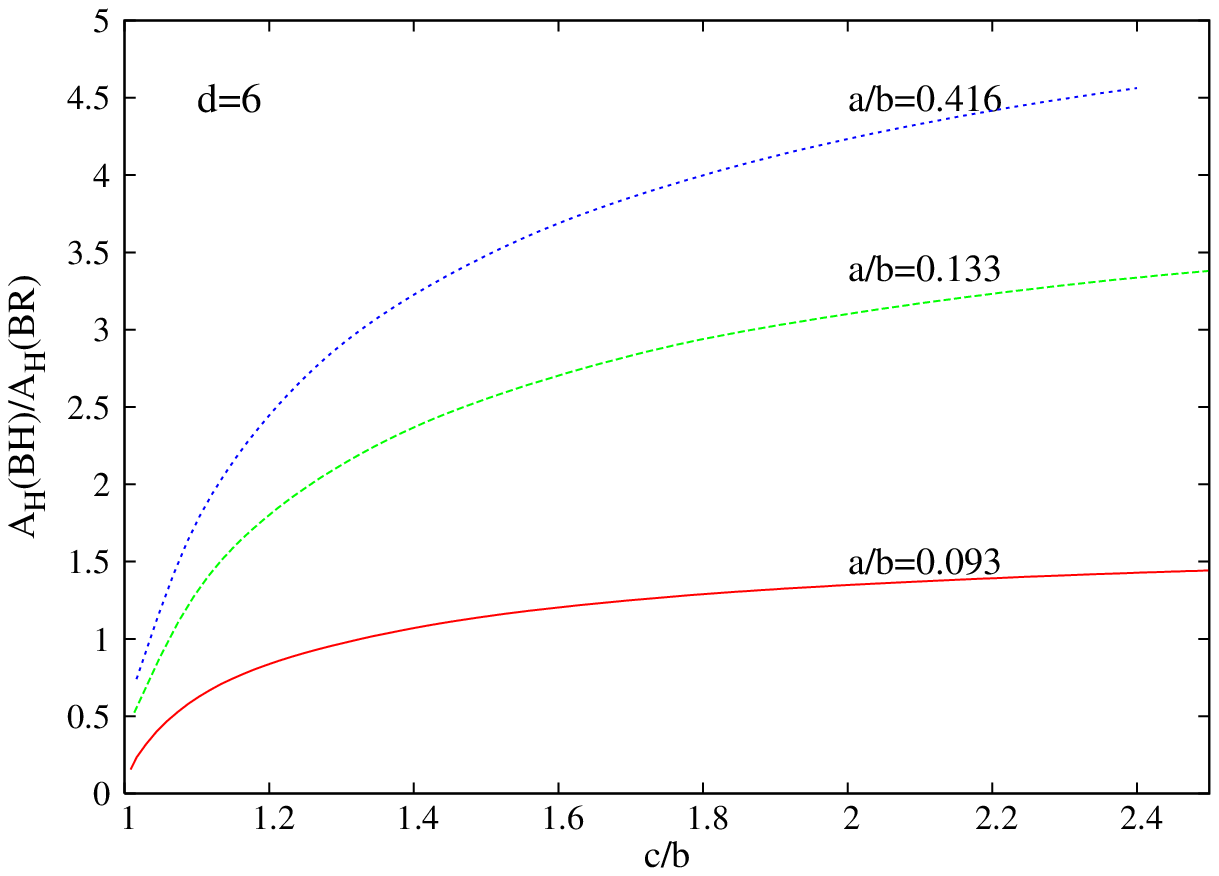,width=7.5cm}}
\put(-1,13){\epsfig{file=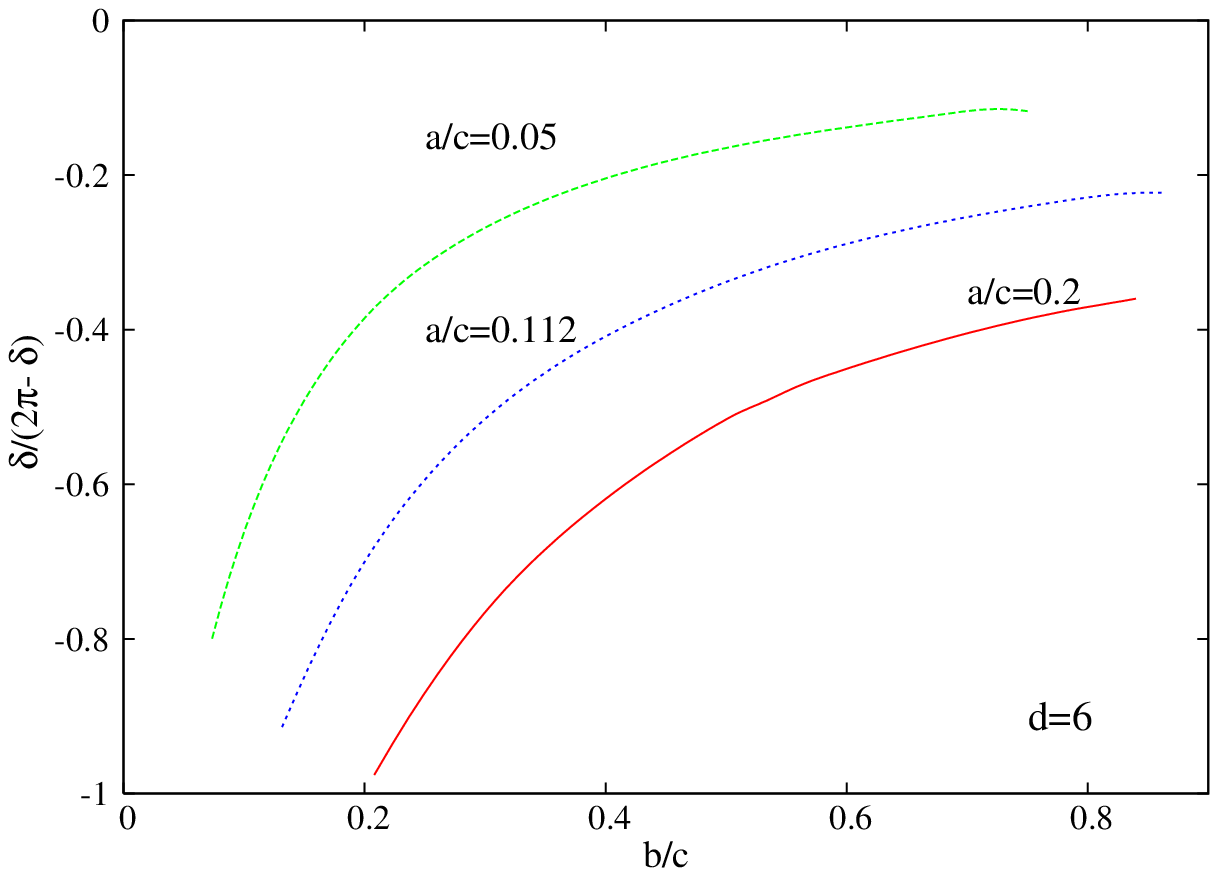,width=7.5cm}}
\put(7,13){\epsfig{file=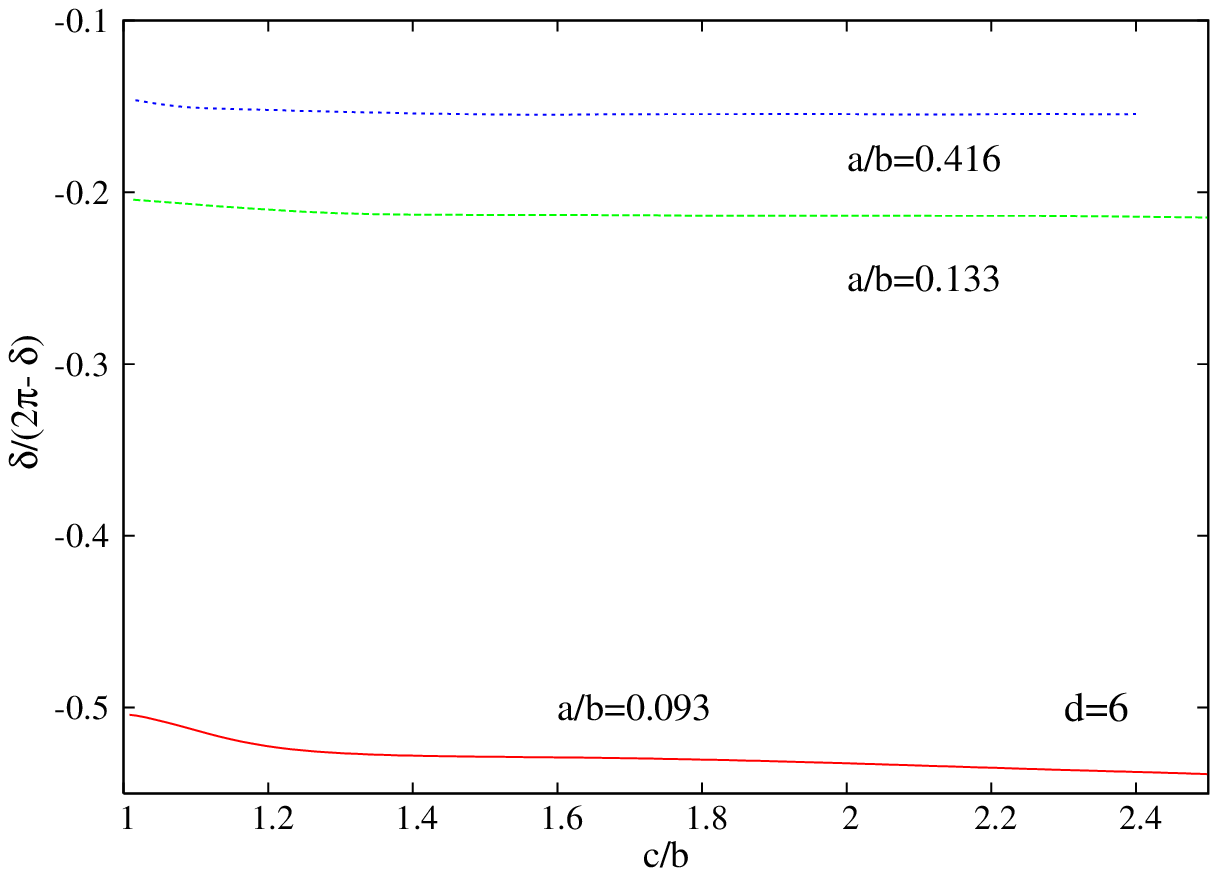,width=7.5cm}} 
\end{picture} 
\\
{\small {\bf Figure 6.}
$d=6$ generalized black Saturn solutions: a number of relevant dimensionless quantities are shown as a function 
of the ratio $b/c$ for several fixed values of $a/c$ (left)
and for a varying ratio $c/b$ for fixed $a/b$ (right).  
In these plots the indices ${\rm BR}$, ${\rm BH}$ stand  for the objects with  $S^2\times S^{d-4}$ and $ S^{d-2}$
horizon topology, respectively.      
}
\\

 As expected, all our solutions have a conical singularity on the rod between the horizons, with $\delta<0$
there, which corresponds to a conical excess (and thus a negative energy density for the 
strut source).

Starting with the dependence of
solutions on the ratio $b/c$ for fixed $a/c$
one can see that the solutions  smoothly interpolate between two limits.
First, for $b/c\to 1$, the horizon with $S^{d-2}$ topology vanishes and the solution reduces to the higher dimensional generalizations \cite{Kleihaus:2009wh} 
of the static black ring  in \cite{Emparan:2001wk}.
Another limit of interest is $b/c  \to a/c$, in which case 
the finite $\psi$-rod vanishes and  a Schwarzschild-Tangherlini configuration is
recovered.

We have studied as well the dependence of the 
solutions on the ratio $c/b$ for fixed $a/b$.
There, for $c/b\to 1$, the horizon with $S^{d-2}$ topology vanishes and the solution reduces to a generalized black ring.
Other interesting limits are $a \to b$ or $a=0$, in which cases the Schwarzschild-Tangherlini configuration is
recovered.

Some results of the numerical integration supporting the above statements 
are shown in Figure 6 for several fixed values of $a/c$  and a varying $b/c$ (left)
and for a varying ratio $c/b$ for fixed $a/b$ (right).
One can see that the generic solutions are not in thermal equilibrium, $T_H^{BR}\neq T_H^{BH}$
(also, we could not find configurations with a vanishing Hawking temperature
for one of the components).
For all solutions, we have noticed a good qualitative agreement of their behaviour
with that found for $d=5$ static black Saturns.

\subsubsection{Generalized di-rings: $(S^2\times S^{d-4}) \times (S^2\times S^{d-4})$ horizon}\label{sec:GBD}
  For  $d=5$, the Einstein equations have an exact solution describing
two concentric black rings  \cite{Iguchi:2007is}, \cite{Evslin:2007fv}.
In the static limit, this asymptotically flat configuration is supported by a strut with positive pressure
and negative energy density. The explicit form of the $d=5$ di-ring solution is given in Appendix A.3.

By using the same approach as in the previous cases, we 
could construct higher dimensional generalizations of this static configuration.
There we have again two black objects, both with
a topology of the horizon $S^2\times S^{d-4}$.
 They are found for the following rod structure:
\begin{itemize}
\item A semi-infinite space-like rod $[-\infty,-d]$
in the $\partial / \partial \psi$ direction  (with $f_2(0,z)=0$),
\item A finite time-like rod $[-d,-c]$
in the $\partial / \partial t$ direction  ($f_0(0,z)=0$),
\item A finite time-like rod $[-c,-a]$
in the $\partial / \partial \psi$ direction  (where $f_2(0,z)=0$ again),
\item A second finite time-like rod $[-a,a]$
in the $\partial / \partial t$ direction  ($f_0(0,z)=0$),
\item A finite space-like rod $[a,b]$
in the $\partial / \partial \psi$ direction ($f_2(0,z)=0$),
\item A semi-infinite space-like rod $[b,\infty]$
in the $\Omega$-direction (with a vanishing $f_3(0,z)$).
\end{itemize}
This rod structure is illustrated in Figure 1d.

Given the presence of four finite rods, finding such solutions is a more difficult problem, and we did not manage
to obtain $d=7$ numerical solutions with reasonable accuracy.
However, we think this is due to the limitations imposed by our approach only.

The metric functions  $f_i$ and the functions $F_i$ used in the numerical calculations
change smoothly with the rod parameters $a$, $b$, $c$ and $d$.
Typical profiles of the solutions are  presented   in Figure 7 as a function of $z$ for several values of $\rho$.

\newpage
\setlength{\unitlength}{1cm}
\begin{picture}(15,20.85)
\put(-1,0){\epsfig{file=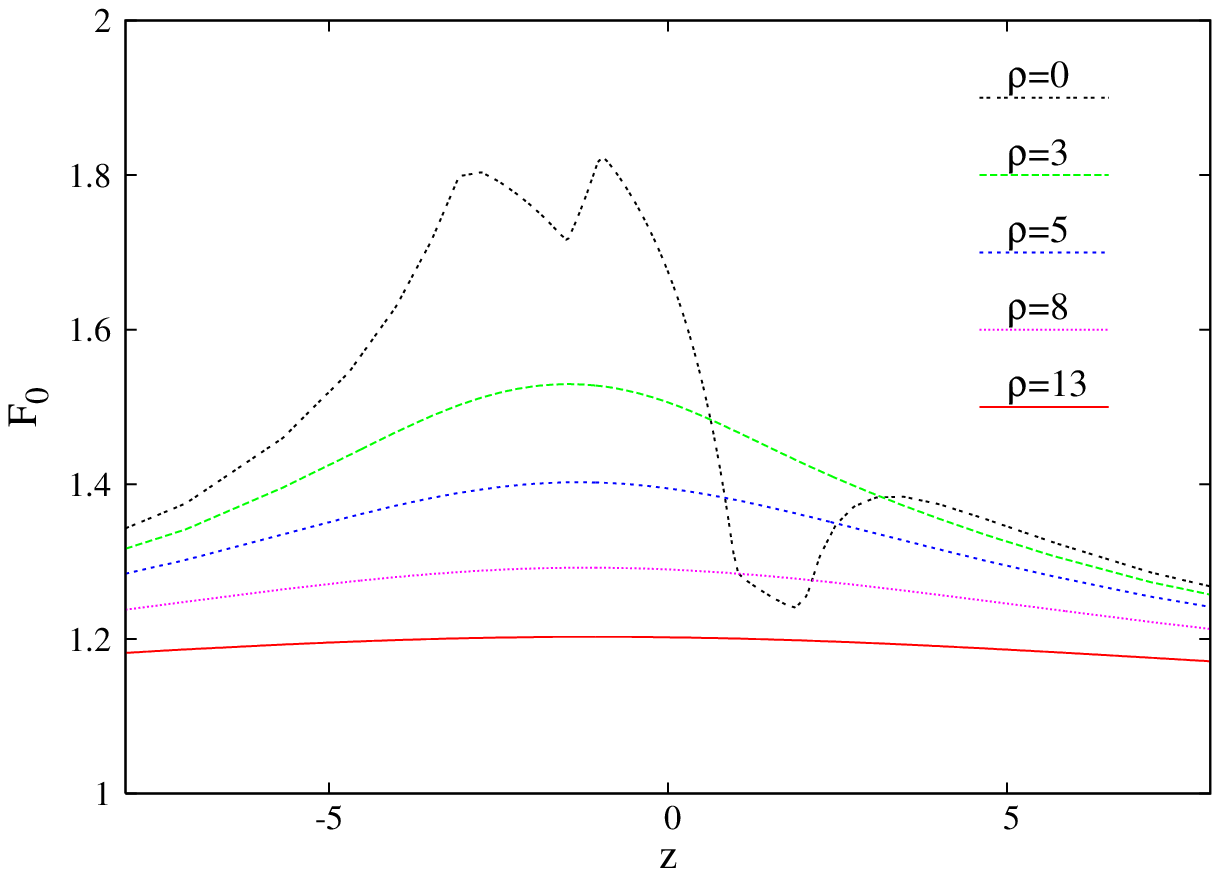,width=7.5cm}}
\put(7,0){\epsfig{file=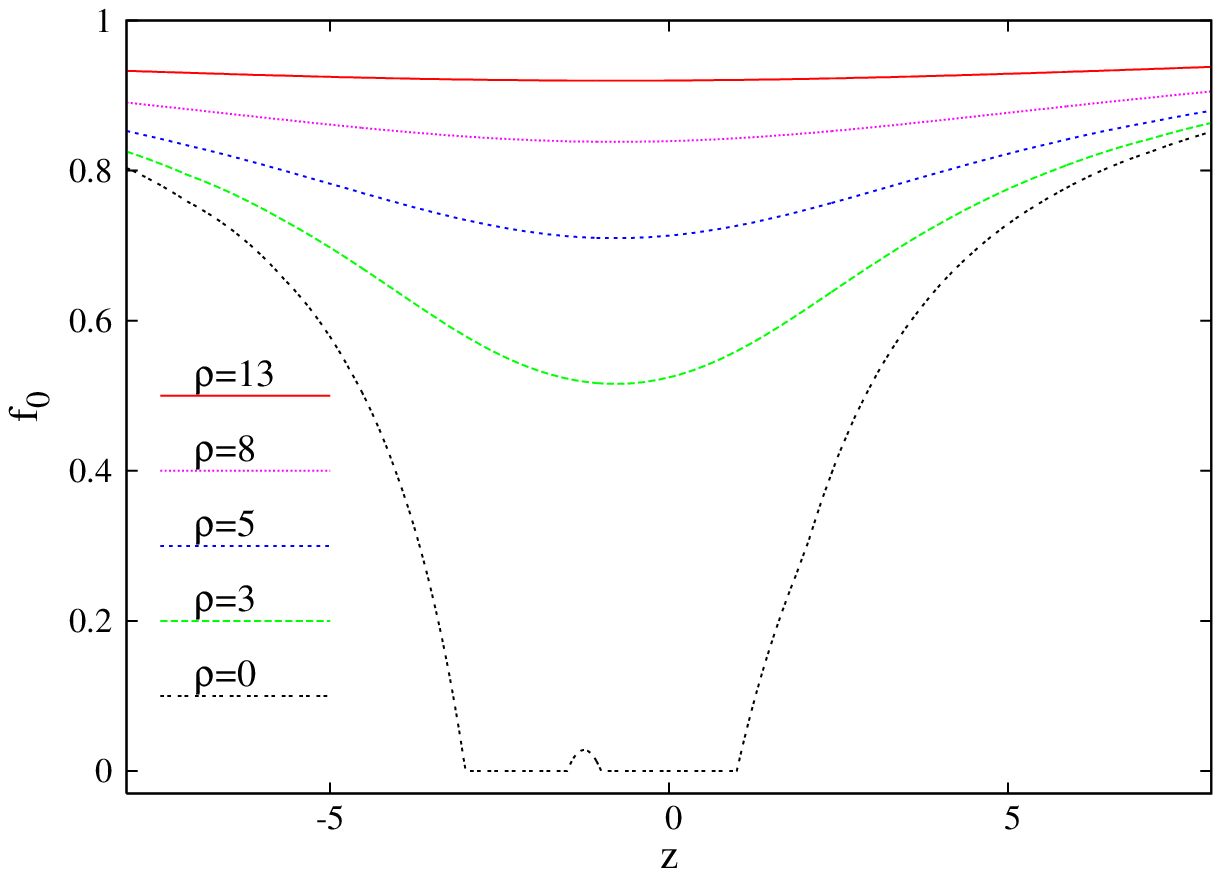,width=7.5cm}}
\put(-1,6){\epsfig{file=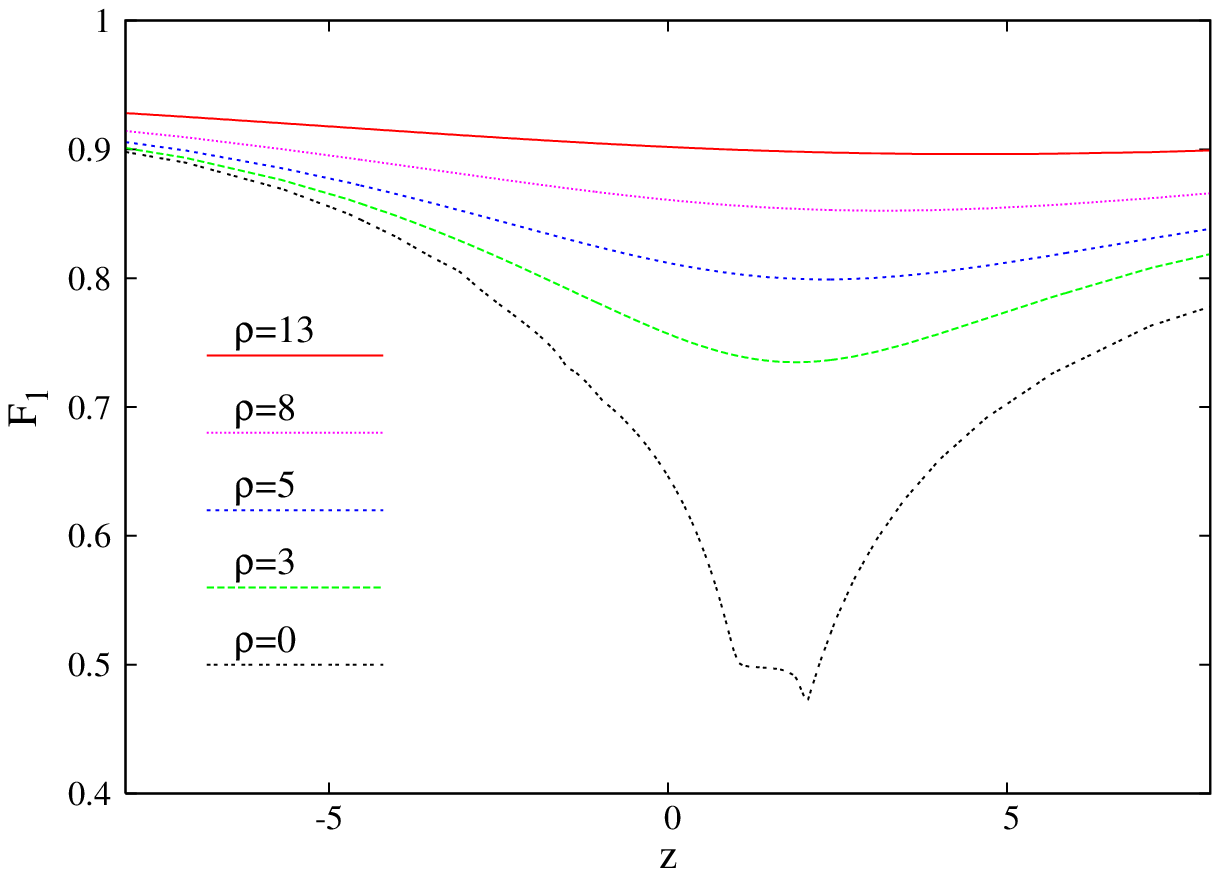,width=7.5cm}}
\put(7,6){\epsfig{file=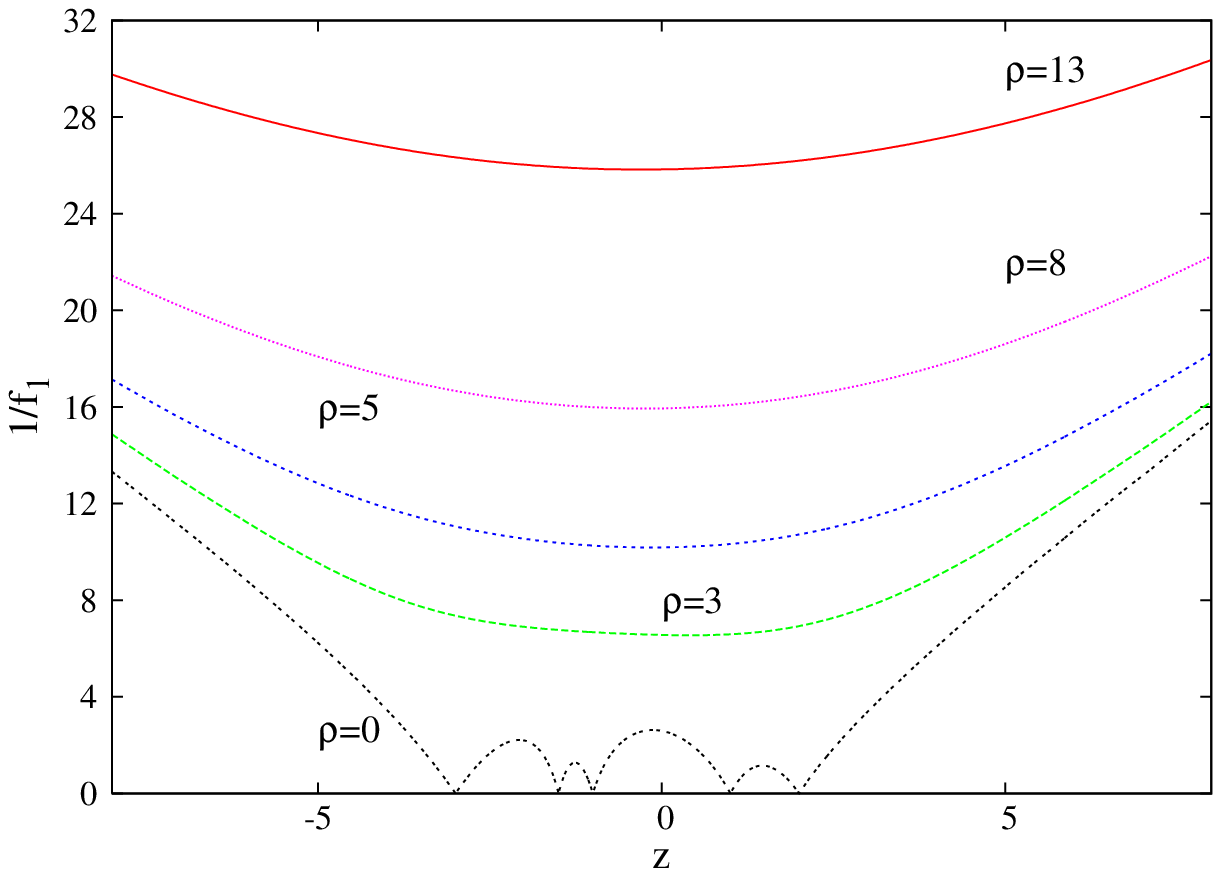,width=7.5cm}}
\put(-1,12){\epsfig{file=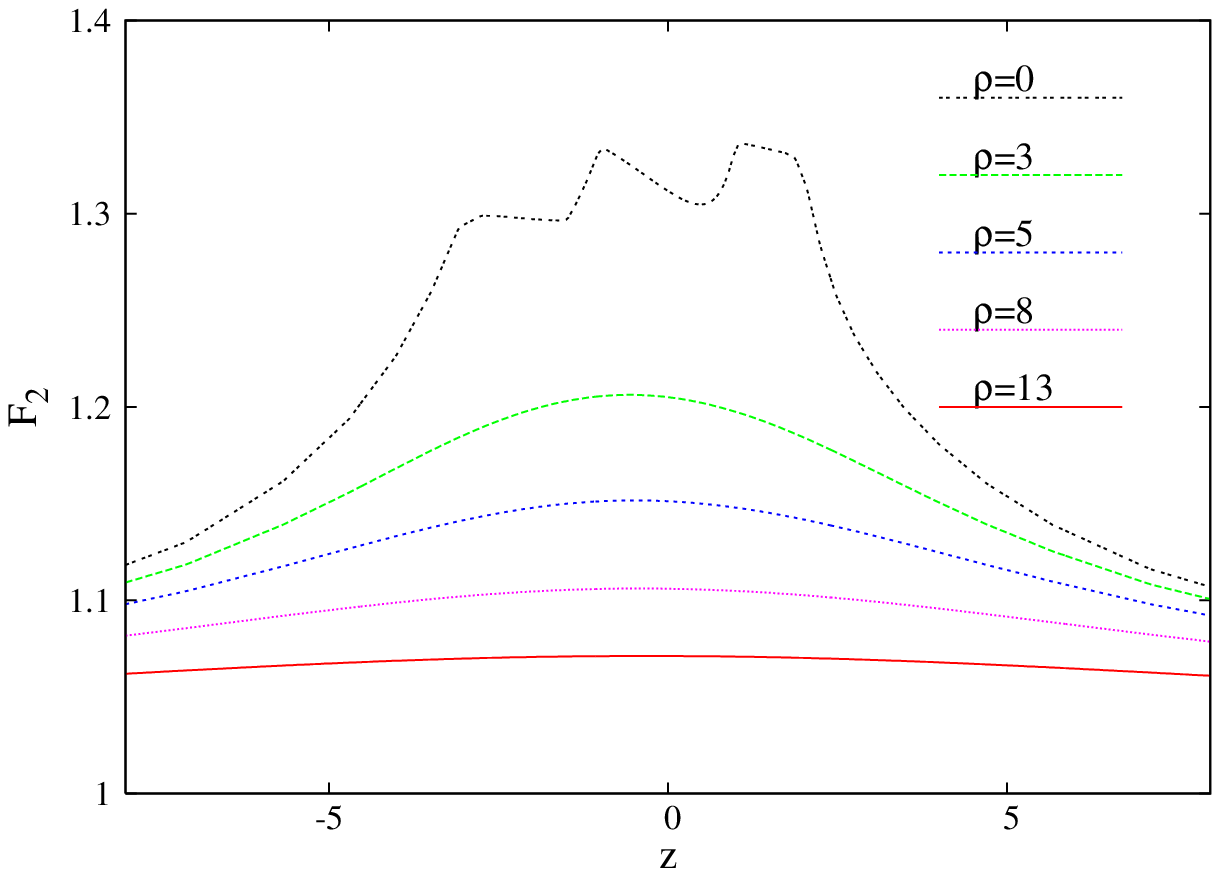,width=7.5cm}}
\put(7,12){\epsfig{file=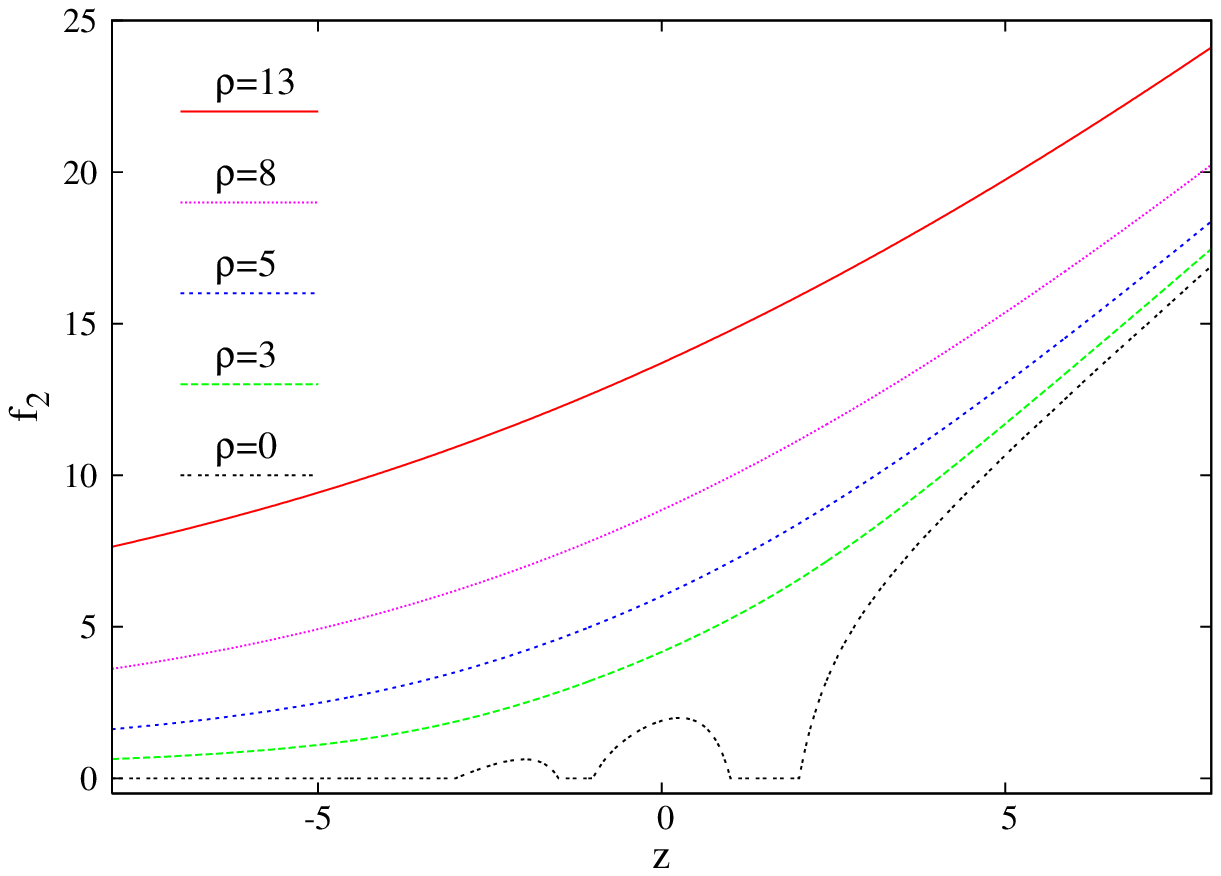,width=7.5cm}}
\put(-1,18){\epsfig{file=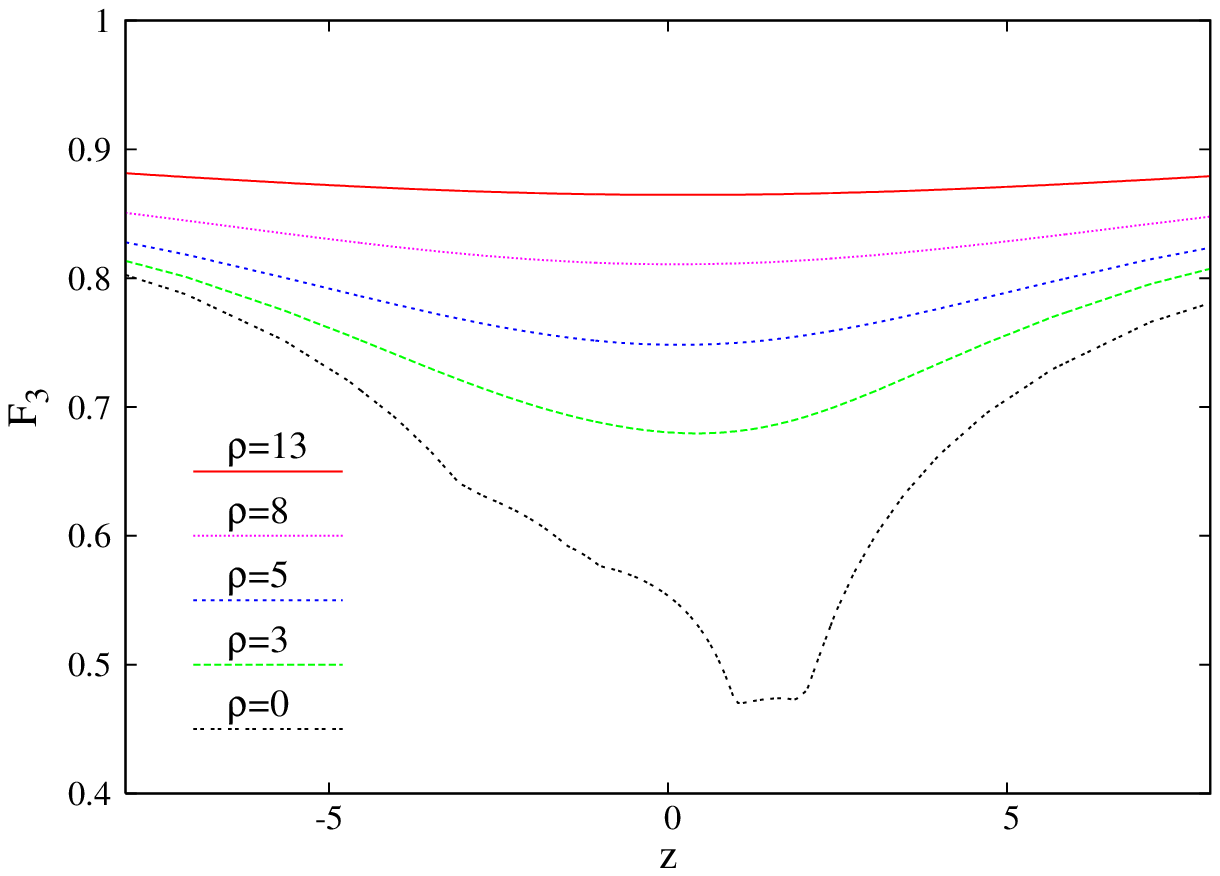,width=7.5cm}}
\put(7,18){\epsfig{file=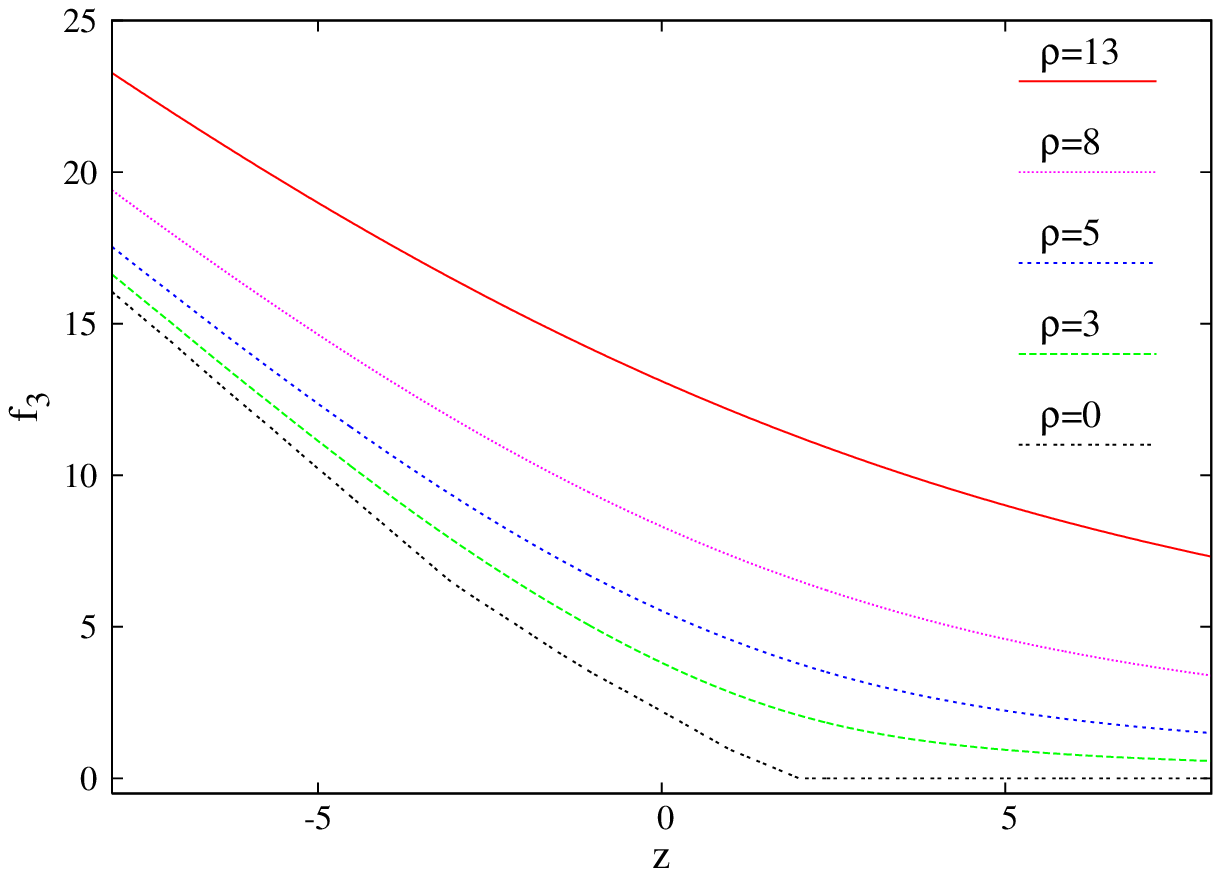,width=7.5cm}}
\end{picture} 
\\
{\small {\bf Figure 7.}
The profiles of the functions $F_i$ used in the numerical calculations and of the metric functions $f_i$ 
are shown for a typical
$d=6$ generalized black diring  solution with $a=1$, $b=2$, $c=1.5$, $d=3$.
} 

\newpage
\begin{figure}[ht]
\hbox to\linewidth{\hss%
	\resizebox{8cm}{6cm}{\includegraphics{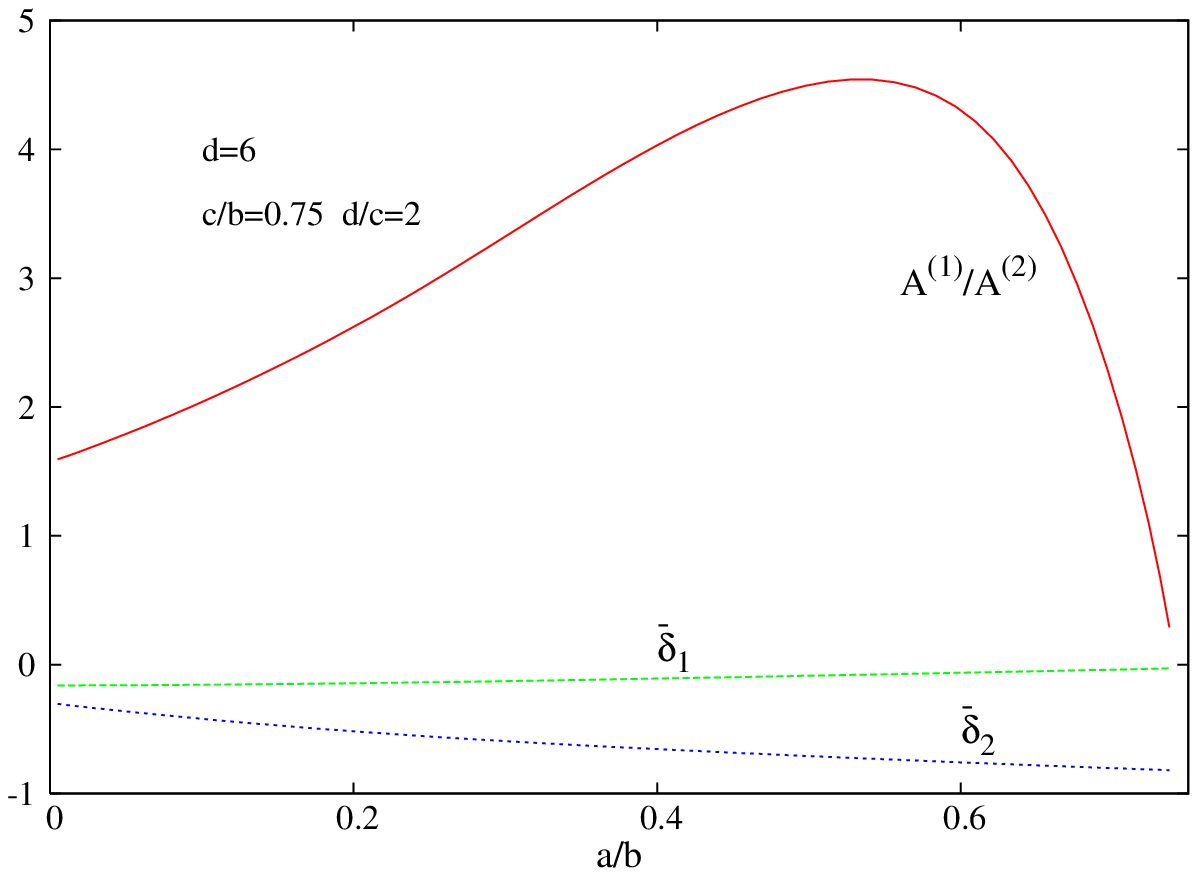}}
\hspace{10mm}%
        \resizebox{8cm}{6cm}{\includegraphics{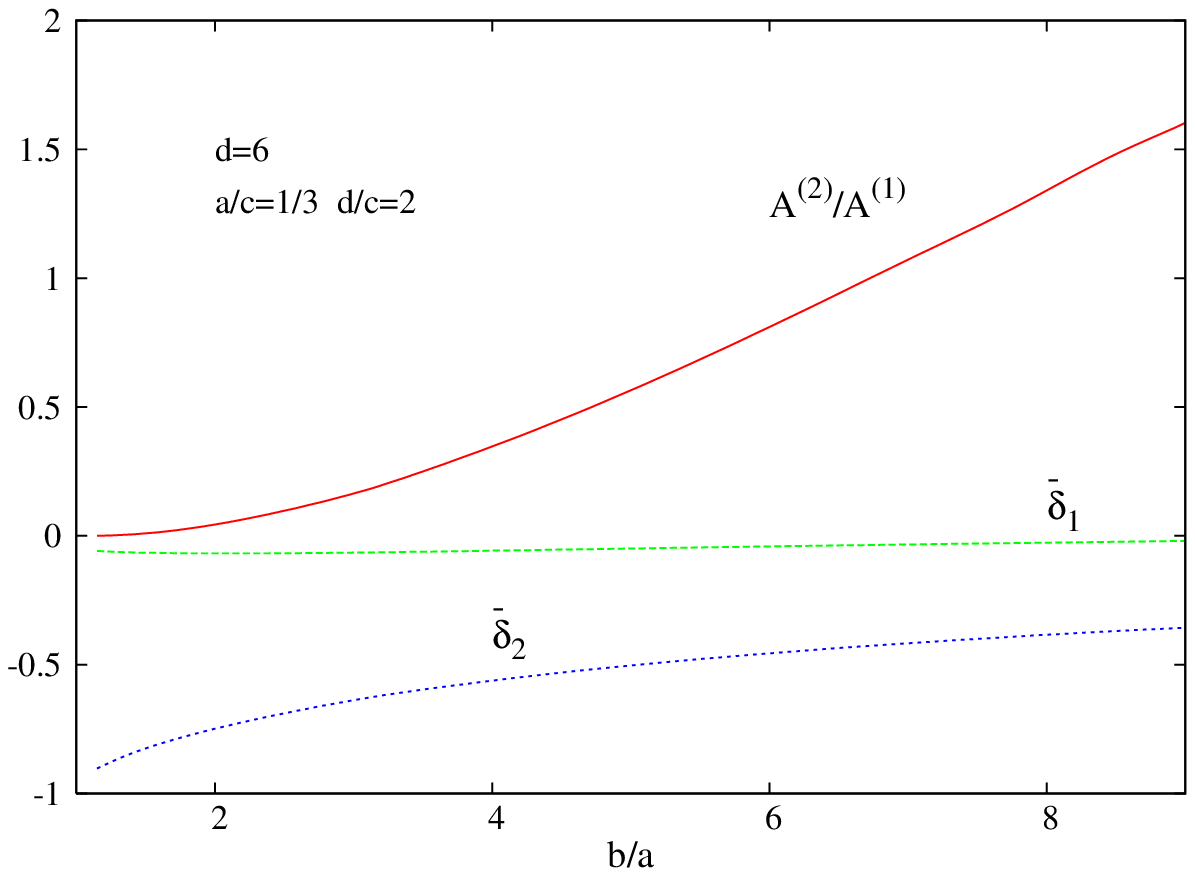}}	
\hss}
 \end{figure}
  \vspace*{-1.cm}
 {\small \hspace*{3.cm}{\it  } }
\begin{figure}[ht]
\hbox to\linewidth{\hss%
	\resizebox{8cm}{6cm}{\includegraphics{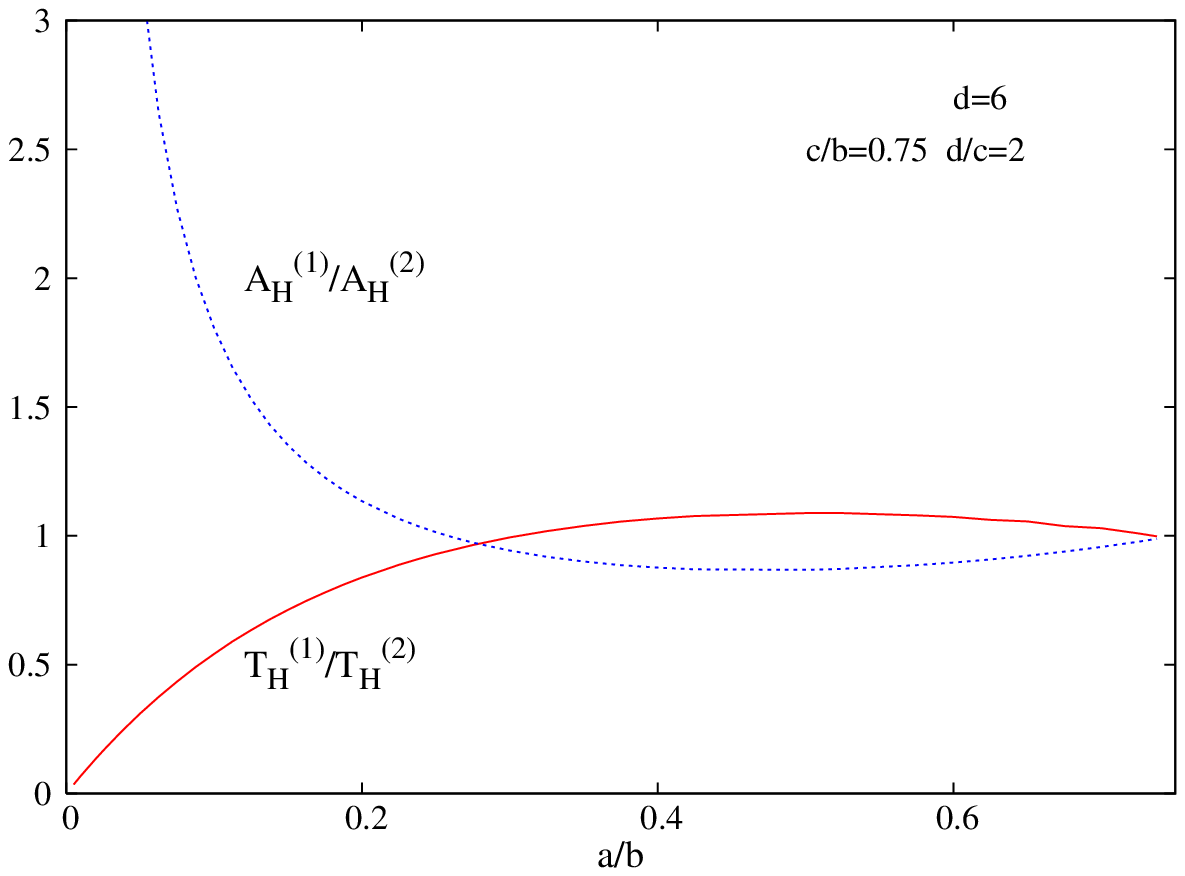}}
\hspace{10mm}%
        \resizebox{8cm}{6cm}{\includegraphics{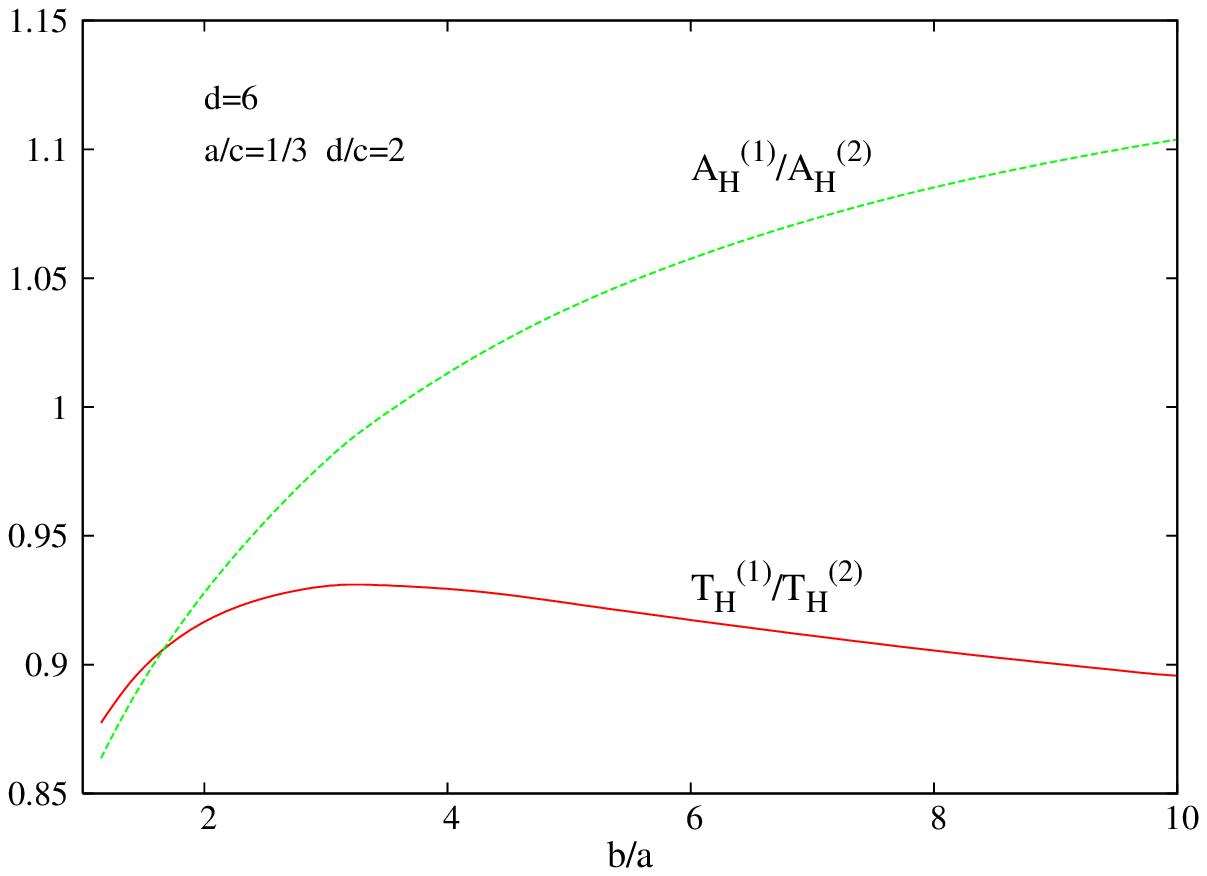}}	
\hss}
\end{figure}
\\
{\small {\bf Figure 8.}
$d=6$ generalized black diring solutions: a number of relevant quantities are shown as a function 
of the ratio $a/b$ for several fixed values of $c/b,~d/c$ (left)
and for a varying ratio $b/a$ for fixed $a/c,~d/c$ (right).  
In these plots the indices $1$, $2$ stand for the objects with an event horizon at $\rho=0,~-d\leq z\leq -c$ and
$\rho=0,~-a\leq z\leq a$, respectively.    } 
 \\
 \\
One can see that the functions $F_i$ are smooth outside of the $z-$axis and show no sign of a singular behaviour
(although they have a complicated behaviour at $\rho=0$).

In this case we have studied the dependence of the solutions on the parameters $a,b,c,d$
for two different situations
(In fact a simple rescaling leads to a dependence on only three dimensionless quantities.).
In the first case (see Figure 8 (left)), we have fixed $b,c,d$ and 
varied the parameter $a$ associated with the length 
of the second timelike rod
(one of the constants $b,c,d$
can be taken to represent the length scale of the problem). 

As $a\to 0$, a generalized black ring is approached since the second horizon 
vanishes, which implies $T_H^{(2)}\to \infty$ and $A_H^{(2)}\to 0$.
As $a/c\to 1$ for fixed $b,c,d$, the first finite
$\psi$-rod vanishes and the two 
horizons coalesce to form a single black object with $S^2\times S^{d-4}$
horizon topology.

 The second case  we have investigated corresponds to fixed $a,c,d$
and a varying $b$  (see Figure 8 (right)).
The relevant limits here are $b/a \to 1$ and $b/a\to \infty$.
As $a/b \to 1$, one finds $\bar \delta_2\to -1$ ($i.e.$ the conical excess $\delta_2\to -\infty$) and
a generalized black Saturn configuration is approached (the second finite $\psi-$rod vanishes).  
The case  $b/a\to \infty$ is more involved
and we have found it more difficult to investigate this limit.
As the second $\psi-$rod extends to infinity, one
expects to recover, after a suitable rescaling,  the 
four dimensional  double-Schwarzschild configuration ($i.e.$ the Bach-Weyl solution \cite{BW}) uplifted to six dimensions. 
This configuration still has a conical singularity  in between the two black holes
which provides the force balance that allows its existence.

Again, we have noticed a good qualitative agreement of this behaviour with that
 found for the $d=5$ static dirings exact solution (see Appendix A3). 
 The generic solutions have different Hawking temperatures for the individual components,  $T_H^{(1)}\neq T_H^{(2)}$,
and thus are not in thermal equilibrium.
Also, it seems that there are no
generalized black diring solutions with a vanishing Hawking temperature
for one of the components.
 
\section{Remarks on rotating black holes 
 with a nonspherical horizon topology}\label{sec:withrotations}
All static solutions with a nonspherical horizon topology discussed above are plagued by conical singularities,
which seems to be an unavoidable feature of all such asymptotically flat   black objects.

 However, for their $d=5$ counterparts, the conical singularities are eliminated by spinning the configurations,  
the rotation providing the centrifugal repulsion that allows a regular solution to 
exist.

On general grounds, one expects the $d>5$ new solutions in this work 
to possess rotating generalizations.
Thus one may hope that by adding (at least) one spin to the system (without changing the 
rod structure)
the configuration will be balanced ($i.e.$ without conical singularities)
for a critical value of the angular momentum.
Unfortunately, there is no simple procedure to spin a given static system. 
Moreover, all known techniques fail for the solutions considered in this work. 
Therefore, we will again use a numerical approach. 

In what follows, we present two different proposals for a metric ansatz
which may describe rotating black holes 
 with a nonspherical horizon topology.
 The first proposal is a straightforward extension of the ansatz (\ref{metric})
 used in the static case,
 and leads to equations with dependence on 
 two variables.
 The second ansatz proposal implies an {\it a priori} dependence of the unknown metric functions
 on three variables.

\subsection{A generalized  metric ansatz }

As discussed in \cite{Harmark:2004rm},
the Weyl coordinates and the
 rod-structure employed to construct $d=5$
static axisymmetric solutions 
can be generalized to the rotating case. 
Here we present arguments that 
the general framework proposed in Section 2 can also be generalized to include rotation.

In the simplest case,  
 one considers a slightly more general metric form than (\ref{metric-canonical}),
with
\begin{eqnarray}
\label{ma1}
ds^2=e^{2\nu(\rho,z)}(d\rho^2+dz^2)+G_{ij}(\rho,z)\,dx^idx^j+H(\rho,z)\,d\Omega_{d-4}^2,
 \end{eqnarray}
and the coordinates $x^i=t,\psi$, which
includes spinning configurations as well (since the metric component $g_{\psi t}$ can be nonzero).  

In what follows we show that, for $d>5$, a rotating black hole with a spherical topology of the horizon 
can be written within this ansatz (then it corresponds to 
the Myers-Perry solution with a 
single   angular momentum). 
 Another interesting  case is represented by a black hole with a $S^2\times S^{d-4}$ topology of the horizon
  rotating  with respect to
 the azimuthal angle $\psi$ (thus with a rotating $S^2$).

To make correspondence with (\ref{metric}), it is convenient to choose the following parametrization of (\ref{ma1})
\begin{eqnarray} 
\label{gen-ans-rot1}
&&ds^2=
-f_0(\rho,z) dt^2
+ f_1(\rho,z ) \left(d \rho^2+dz^2\right)
+ f_2(\rho,z ) (d\psi+W(\rho,z)dt)^2
+f_3(\rho,z) d\Omega_{d-4}^2.~~~{~~~~~~}
\end{eqnarray}
The resulting equations for $f_i,~W$ still have a dependence on only  two variables,
their structure being quite similar to that found in the static case (and thus we shall not write them here). 

The rod structure as defined in Section 2 still holds for rotating solutions
($e.g.$ $f_0(0,z)=0$ still defines the position of a horizon).
Moreover, for a generic configuration, the metric functions $f_i$
admit the same expansion at $\rho=0$
as in the static case.
As $\rho\to 0$, the new metric function $W(\rho,z)$  has the following form:
$W(\rho,z)=w_0(z)+\rho^2w_2(z)+O(\rho^4)$,
with $w_0(z)=\Omega_H=const.$ on a timelike rod, $\Omega_H$ being the event horizon velocity (note that the Killing vector
$\partial/\partial t+\Omega_H \partial/\partial \psi$ is null at the horizon).
At infinity one imposes again the same asymptotic behaviour for $f_i$,
while $W\to 8 \pi J/V_{d-2}(\rho^2+z^2)^{(d-1)/4}$, with $J$ the angular momentum of the solutions.
The mass of the solutions is read again from the asymptotic expression of $f_0$.

However, other kinds of black holes, such as the spinning Myers-Perry black holes
with multiple angular momenta, and presumably also the spinning 
balanced black holes with
$S^2\times S^2$ horizon topology, do not fit within the conformal ansatz we use. 
For example, in the cases we studied, the conical sigularities of the spinning 
$S^2\times S^2$ black hole, could not be eliminated. This should not be taken as a 
sign of the absence of regular $S^2\times S^2$ black holes, but rather a consequence 
of the very restrictive form of the metric ansatz we employ. 

Another argument comes from the recently found new extremal near horizon geometries in Ref. \cite{Kunduri:2010vg},
that are similar to the near horizon geometries of the doubly equaly spinning extremal 
 Myers-Perry black hole. 
These near horizon geometries were found as deformations of the doubly spinning extremal 
Myers-Perry black hole with equal spins and have angular cross terms that we do not consider in (\ref{ma1}).
 So, taking into account the similarities between these near horizon geometries, 
 we expect that balanced rotating $S^2\times S^2$ black hole or even more exotic, 
 less symmetric, rotating black hole solutions not to fall in the (\ref{ma1}) category.

\subsubsection{The singly spinning Myers-Perry black hole}\label{sec:MP}
One can show that a Myers-Perry  black hole
with a single nonvanishing angular momentum can  be written with the
line element (\ref{ma1}).
This solution  is usually written in the following form \cite{Myers:1986un} 
\begin{eqnarray}
\label{MPmetric}
ds^2&=& -dt^2 + \frac{\mu}{r^{d-5}\Sigma}\left( dt-a\sin^2\theta
\,d\psi\right)^2 +{\frac{\Sigma}{\Delta}}dr^2+\Sigma d\theta^2
\\
&&
{~~~~~~~~~~~~~~~~~~~~~~~~~~~~}~
\nonumber
+(r^2+a^2)\sin^2\theta\, d\psi^2 
+ r^2\cos^2\theta\, d\Omega^2_{d-4}\,,~~{~~~~}
\end{eqnarray}
where
\begin{eqnarray}
\Sigma=r^2+a^2\cos^2\theta\,,\qquad
\Delta=r^2+a^2-\frac{\mu}{r^{d-5}}\,, 
\end{eqnarray}
such that at infinity the line element (\ref{m1}) is approached.
Employing the same change of coordinates (\ref{eq:coordchange}) as in the static case, where now
\begin{eqnarray}
G(r)=2 \int\frac{1}{\sqrt{\Delta }}\, dr,
\end{eqnarray}
we can get the metric to be in the form (\ref{ma1}). 
Similar to the Schwarzschild-Tangherlini case, the above integral has a particularly 
simple form for $d=5$  \footnote{Note that also the expression for $\rho= \frac{1}{2}  
\sin 2\theta \left(\frac{2 r \sqrt{a+r^2-\mu }}{a-\mu }\right),\, 
z=  \frac{1}{2} \cos 2\theta \left(1+\frac{2 r^2}{a-\mu }\right)$ are simple in this case.}
\begin{eqnarray}\label{eq:coordchangefiveROT}
G(r)&=&2 \log\left[2 \left(r+\sqrt{a+r^2-\mu }\right)\right]\,, 
\end{eqnarray}
and for $d=7$, where
\begin{eqnarray}\label{eq:coordchangesevenROT}
G(r)=\log\left[a+2 \left(r^2+\sqrt{a r^2+r^4-\mu }\right)\right]\,.
\end{eqnarray}
Therefore the explicit form of the metric functions $\nu$, $G_{ij}$ and $H$ in (\ref{ma1})
can be derived from (\ref{MPmetric}).
However, their expression is very complicated and we shall not present it 
here\footnote{Their explicit form for $d=5$
is given $e.g.$ in \cite{Harmark:2004rm}.}.

  One may wonder whether more general Myers-Perry solutions can also be written within the ansatz (\ref{ma1}).
For example, when $n \in N$ angular momenta are equal, the Myers-Perry
 black hole exhibits a symmetry enhancement to $U(n)\times U(1) \times R_t$ symmetry. 
However, we have verified
that such black holes require a more general metric form than (\ref{gen-ans-rot1}). 

\subsubsection{A rotating $S^2$: $d=6$ black holes with $S^2\times S^2$
topology of the horizon}

The Myers-Perry black hole has a $S^{d-2}$ horizon topology.
However, in principle, solutions with a more complicated horizon 
topology can also be written within the  ansatz (\ref{ma1}).

 For example,
by employing the same methods as in the static case, 
we have constructed  $d=6$ rotating black holes with a $S^2\times S^2$
topology of the event horizon (note that only one of these two spheres has a round shape).
They are found by starting from a static  
solution and increasing the value of the angular velocity $\Omega_H$ of the event horizon,
which enters the boundary conditions at $\rho=0$.
Then, by varying also the second  parameter $a/b$,
associated with the position of the rods, the
full set of 'generalized black rings'  with a rotating $S^2$ can be explored in principle.
As the second $\psi-$rod extends to infinity ($a/b\to 0$), the 
radius  of the horizon of the round $S^{2}$-sphere increases and asymptotically it
becomes a two-plane.
Here one
expects to recover, after a suitable rescaling,  the 
four dimensional Kerr black hole uplifted to six dimensions.

Although  we did not yet investigate the full parameter space,
all solutions we have constructed so far possess a conical singularity on the finite $\psi$-rod.
This is not unexpected, 
since these configurations are natural higher dimensional counterparts of the 
 $d=5$ black ring  with 
a rotating $S^2$ found in \cite{Figueras:2005zp}.
Different from the balanced  black ring  in \cite{Emparan:2001wn} (which has $g_{\psi t}=0,~g_{\varphi t}\neq 0$),
the  solution in \cite{Figueras:2005zp} has a conical singularity
inside the ring for any allowed value of the angular momentum\footnote{The
  thermodynamical properties  of this solution are examined in \cite{Herdeiro:2010aq}.}. Balanced black rings with a rotating $S^2$ 
exist only if they rotate  along $S^1$ as well \cite{Pomeransky:2006bd}.
Therefore, we expect a similar result also for $d>5$ solutions.
However, this class of configurations would have a rotating $S^{d-4}$
and then would not be described by the ansatz (\ref{ma1}).

A description of the  spinning $d=6$ black holes with $S^2\times S^2$
topology of the horizon, together with generalizations for $d=7$ and multi-black hole objects
will be presented elsewhere.

\subsection{A rotating $S^{d-4}$}\label{sec:S4}

\subsubsection{A general metric ansatz}
Heuristically, to provide a centrifugal force which may balance 
a system with a nonspherical horizon topology, 
one needs to rotate the $S^{d-4}$ sphere\footnote{For example, in $d=5$ dimensions, the 
balanced ring is rotating with respect to the $S^1$ direction.}.
In principle, the simplest case corresponds to 
  black objects with a single angular momentum with respect to a direction on $S^{d-4}$.
  For $d>5$, a possible generalization of the static ansatz (\ref{metric}) 
  to this case has a dependence  on $\rho,z$ and 
an angular variable
 $\theta$ on $S^{d-4}$, with
\begin{eqnarray} 
\label{gen-ans-rot0}
&&ds^2= g_{ij}(\rho,z,\theta)\,dz^idz^j+G_{ij}(\rho,z,\theta)\,dx^idx^j+H(\rho,z,\theta)\,d\Omega_{d-6}^2,
\end{eqnarray} 
 where $z^i=\rho,z,\theta$ and $x^i=t,\psi,\varphi$
  (note that the metric on $S^{d-4}$ is 
 written in terms of a warped product of $S^2$ and $S^{d-6}$, with
 $d\Omega_{d-4}^2=d\theta^2+\sin^2\theta d\varphi^2+\cos^2\theta d\Omega_{d-6}^2$
 in the absence of rotation).
  
An explicit parametrization of the above line element which makes contact with the static ansatz (\ref{metric}),
proposing also a choice of the gauge in the $z^i$-sector, is
\begin{eqnarray} 
\label{gen-ans-rot}
&&ds^2=
-f_0(\rho,z,\theta) dt^2
+ f_1(\rho,z,\theta) \left(d\rho^2+S_1 (\rho,z,\theta)dz^2\right)
+ f_2(\rho,z,\theta) d\psi^2
\\
\nonumber
&&{~~~~~~~~}
+f_3(\rho,z,\theta) \bigg( d\theta^2
+S_2 (\rho,z,\theta) \sin^2\theta (d\varphi+W(\rho,z,\theta)dt)^2
+S_3 (\rho,z,\theta)\cos^2\theta d\Omega_{d-6}^2 \bigg),
\end{eqnarray}
with $S_i=1$ and $W=0$ in the static case
(note  that $S_3=0$ for $d=6$; also we did not consider solutions with rotation on $S^2$, 
$i.e.$ $g_{\psi t}=0$).
Therefore finding rotating solutions with a nonspherical topology of the horizon 
reduces to solving a set of  coupled partial differential equations for $f_i,S_i,W$,
with suitable boundary conditions.
However, this is a very difficult numerical problem, since the equations depend 
on three variables.

The only configurations we have attempted  to construct within the above ansatz
correspond  to $d=6$
black holes with $S^2\times S^2$ topology of the horizon (this time both spheres deviate from sphericity).
Again, one starts with static  solutions and increases the value of the angular velocity $\Omega_H$.
On general grounds, balanced solutions are expected to exist for a critical value of $\Omega_H$.

However, although in principle all methods and the software used in the static case
can also be applied for this 3D problem,
 so far we could not make any progress in this direction.
 The main problem is that we could not assure 
 the convergence of the numerical process in the presence of rotation, even for small values of $\Omega_H$. 
This problem may be related to the issue of the gauge choice in this case\footnote{Note also that the complexity of 
the equations increases tremendously as compared to the static limit (\ref{eqU2}), (\ref{eqnu}).}.
For the version (\ref{gen-ans-rot}) of the generic  ansatz (\ref{gen-ans-rot0}), we have fixed the metric gauge by setting to zero
the extradiagonal components $g_{\rho z},g_{\rho \theta},g_{z\theta}$.
This seems to be a natural generalization 
of the "conformal gauge" employed in (\ref{metric}) ($i.e.$ $g_{\rho \rho}=g_{zz}$ and $g_{\rho z}=0$), which is the most convenient
choice in a numerical approach.
However, different from the static case discussed above,
we could not prove the consistency of the proposed metric ansatz (\ref{gen-ans-rot}).
In other words, for rotating solutions  there is no obvious way to
prove that the constrained equations
are solved automatically via Bianchi identities (plus suitable boundary conditions), 
as for the case of solutions with dependence on $\rho,z$ only (see the discussion in Section 2.1).
This problem survives for other metric gauge choices 
we have considered.

One should also remark that $d>5$ balanced black rings with a single angular momentum can be constructed
in principle
by using the ansatz (\ref{gen-ans-rot}).
However, all technical issues pointed out above apply for ring solutions as well.
Moreover, higher dimensional black rings can also have a rotating $S^2$.   
Again, the exact solution is known only in five dimensions
\cite{Pomeransky:2006bd}. For $d>5$, we expect such solutions to have non trivial cross term metric components on the $S^{d-3}$ part. 
With the $S^{d-3}$-spins turned on, the solutions will presumably have $R\times U(1)^{n+1}$ symmetry and be cohomogeneity $d - (n + 2)$
(with $n=[(d-2)/2])$. 
And, even when $n$  angular momenta are set to be equal (and non-zero), 
the enhancement to a $U(n) \times U(1)$ rotational symmetry would not 
 be enough  
to lead to a simple metric ansatz as it was the case for the  Myers-Perry black hole with equal angular momenta
(see $e.g.$ \cite{Kunz:2006eh}).

 \subsubsection{$d=7$ rotating solutions with $S^2\times S^3$ horizon topology}\label{sec:d7}
The only possibility we have found so far to construct nonperturbative
spinning black objects with rotation on $S^{d-4}$ and possessing a nontrivial
topology of the horizon, 
corresponds to the case $d-4=2k+1$ (with $k=1,2,\dots$).
There the problem can be greatly simplified, when the
${\it a\ priori}$ independent $(d-3)/2$ angular momenta on  $S^{d-4}$
are chosen to have equal magnitude,
since this factorizes the angular dependence \cite{Kunz:2006eh}.
The problem then reduces to studying the solutions of a set of five partial
differential equations with dependence only
on the  variables $\rho,z$.

In what follows, we present some partial results for the 
simplest case $d=7$.
The metric ansatz in this case is a straightforward reduction of (\ref{gen-ans-rot0}), with
\begin{eqnarray}
\label{metric-7d} 
&&ds^2=-f_0(\rho,z)dt^2+f_1(\rho,z)(d \rho^2+dz^2)+f_2(\rho,z)d\psi^2+f_3(\rho,z)d\theta^2
\\
\nonumber
&&{~~~~~~~}+f_4(\rho,z)\big( \sin^2\theta(d\varphi_1-W(\rho,z)dt)^2+\cos^2\theta(d\varphi_2-W(\rho,z)dt)^2 \big)
\\
\nonumber
&&{~~~~~~~}-(f_4(\rho,z)-f_3(\rho,z))\sin^2\theta \cos^2\theta (d\varphi_1-d\varphi_2)^2
,
 \end{eqnarray} 
 where $0\leq \theta\leq \pi/2$, $0\leq \varphi_1,\varphi_2\leq 2 \pi$.
The static ansatz (\ref{metric}) is recovered for $W=0$ and $f_4=f_3$.
 
A suitable combination 
of the Einstein equations,
 $G_t^t=0,~G_\rho^\rho+G_z^z=0$, $G_{\theta}^{\theta}=0$, $G_{\psi}^{\psi}=0$,
 $G_{\varphi_1}^t=0$  and  $G_{\varphi_1}^{t}=0$,
yields the following set of equations for the metric functions:
\begin{eqnarray}
\nonumber
&&\nabla^2 f_0-\frac{1}{2f_0}(\nabla f_0)^2+\frac{1}{2f_2}(\nabla f_0)\cdot( \nabla f_2)+\frac{1}{ f_3}(\nabla f_0)\cdot( \nabla f_3)
+\frac{1}{2f_4}(\nabla f_0)\cdot( \nabla f_4)-f_4(\nabla W)^2=0, 
\\
\nonumber
&&\nabla^2 f_1-\frac{1}{f_1}(\nabla f_1)^2- \frac{f_1}{2f_3^2}(\nabla f_3)^2
- \frac{f_1f_4}{2f_0}(\nabla W)^2-\frac{f_1}{ f_0f_3}(\nabla f_0)\cdot( \nabla f_3)
- \frac{ f_1}{ f_3f_4}(\nabla f_3)\cdot( \nabla f_4)
\\
\nonumber
&&{~~~~~~~~} - \frac{ f_1}{ f_2f_3}(\nabla f_2)\cdot( \nabla f_3)
- \frac{ f_1}{2 f_0f_2}(\nabla f_0)\cdot( \nabla f_2)
- \frac{ f_1}{ 2f_0f_4}(\nabla f_0)\cdot( \nabla f_4)
\\
\nonumber
&&{~~~~~~~~}- \frac{ f_1}{ 2f_2f_4}(\nabla f_2)\cdot( \nabla f_4)
+\frac{2f_1^2}{ f_3^2}(4f_3-f_4)=0,
\\
\label{eqs1} 
&&\nabla^2 f_2-\frac{1}{2f_2}(\nabla f_2)^2+\frac{1}{2f_0}(\nabla f_0)\cdot( \nabla f_2)
+\frac{1}{ f_3}(\nabla f_2)\cdot( \nabla f_3)+\frac{1}{ 2f_4}(\nabla f_2)\cdot( \nabla f_4)=0,
\\
\nonumber
&&\nabla^2 f_3
+ \frac{1}{2f_0}(\nabla f_0)\cdot( \nabla f_3)
+\frac{1}{2f_2}(\nabla f_2)\cdot( \nabla f_3)
+\frac{1}{2f_4}(\nabla f_3)\cdot( \nabla f_4)
+4f_1(\frac{f_4}{f_3}-2)=0,
\\
\nonumber
&&\nabla^2 f_4
-\frac{1}{2f_4}(\nabla f_4)^2
+\frac{f_4^2}{f_0}(\nabla W)^2
+\frac{1}{2f_0}(\nabla f_0)\cdot( \nabla f_4)
+\frac{1}{2f_2}(\nabla f_2)\cdot( \nabla f_4)
\\
\nonumber
&&{~~~~~~~~}
+\frac{1}{ f_3}(\nabla f_3)\cdot( \nabla f_4)
-\frac{4f_1 f_4^2}{f_3^2}=0,
\\
\nonumber
&&\nabla^2 W
- \frac{1}{2f_0}(\nabla f_0)\cdot( \nabla W)
+\frac{1}{2f_2}(\nabla f_2)\cdot( \nabla W)
+\frac{1}{ f_3}(\nabla f_3)\cdot( \nabla W)
+ \frac{3}{2f_4}(\nabla f_4)\cdot( \nabla W)
=0.
\end{eqnarray} 
All other Einstein equations except for  $G_\rho^z=0$ and $G_\rho^\rho-G_z^z=0$
are linear combinations of those used to derive
the above equations or are identically zero.
Moreover, a similar reasoning to that presented for static solutions
 assures that the constraints $G_\rho^z=0$ and $G_\rho^\rho-G_z^z=0$
vanish  identically via Bianchi identities plus suitable boundary conditions.

The only rotating solutions we have attempted to construct within this ansatz
have an $S^2\times S^3$
topology of the horizon ($i.e.$ $d=7$ generalized black rings) and 
a rod structure  similar to that 
discussed in Section 2 ($e.g.$ the horizon is located at $\rho=0$ and $-a\leq z\leq a$, where $f_0=0$).
Moreover, the expansion at $\rho=0$ of the metric functions $f_i$ ($i=0,\dots,3$) 
is similar to that presented in the static case. 
For $\rho\to 0$ and $z\leq b$, the new functions $f_4, W$
have  the following expansion
\begin{eqnarray}
f_4(\rho,z)= f_{40}(z)+\rho^2f_{42}(z)+\rho^4f_{44}(z)+\dots,~~W(\rho,z)=w_{0}(z)+\rho^2w_{2}(z)+\dots,
\end{eqnarray}
where $w_{0}(z)=const.$ for $-a\leq z \leq a$ ($i.e.$ on the event horizon).
 The expansion of $f_4(\rho,z)$ is different for the $\Omega-$rod ($i.e.$ $\rho=0$ and a $z$-interval $[b,\infty]$), where
$ f_4(\rho,z)= \rho^2f_{10}(z)+\rho^4f_{44}(z)+\dots$.
The obvious boundary conditions for large $\rho,z$ are that $f_i$ approach the Minkowski background
functions (\ref{Mink}), while $W=0$.

The Killing vector  
$\chi=\partial/\partial_t+  \Omega_1 \partial/\partial \varphi_1+  \Omega_2 \partial/\partial \varphi_2 $
is orthogonal to and null on the horizon. 
For the solutions within the ansatz (\ref{metric}),
the event horizon angular velocities are equal, $\Omega_1=\Omega_2=W(0,z)|_{-a\leq z\leq a}=\Omega_H$.

As in the case of Myers-Perry black holes,
these rotating black holes have an ergosurface
inside of which
observers cannot remain stationary, and will 
move in the direction of the rotation. 
\newpage
\begin{figure}[ht]
\hbox to\linewidth{\hss%
	\resizebox{8cm}{6cm}{\includegraphics{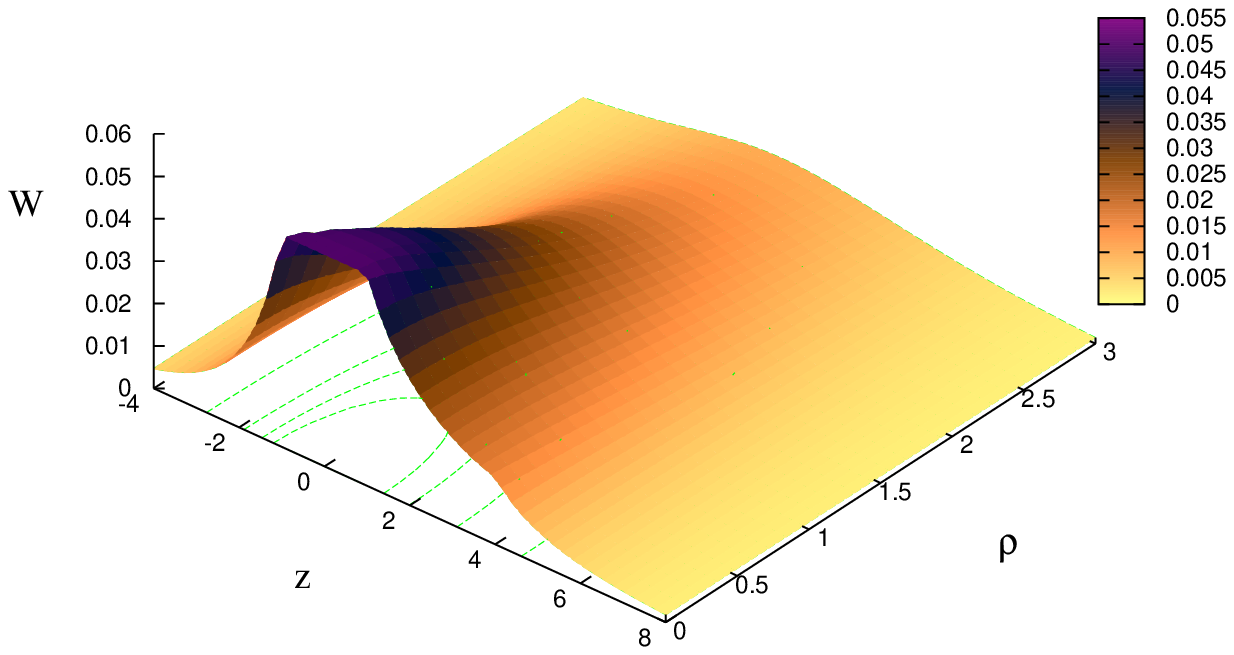}}
\hspace{15mm}%
        \resizebox{8cm}{8cm}{\includegraphics{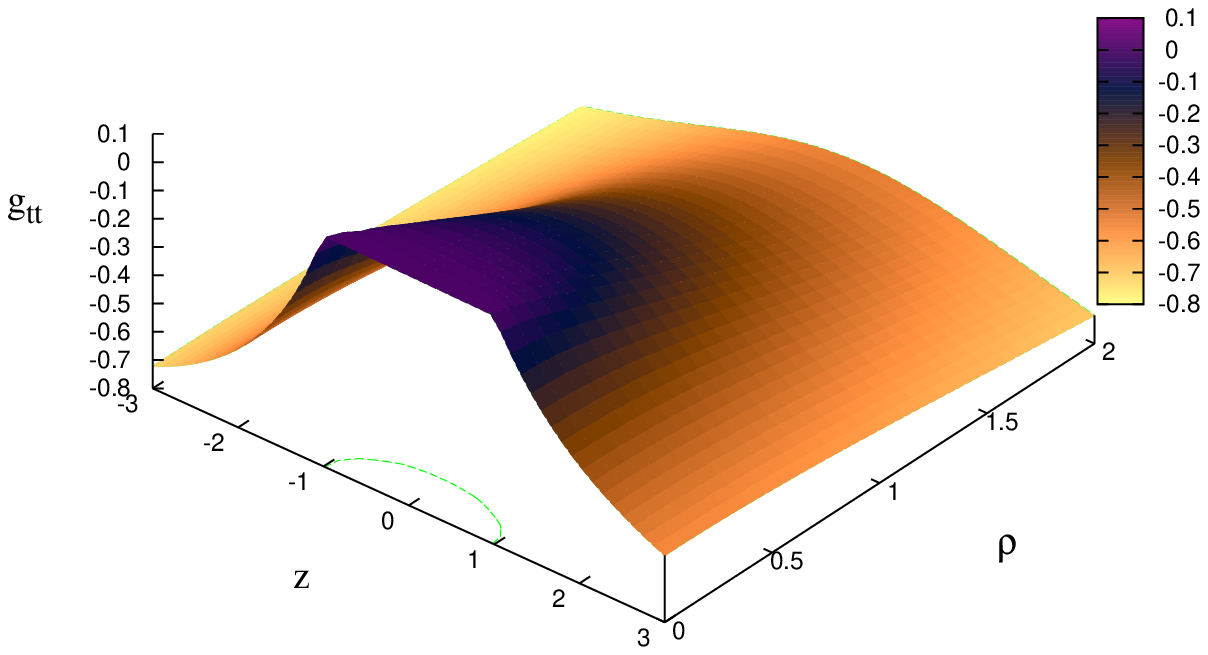}}	
\hss}
 \label{rotfig}
\end{figure}
\vspace*{0.7cm}
{\small {\bf Figure 9.}
The metric function $W$ (left) and $g_{tt}=-f_0+f_4 W^2$ (right) are shown close to the 
horizon for a $d=7$ rotating solution with $S^2\times S^3$ event horizon topology.
The input parameters here are $a=1$, $b=4$ and $\Omega_H=0.05$.
The contour line in the right panel indicates the ergo-region.
}
\\
\\
The ergosurface is located at $g_{tt}=0$, $i.e.$
\begin{eqnarray} 
\label{er}  
-f_0+f_4 W^2=0 ,
\end{eqnarray} 
and does not intersect the horizon.

The area $A_H$ and the Hawking temperature of the black hole can be expressed as
\begin{eqnarray}
\label{A} 
A_H= \frac{1}{2}\Delta \psi \int_{-a}^a dz f_3\sqrt{f_1f_2f_4}\bigg|_{\rho=0},~~~
T_H=\frac{1}{2\pi}\lim_{\rho\to 0} \sqrt{\frac{f_{0}(\rho,z)}{\rho^2 f_{1}(\rho,z)}}\bigg|_{-a\leq z\leq a}.
\end{eqnarray}
The ADM mass $M$ and the angular momenta $J_1=J_2=J$ of the solutions can be read from   the asymptotic
expression of $f_0,W$:
\begin{eqnarray}
f_0\sim 1-\frac{16 \pi M}{5V_{5}(\rho^2+z^2)}+\dots,~~W\sim \frac{8 \pi  J}{ V_{5}(\rho^2+z^2)^{3/2}}.
\end{eqnarray}
These spinning black objects satisfy the Smarr formula  
\begin{eqnarray}
\label{smarrform} 
\frac{4}{5}M =\frac{1}{4}T_HA_H+\frac{5}{2} \Omega_H J~.
\end{eqnarray}
The rotating solutions are found by starting with the static configurations with $S^2\times S^3$
topology
of the horizon discussed in Section \ref{sec:GBR} and increasing the horizon velocity $\Omega_H$.
We have found that the absolute value of the conical excess $\delta$
decreases with $\Omega_H$, which suggests the existence of a critical value of the horizon
angular velocity such that $\delta=0$
 (this is the case for the general $d=5$ Emparan-Reall  rotating black ring \cite{Harmark:2004rm}, see Appendix B).
Unfortunately, the accuracy is lost and the numerical process diverges before
approaching a balanced configuration, without being possible to identify a clear origin of this behaviour.
However, a more sophisticated numerical approach
may be able to find balanced black holes with $S^2\times S^3$
topology
of the horizon.

As an example, we show in Figure 9 the metric function $W(\rho,z)$ 
and $g_{tt}(\rho,z)$
for the rotating solution with parameters $a=1$, $b=4$ and $\Omega_H=0.05$
(the shape of the other metric functions is similar to that found in the static case).
In the right panel, one notices the existence  of a region in the $(\rho,z)$ plane
with $g_{tt}<0$ and of an ergosurface where $f_0=f_4 W^2$.

\section{Conclusions}
 The main purpose of this work was to present a general framework for the
 nonperturbative  construction of a class of  $d\geq 5$ static black objects 
with a nonspherical topology of the horizon.
The solutions are found by solving numerically a set of four partial differential
equations with suitable boundary conditions.
Such an approach may be viewed as complementary to the approximate
construction of such black objects developed recently in \cite{Emparan:2007wm},
\cite{Emparan:2009vd}, since it may work well if the length scales involved are not widely separated.
Also, this made possible to consider some black object topologies that are not captured within the blackring/fold approach ($e.g.$ $S^2\times S^2$ in $d=6$). 

As a concrete application of the proposed formalism, we have presented numerical
evidence for the existence of several $d>5$
black objects with a nonspherical
topology of the horizon.
These solutions represent generalizations of the $d=5$ static black rings, dirings and Saturns, with
similar basic properties.
Without entering into details, we mention that the double analytic continuation 
$\psi \to i T$, $t\to i \tau $ in the line element (\ref{metric}) leads to the more exotic 
interpretation of the solutions
in this work as bubble-black hole sequences in a Kaluza-Klein theory.
 For example, the black hole with a $S^2\times S^{d-4}$ topology of the horizon
 becomes a pair of  black objects (with one accelerated horizon) sitting on a bubble.
 
Not completely unexpected, our static solutions always possess conical singularities.
The only way to achieve balance seems to be to rotate the solutions,
 no other mechanism being known at this moment.
 For example, the arguments in \cite{Kleihaus:2009wh} put forward for generalized black rings
 apply directly to all static solutions in this work and one can show that  the conical singularities plague 
 also the Einstein-Maxwell-dilaton generalizations of the solutions in this paper\footnote{Static balanced black objects
 with a nonspherical topology of the horizon may exist, however, if the gauge fields
 are not vanishing at infinity, $i.e.$ for an asymptotic Melvin structure of spacetime \cite{Kleihaus:2009wh}.
 Such solutions are known in closed form in $d=5$ dimensions, see $e.g.$ \cite{Kunduri:2004da}.}.
 Moreover, the results in \cite{Kleihaus:2009dm} show that the Gauss-Bonnet corrections
 to Einstein gravity cannot eliminate the conical singularity of
 a $d=5$ static black ring,
 and we expect a similar result to hold also for the higher dimensional configurations discussed in this work.
 
 As argued in Section 4, the construction of the rotating balanced version of the solutions we have considered is a 
 much more difficult task.
 However, based on the experience with $d=5$ exact solutions,
it is likely that some of the qualitative features of the 
 static configurations will hold also in the spinning case.

A further generalization of the solutions may be along the lines
of Ref.~\cite{Schwartz:2007gj}, where (apparent) horizons
of topology $S^n \times S^{m+1}$, $n,m \ge 1$ were considered.
 
 Thus, we expect that the new configurations 
  discussed in this work represent just 'the tip of the iceberg' and a variety of new $d>5$
 black objects with nonspherical topology of the horizon are likely to be discovered 
 within a nonperturbative approach.
 In any such attempt, the rod structure of the solutions (or a suitable generalization of it) 
 would represent an important ingredient, as a tool to fix 
 the topology of the horizon.
 For example, it would be interesting to adapt the numerical methods in this work
 for the domain structure approach introduced recently in \cite{Harmark:2009dh}.
 
 In our opinion, any progress in this direction would require the development of a consistent  numerical
 scheme capable to solve as a boundary value problem the Einstein equations with
 a dependence on at least three coordinates.

\section*{Acknowledgements}
We are grateful to Roberto Emparan for clarifying comments
on the singularity structure of $d>5$-dimensional black rings,
and for bringing Ref.~\cite{Schwartz:2007gj} to our attention,
pointing out a possible extension of our current results.
M.J.R. wants to thank Oscar Varela for helpful discussions.
B.K. gratefully acknowledges support by the DFG. 
The work of E.R. was supported by a fellowship from the Alexander von Humboldt Foundation and the
Science Foundation Ireland (SFI) project RFP07-330PHY.

\appendix
\setcounter{equation}{0}
\section{Five-dimensional seeds}
For completeness and comparison with the higher dimensional counterparts,
we present in what follows the expression (in a form
suitable for numerical calculations)
and some basic properties of the
five dimensional seed solutions\footnote{To the best of our knowledge, this study is missing in the literature and thus
may be  useful for future studies.}.
Also, the functions $f_i$ below are used as background functions for the corresponding
higher dimensional solutions.

Moreover, based on the results below, one can easily construct $e.g.$ the $d=5$
counterparts of the Figures 6 and 8,
which clearly show that the $d=5$
pattern repeats 
in higher dimensions.

\subsection{The static black ring}
The  metric functions $f_i$ of the static black ring  are given  by \cite{Emparan:2001wk},\cite{Harmark:2004rm}
\begin{eqnarray}
\label{BR5d}
&f_0=\frac{P_2+2\xi_2}{P_1+2\xi_1},~
f_1= \frac{(P_1+2\xi_1+P_2 )(P_1+P_2+ \xi_1+\xi_2+c(-\xi_1+\xi_2+2\xi_3-P_1+P_2+2P_3)}
{8(1+c )(P_1+\xi_1)(P_2+\xi_2)(P_3+\xi_3)},~~~{~~~~}
\\
\nonumber
&f_2=\frac{ P_2}{P_1 } (P_3+2\xi_3),~~f_3=P_3,
\end{eqnarray}
  where $\xi_i=z-z_i,$
\begin{eqnarray}
\label{rel1}
P_i=\sqrt{\rho^2+(z-z_i)^2}-(z-z_i),
\end{eqnarray}
and
\begin{eqnarray}
z_1=-a,~~z_2=a,~~z_3=b,
\end{eqnarray}
$a$ and $b$ being two positive constants, with $a<b$ and $c=a/b$.

The leading order expansion as $\rho\to 0$ of these functions is:
\begin{eqnarray}
\nonumber
&&f_0=\frac{z+a}{z-a},~~
f_1=-\frac{a-z}{2(b-z)(a+z)},~~
f_2=\frac{\rho^2(z-a)}{2(b-z)(a+z)},~~
\\
\nonumber
&&{~~~~~~~~~~~~~~~~~~~~~~~~~~~~~~~~~~~~~~~~~~~~~~~~~~}
f_3=2(b-z),~~~~{\rm for~~} -\infty<z\leq-a,
\\
&&f_0=\frac{\rho^2}{4(a^2-z^2)},~~
f_1= \frac{2a^2}{(a+b)(a^2-z^2)},~~
f_2=\frac{2(a^2-z^2)}{(b-z)},~~
\\
\nonumber
&&{~~~~~~~~~~~~~~~~~~~~~~~~~~~~~~~~~~~~~~~~~~~~~~~~~~}
f_3=2(b-z),~~~~{\rm for~~} -a\leq z\leq a,~~~~{~~~~~}
\\
\nonumber
&&f_0=1-\frac{2a}{z+a},~~
f_1= \frac{(b-a)(z+a)}{2(b+a)(z-a)(b-z)},~~
f_2=\frac{\rho^2(z+a)}{2(z-a)(b-z)},~~
\\
\nonumber
&&{~~~~~~~~~~~~~~~~~~~~~~~~~~~~~~~~~~~~~~~~~~~~~~~~~~~~~~~~~~~~~~}
f_3=2(b-z),~{\rm for~} a\leq z\leq b,
\\
\nonumber
&&f_0=1-\frac{2a}{z+a},~~
f_1= \frac{1}{2 (z-b)},~~
f_2=\frac{ 2(z-b)(z+a)}{(z-a)},~~
f_3=\frac{\rho^2}{2(z-b)},~~~~{\rm for~~} b\leq z<\infty.
\end{eqnarray}
The mass, event horizon area and temperature of the $d=5$ static black ring are:
\begin{eqnarray}
M^{(5)} =\frac{3aV_2}{4\pi  },~~
A_H^{(5)} =8a^2\sqrt{\frac{2}{a+b}} {V_2} ,~~
T_H^{(5)} =\frac{1}{4 \pi a}\sqrt{\frac{a+b}{2}}.
\end{eqnarray}
These black rings have a conical deficit for the finite $\psi$-rod, with
\begin{eqnarray}
\delta=2\pi \left(1-\sqrt{\frac{b+a}{b-a}}\right).
\end{eqnarray}

\subsection{The static Saturn}

 The metric functions are given in this case by   \cite{Elvang:2007rd}
\begin{eqnarray}
\label{BS5d}
&&f_0 = \frac{P_1 P_3}{P_2 P_4},~~
f_1 =  \frac{(\rho^2+P_1P_2)^2(\rho^2+P_2P_3)^2(\rho^2+P_1P_4)(\rho^2+P_3P_4)P_4}
{(\rho^2+P_1^2)(\rho^2+P_2^2)(\rho^2+P_3^2)(\rho^2+P_4^2)(\rho^2+P_1P_3)^2(\rho^2+P_2P_4)},~~~{~~~~}
\\
\nonumber
&&f_2 =\frac{\rho^2P_2}{P_1P_3} ,~~
f_3 =  P_4,
\end{eqnarray}
  where $P_i$ is  given by (\ref{rel1})
and
\begin{eqnarray}
z_1=-a,~~z_2=a,~~z_3=b,~~~~z_4=c,
\end{eqnarray}
$a,b$ and $c$ being three positive constant, with $a\leq b\leq c$.
The notation here is somehow arbitrary and 
has been chosen to make contact with the results in \cite{Kleihaus:2009wh}.

This describes a multi-black hole solution, with a black ring with horizon topology $S^2\times S^1$ around an $S^3$ black hole.

The leading order expansion as $\rho\to 0$ of these functions is:
\begin{eqnarray}
\nonumber
&&f_0 =\frac{(z+a)(z-b)}{(a-z)(c-z)},~
f_1 = \frac{z-a}{2(b-z)(a+z)},~
f_2 =\frac{\rho^2(z-a)}{2(b-z)(a+z)},~
\\
\nonumber
&&{~~~~~~~~~~~~~~~~~~~~~~~~~~~~~~~~~~~~~~~~~~~~~~~~~~~~~~~~~~~~}
f_3 =2(c-z),~{\rm for~} -\infty<z\leq-a,
\\
\nonumber
&&f_0 =\frac{\rho^2(b-z)}{4(c-z)(a^2-z^2)},~
f_1 = \frac{2a^2(a+c)(b-z)}{(a+b)^2(c-z)(a^2-z^2)},~
f_2 =\frac{2(a^2-z^2)}{(b-z)},~~
\\
\nonumber
&&{~~~~~~~~~~~~~~~~~~~~~~~~~~~~~~~~~~~~~~~~~~~~~~~~~~~~~~~~~~~~}
f_3 =2(c-z),~{\rm for~~} -a\leq z\leq a,
\\
&&f_0 = \frac{(a-z)(z-b)}{(c-z)(z+a)},~
f_1 = \frac{(a-b)^2(a+c)(a+z)}{2(a+b)^2(a-c)(a-z) (b-z)},~
\\
\nonumber
&&{~~~~~~~~~~~~~~~~~~~~~~~~~~~~~~~~~~~~~~~~~~~~~~~~~~~}
f_2 =\frac{\rho^2(z+a)}{2(a-z)(z-b)},~
f_3 =2(c-z),~{\rm for~} a\leq z\leq b,
\\
\nonumber
&&f_0 = \frac{(a-z)\rho^2}{4(b-z)(c-z)(z+a)},~
f_1 = \frac{ (a+c)(a-z)}{2(a-c)(b-z) (a+z) },~
\\
\nonumber
&&{~~~~~~~~~~~~~~~~~~~~~~~~~~~~~~~~~~~~~~~~~~~~~~~~~~~~}
f_2 =\frac{2(a+z)(z-b)}{(z-a) },~
f_3 =2(c-z),~{\rm for~} b\leq z\leq c,
\\
\nonumber
&&f_0 = \frac{(a-z)(c-z)}{(z+a)(z-b)},~
f_1 = \frac{1}{2(z-c)},~
f_2 =\frac{ 2(b-z)(z+a)}{(a-z)},~
f_3 =\frac{\rho^2}{2(z-c)},~{\rm for~} c\leq z<\infty.
\end{eqnarray}
The event horizon area and temperature of the black hole with  horizon topology $S^2\times S^1$ are:
\begin{eqnarray}
A^{(5)}_{BR}=4 \pi^2\frac{4\sqrt{2}a^2\sqrt{a+c}}{a+b},~~
T^{(5)}_{BR}=\frac{a+b }{ 4\sqrt{2}\pi a \sqrt{a+c} } .
\end{eqnarray}
The same quantities for the black hole with  horizon topology $S^3$ are:
\begin{eqnarray}
A^{(5)}_{ST}= 4 \pi^2\sqrt{2}(c-b)\sqrt{\frac{(a+c)(c-b)}{c-a}},~~
T^{(5)}_{ST}=\frac{1}{\sqrt{2}\sqrt{2}\pi }\sqrt{\frac{c-a}{(a+c)(c-b)}}.
\end{eqnarray}
These solutions have a conical deficit for the finite $\psi$-rod ($i.e.$ $\rho=0,$ $-a\leq z \leq a$), with
\begin{eqnarray}
\delta=2\pi \left(1-\frac{b+a}{b-a}\sqrt{\frac{c-a}{c+a}}\right),
\end{eqnarray}
which prevents the configuration from collapsing.

The ADM mass of this system, as measured at infinity is:
\begin{eqnarray}
M=\frac{3 \pi}{4 }(2a-b+c).
\end{eqnarray} 

\subsection{The static di-ring}
The metric functions are given in this case by
\begin{eqnarray}
\label{d=5}
&f_0 = \frac{P_1 P_3}{P_2 P_4},~~f_2 =\frac{\rho^2P_2P_4}{P_1P_3P_5} ,~~
f_3 =  P_5,
\\
\label{diring5d}
&
\nonumber
f_1 =  \frac{P_5(\rho^2+P_1P_2)^2(\rho^2+P_1P_3)^2(\rho^2+P_1P_4)^2(\rho^2+P_3P_4)^2(\rho^2+P_2P_5)(\rho^2+P_4P_5)}
{(\rho^2+P_1^2)(\rho^2+P_2^2)(\rho^2+P_3^2)(\rho^2+P_4^2)(\rho^2+P_5^2)(\rho^2+P_1P_3)^2(\rho^2+P_2P_4)^2(\rho^2+P_1P_5)(\rho^2+P_3P_5)},
\end{eqnarray}
  where $P_i$ is still given by (\ref{rel1})
and
\begin{eqnarray}
z_1=-d,~~z_2=-c,~~z_3=-a,~~~~z_4=a,~~z_5=b,
\end{eqnarray}
$a,b,c$ and $d$ being three positive constants, with $d>c>a$ and $a\leq b$. 
This describes a configuration consisting of two concentric black rings with horizon topology $S^2\times S^1$. 

The leading order expansion as $\rho\to 0$ of the metric functions is:
\begin{eqnarray}
\nonumber
&&f_0 =\frac{(z+a)(d+z)}{(z-a)(c+z)},~
f_1 = \frac{(a-z)(c+z)}{2(a+z)(z-b)(d+z)},~
\\
\nonumber
&&{~~~~~~~~~~~~~~~~~~~~~~}
f_2 =\frac{\rho^2(a-z)(c+z)}{2(a+z)(z-b)(d+z)},~
f_3 =2(b-z),
~{\rm for~} -\infty<z\leq-d,
\\
\nonumber
&&f_0 =\frac{\rho^2(a+z)}{4(a-z)(c+z)(d+z)},~
f_1 = \frac{(c-d)^2(a+d)^2 }{2(a-d)^2(b+d) }\frac{(a+z)  }{(a-z)(c+z)(d+z) },~
\\
\nonumber
&&{~~~~~~~~~~~~~~~~~~~~~~}
f_2 =\frac{2(a-z)(c+z)(d+z)}{(b-z)(a+z)},~~
f_3 =2(b-z),~{\rm for~} -d\leq z\leq -c,
\\
\nonumber
&&f_0 = \frac{(a+z)(c+z)}{(z-a)(z+d)},~
f_1 = \frac{(a+d)^2 }{2(a-d)^2(b+d) }\frac{(a+z)(d+z)}{(a-z)(c+z)},~
\\
\nonumber
&&{~~~~~~~~~~~~~~~~~~~~~~}
f_2 =\frac{\rho^2(a-z)(d+z)}{2(a+z)(z-b)(c+z)},~
f_3 =2(b-z),~{\rm for~} -c\leq z\leq -a,~~~~{~~}
\\
&&f_0 = \frac{(~c+z)\rho^2}{4(d+z)( a^2-z^2)},~
f_1 = \frac{(b+c)(a+d)^2 }{2(a+c)^2(b+d) }\frac{(a-z)(c+z)}{(a+z)(z-b)(d+z)},~
\\
\nonumber
&&{~~~~~~~~~~~~~~~~~~~~~~}
f_2 =\frac{2(a^2-z^2)(d+z)}{(b-z)(c+z) },~
f_3 =2(b-z),~{\rm for~} -a\leq z\leq a,~~~~
\end{eqnarray}
\begin{eqnarray}
\nonumber
&&f_0 = \frac{(c+z)(z-a)}{(z+a)(z+d)},~
f_1 = \frac{(b+c)(a+z)(d+z)}{2(a+b)(b+d)(z-a)(c+z)},~
\\
\nonumber
&&{~~~~~~~~~~~~~~~~~~~~~~}
f_2 =\frac{ \rho^2(a+z)(d+z) }{2(a-z)(z-b)(c+z) },~
f_3 =2(b-z),~{\rm for~} a\leq z<b,
\\
\nonumber
&&f_0 = \frac{(c+z)(z-a)}{(z+a)(z+d)},~
f_1 = \frac{1}{2(z-b)},~
\\
\nonumber
&&{~~~~~~~~~~~~~~~~~~~~~~}
f_2 =\frac{2(a+z)(z-b)(d+z) }{(z-a)(c+z) },~
f_3 =\frac{\rho^2}{2(z-b)},~{\rm for~} b\leq z<\infty.
\end{eqnarray}
The event horizon area and temperature of the "left" black ring (with the horizon located
at $\rho=0$, $-d\leq z\leq -c$) are:
\begin{eqnarray}
A^{(5)}_{L}=4 \pi^2\frac{ \sqrt{2} (c-d)^2(a+d)}{(d-a) }\frac{1}{\sqrt{b+d}},~~
T^{(5)}_{L}=\frac{1}{ 2\sqrt{2}\pi  } \frac{d-a}{d+a}\frac{\sqrt{b+d}}{d-c}.
\end{eqnarray}
The same quantities for the second black ring are:
\begin{eqnarray}
A^{(5)}_{R}= 16 \pi^2\sqrt{2} \frac{a^2(a+d)}{(a+c)}\sqrt{\frac{b+c}{(a+b)(b+d)}},~~
T^{(5)}_{R}=\frac{1}{4\sqrt{2}\pi }\frac{a+c}{a(a+d)} 
\sqrt{\frac{(a+b)(b+d)}{b+c}}.~~~{~~~~}
\end{eqnarray}
These solutions have a conical deficit for the third $\psi$-rod with $a\leq z \leq b$, with
\begin{eqnarray}
\delta=2\pi \left(1- \sqrt{\frac{(a+b)(b+d)}{(b-a)(b+c)}}\right),
\end{eqnarray}
which prevents the configuration from collapsing.

The ADM mass of this system, as measured at infinity is:
\begin{eqnarray}
M=\frac{3 \pi}{4}(2a-c+d).
\end{eqnarray}

\section{ 'Isotropic' coordinates and new diagrams }
\subsection{A coordinate transformation }

The coordinates $(\rho,z)$ defined in Section 2 have the advantage to make contact with
$d=5$ Weyl coordinates and to make possible to visualise some basic properties of the solutions
in terms of rod diagrams.
However, although the domain of integration has a rectangular shape, 
the range of both $\rho$ and $z$ is unlimited and, within our numerical scheme,
it is rather difficult to construct suitable meshes, especially in the $z-$direction.

In practice, we have found another coordinate system which has proven useful in the construction of
some black holes with a nonspherical topology of the horizon. 
The  transformation between $(\rho,z)$ and the new coordinates $(r,\theta)$ goes as follows.
Starting with one of the diagrams in Figure 1, let us choose a finite rod  there as 'central rod'
($i.e.$ that extends from $-u\leq z \leq u$).
Then we introduce the coordinate transformation
\begin{eqnarray}
\label{new-coord}
&&\rho(r,\theta)=\frac{1}{2}\frac{r^4-r_0^4}{r^2}\sin 2\theta,
~~~~
z(r,\theta)=\frac{1}{2}\frac{r^4+r_0^4}{r^2}\cos 2\theta,
\end{eqnarray}
with $r_0^2= u$ and $r_0\leq r<\infty$, $0 \leq \theta \leq \frac{\pi}{2}$,
such that   the 'central rod' is located at $r=r_0$ and all $\theta$-interval.

Then all static solutions in this work (including the $d=5$ configurations\footnote{This approach
can be extended to $d=4$ axisymmetric solutions.
However, in this case the transformation between the Weyl coordinates $\rho,z$ and
the spherical coordinates $r,\theta$ is not given by (\ref{new-coord}).}) can be studied with a metric ansatz akin to (\ref{metric}), 
with 
\begin{eqnarray}
\label{re1metric}
ds^2=-f_0(r,\theta)dt^2+f_1(r,\theta)(dr^2+r^2d\theta^2)
+f_2(r,\theta)d\psi^2+f_3(r,\theta)d\Omega_{d-4}^2.
\end{eqnarray}
For $r_0=0$ (and only two semi-infinite rods), one recovers the flat spacetime metric with $f_0=f_1=1$, 
$f_2=r^2\cos^2\theta$ and $f_3=r^2\cos^2\theta$.
The expression of $f_i$ for a more complicated rod structure can easily be derived once we know the
solutions in $(\rho,z)$ coordinates (note that  $f_1$ in (\ref{re1metric}) does not coincide with $f_1$
in (\ref{metric}), since the Jacobian of the transformation (\ref{new-coord}) enters there also).

Also, it may be interesting to remark that $r$ and $\theta$ can be viewed as 'generalized isotropic coordinates'\footnote{
It may be interesting to notice that a version of the isotropic
coordinates has been used in most of the previous numerical studies
on $d=4$ asymptotically flat axisymmetric solutions, 
see $e.g.$ \cite{kk}.}.
This is justified by the observation that
for the simplest case of a single black hole with spherical
topology of the horizon, the Schwarzschild-Tangherlini solution in isotropic coordinates 
is recovered (the results for $d=5$ below can easily be generalized to higher dimensions).

One of the advantages in the numerics of these coordinates is that the range of $\theta$
is finite. Moreover, the coordinate singularities in the metric functions are easier to handle in this case.
For example, we have constructed in this way the full set of $d=6,7$ black objects with a $S^2\times S^{d-4}$
topology of the horizon, with a better accuracy than that obtained for the $(\rho,z)$ coordinate system.

The new coordinate system leads also to a new type of diagrams, which is the counterpart in $(r,\theta)$
coordinates of the rod-diagrams in Figure 1.
As one can see from (\ref{new-coord}), $\rho=0$ corresponds to $r=r_0$ or $\theta=0,\pi/2$.
This suggests to show the domain of integration  together with the boundary conditions satisfied by the metric functions
$f_0,f_2$ and $f_3$.
In our conventions,
a wavy line indicates a horizon $f_0=0$, a  thick line  means $f_2=0$ ($i.e.$ a $\psi-$rod)
 and a double thin line stands for an $\Omega-$rod, $f_3=0$, see Figure 10 (the generalization of the $d=5$ diagrams there 
to higher dimensions is straightforward). 
The horizon topology can also easily be read from that figure:
a spherical horizon continues with rods of different directions, while for a black ring, the horizon
continues with $\psi-$rods only.

These features are clearly illustrated by a number of $d=5$ exact solutions which we shall present in what follows.

\subsection{$d=5$ static solutions in 'isotropic' coordinates }
We shall start with the simplest example, corresponding to a Schwarzschild-Tangherlini black hole.
The metric functions in this case read
\begin{eqnarray}
\label{ST-rt}
&&f_0(r)=(\frac{r^2-r_0^2}{r^2+r_0^2})^2 ,~~f_1(r)=\left(1+ \frac{ r_0^2}{r^2 } \right)^2,~~
f_2(r,\theta)=\left(1+ \frac{ r_0^2}{r^2 } \right)^2\cos^2\theta,
\\
\nonumber
&&f_3(r,\theta)=\left(1+ \frac{ r_0^2}{r^2 } \right)^2\sin^2\theta,~~~{~~~~}
 \end{eqnarray}
 with an event horizon at $r=r_0$,
 the corresponding diagram being shown in Figure 10a.

The Emparan-Reall static black ring has also a relatively simple expression in these coordinates,
\begin{eqnarray}
\nonumber
&&f_0(r)=(\frac{r^2-r_0^2}{r^2+r_0^2})^2 ,~~
f_2(r,\theta)=\left( \frac{r^2+ r_0^2}{r  } \right)^4\frac{\sin^2\theta \cos^2 \theta}{f_3(r,\theta)},
\\
\label{ring-rt}
&&f_3(r,\theta)= \frac{1}{2}
\bigg(
2 R_3+ {r_b^2} \bigg(
1+\left(\frac{r_0}{r_b}\right)^4
-\left(\frac{r_0}{r_b}\right)^2\cos 2\theta 
\left[\left(\frac{r_0}{r}\right)^2+\left(\frac{r}{r_0}\right)^2\right]
\bigg)
\bigg),
\\
\nonumber
&&f_1(r,\theta)=\frac{r^2}{R_3}\left(\frac{1+\frac{r_0^2}{r^2}}{1+\frac{r_0^2}{r_b^2}}\right)^2
\left[(1+(\frac{r_0}{r_b})^4)(1+(\frac{r_0}{r})^4)+2(\frac{r_0}{r_b})^2( \frac{R_3}{r^2})
-2(\frac{r_0}{r})^2\cos 2\theta \right],
 \end{eqnarray}
where
\begin{equation}
R_3= \frac{r^2}{2}\left[
\left(1+\left(\frac{r_b}{r}\right)^4
         -2 \cos 2\theta \left(\frac{r_b}{r}\right)^2\right)
\left(1+\left(\frac{r_0^2}{r r_b}\right)^4
         -2\cos 2\theta  \left(\frac{r_0^2}{r r_b}\right)^2\right)	 
\right]^{1/2} \ .
\nonumber
\end{equation}

\newpage
\begin{figure}[ht]
\hbox to\linewidth{\hss%
	\resizebox{8cm}{6cm}{\includegraphics{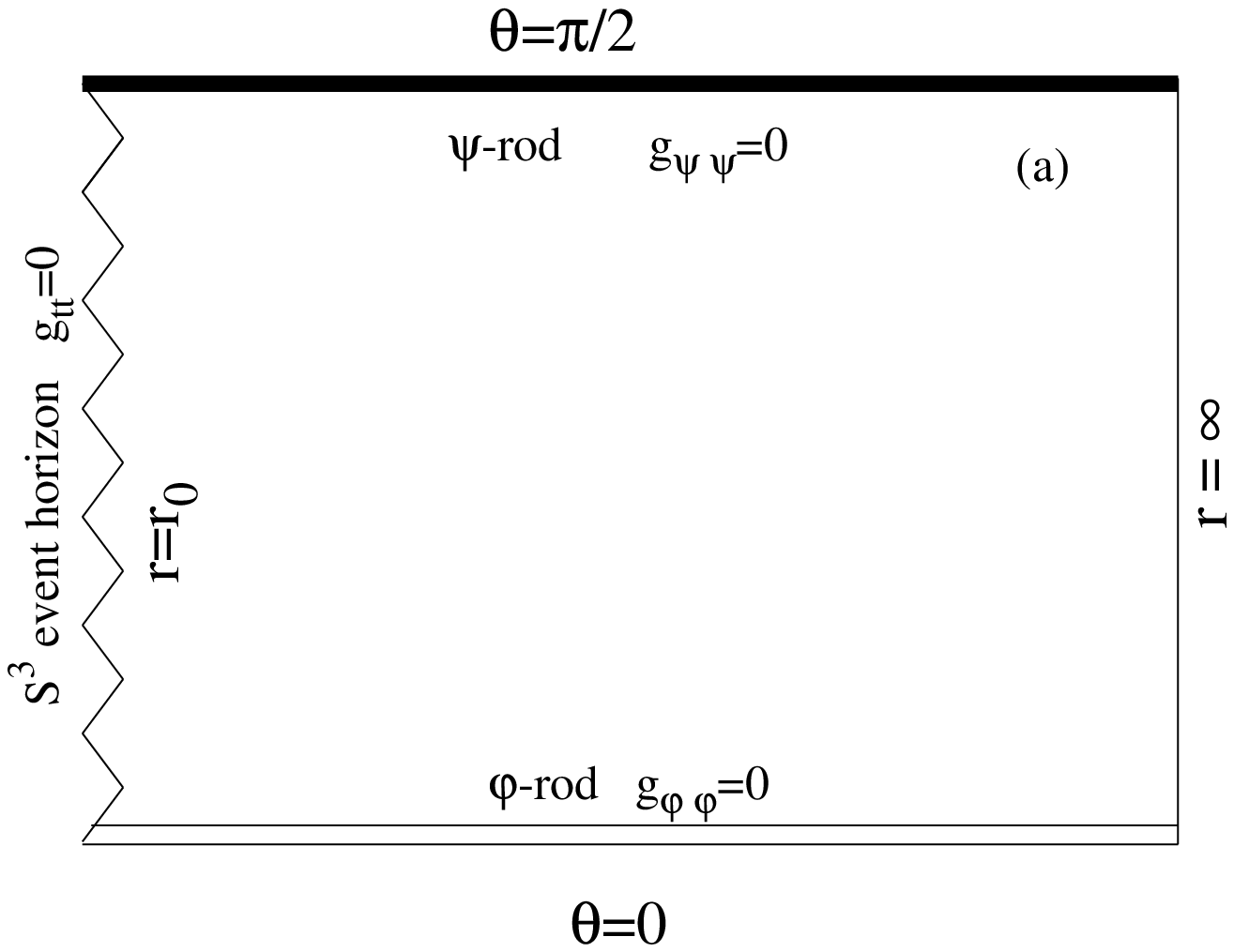}}
\hspace{15mm}%
        \resizebox{8cm}{6cm}{\includegraphics{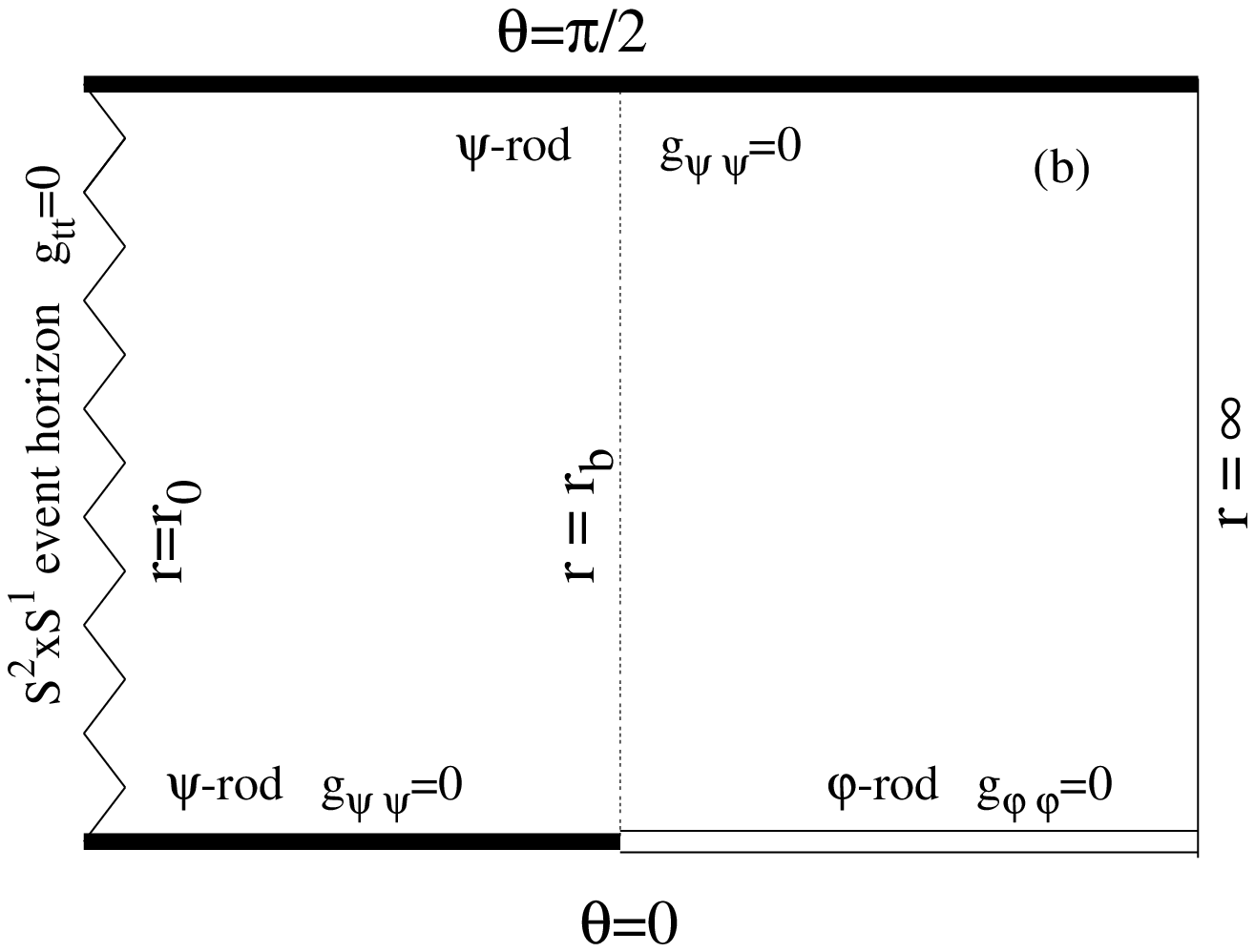}}	
\hss}
 \end{figure}
\vspace*{0.3cm}
 {\small \hspace*{3.cm}{\it  } }
\begin{figure}[ht]
\hbox to\linewidth{\hss%
	\resizebox{8cm}{6cm}{\includegraphics{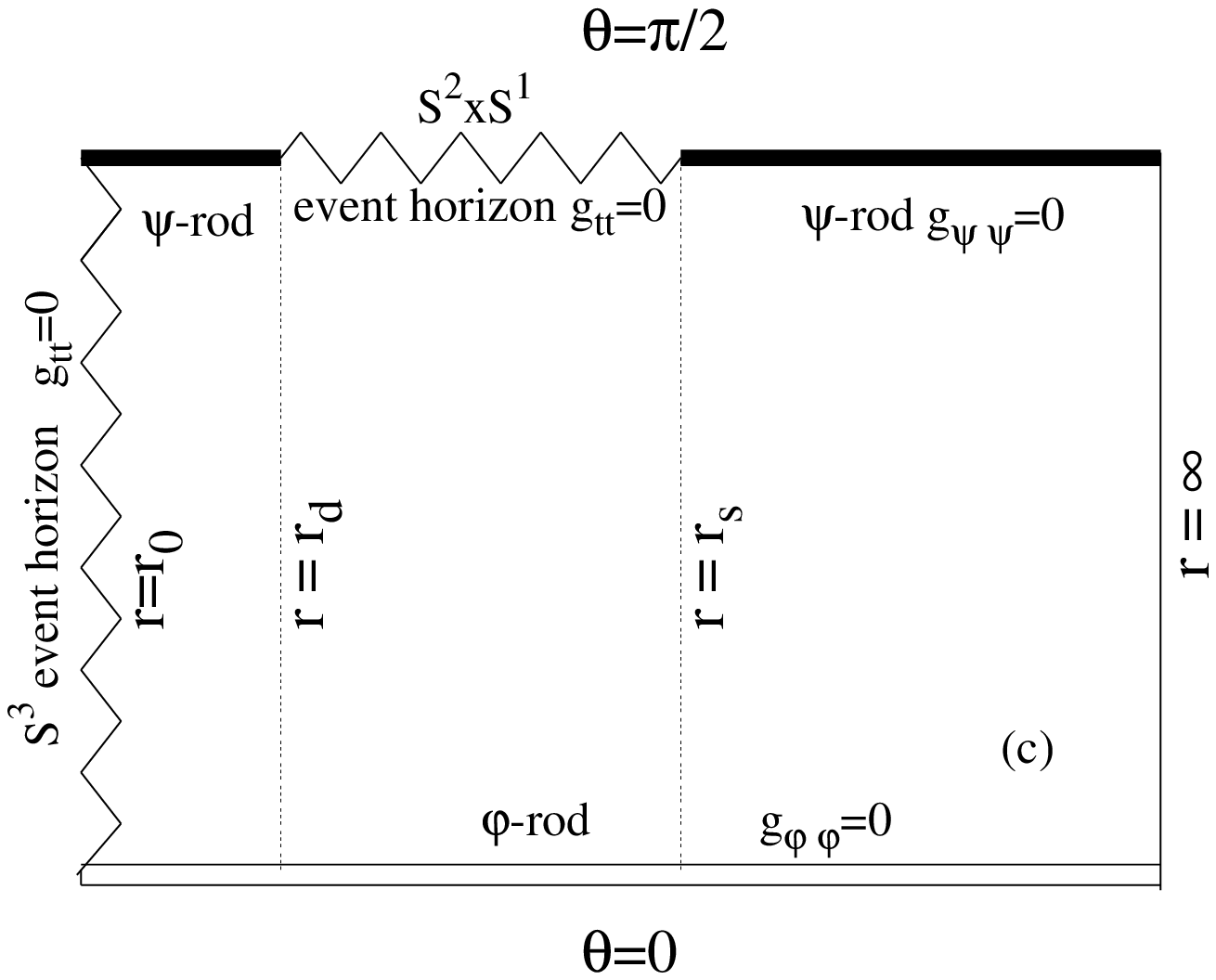}}
\hspace{15mm}%
        \resizebox{8cm}{6cm}{\includegraphics{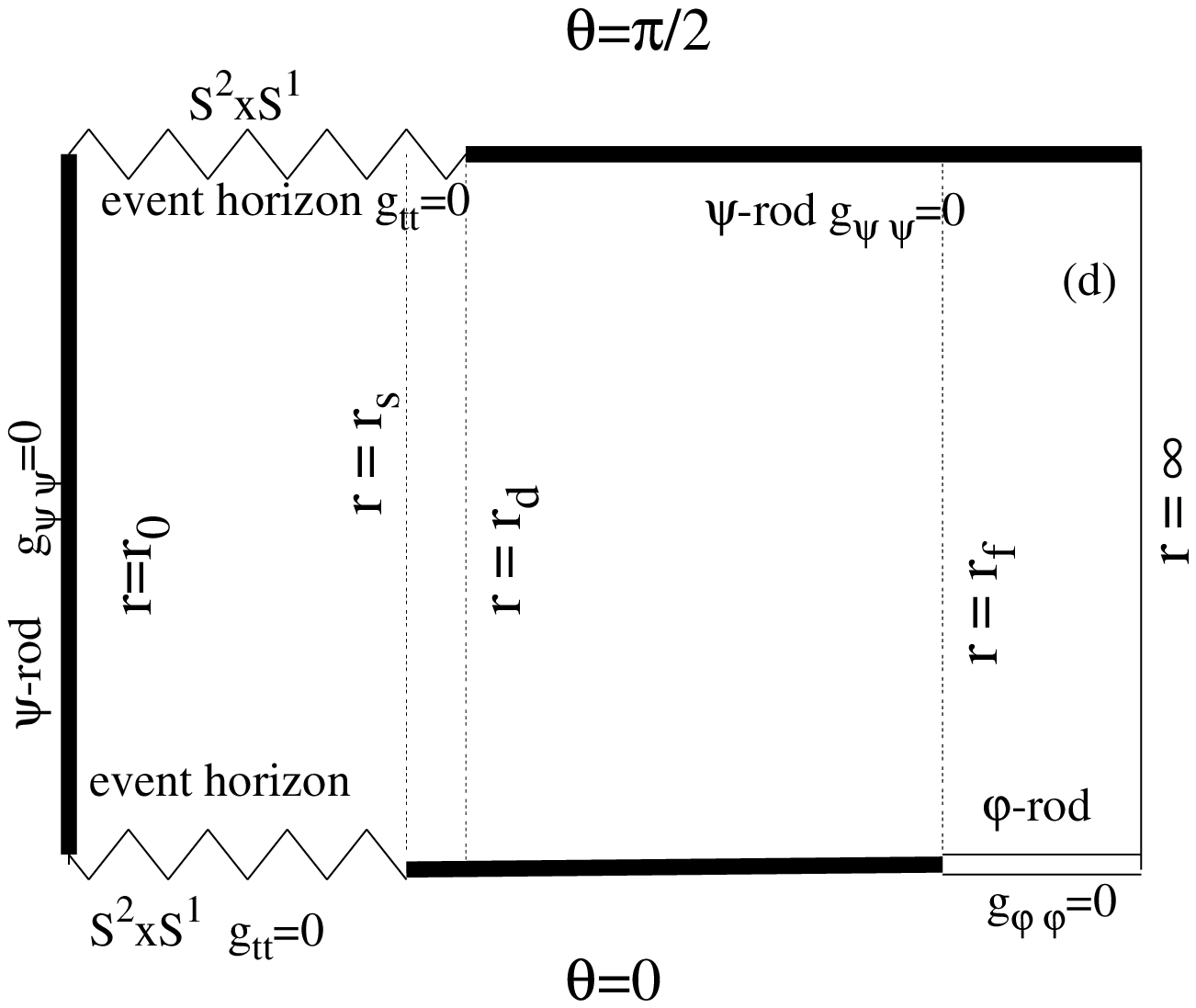}}	
\hss}
\end{figure}
\\
\\
{\small {\bf Figure 10.}
The domain of integration for the
 coordinate system (\ref{re1metric})   is shown for
 a Schwarzschild black hole, a static black ring, a black Saturn
 and a diring in $d=5$ dimensions. } 
 \\
 \\
(Note that the metric function $g_{tt}$ has no angular dependence.)
The new type of rod-diagram for a static black ring is shown in Figure 10b.
The horizon is  again located at $r=r_0$, the finite $\psi$-rod with an
angular excess being at $\theta=0$, $r_0\leq r\leq r_b$. 


The black Saturn can also be written in $(r,\theta)$-coordinates, with the 
following expression of the metric functions
\begin{eqnarray}
\label{saturn-rt}
&&f_0(r,\theta)=(\frac{r^2-r_0^2}{r^2+r_0^2})^2\frac{R_1}{R_2 },~~~
f_2(r,\theta)= \frac{(r^2+r_0^2)^2}{r^2}\frac{R_2}{R_1 }\cos^2\theta,~~~
f_3(r,\theta)=\frac{(r^2+r_0^2)^2}{r^2}\sin^2\theta,
\\
\nonumber
&&f_1(r,\theta)=
\left(1+\frac{r_0^2}{r^2}\right)^2
\left(
\frac{R_2 +\frac{(r^2+r_0^2)^2}{r^2}\cos^2\theta}
{R_1+\frac{(r^2+r_0^2)^2}{r^2}\cos^2\theta}
\right)^2
\frac{R_1 +\frac{(r^2-r_0^2)^2}{r^2}\cos^2\theta}
{R_2 +\frac{(r^2-r_0^2)^2}{r^2}\cos^2\theta}
\frac{(\rho^2(r,\theta)+R_1 R_2 )^2}{(\rho^2(r,\theta)+R_1^2)(\rho^2(r,\theta)+R_2^2)},
\end{eqnarray}
where we have introduced the auxiliary functions
\begin{eqnarray}
\nonumber
&&R_1(r,\theta)=\frac{1}{2r^2 r_s^2}
\bigg(
-r^2(r_0^4+r_s^4)-r_s^2(r^4+r_0^4)\cos 2\theta
\\
\nonumber
&&{~~~~~~~~~~~~~~~~~~~~~~~~~~}+\sqrt{ (r^4+r_s^4+2r^2r_s^2\cos 2\theta)(r_0^8+r^4 r_s^4+2r^2r_0^4r_s^2\cos 2\theta)}
\bigg),
\\
\nonumber
&&R_2(r,\theta)=\frac{1}{2r^2 r_d^2}
\bigg(
-r^2(r_0^4+r_d^4)-r_d^2(r^4+r_0^4)\cos 2\theta~
\\
\nonumber
&&{~~~~~~~~~~~~~~~~~~~~~~~~~~}
+\sqrt{ (r^4+r_d^4+2r^2r_d^2\cos 2\theta)(r_0^8+r^4 r_d^4+2r^2r_0^4r_d^2\cos 2\theta)}
\bigg),
\end{eqnarray}
and $\rho(r,\theta)$ as given by (\ref{new-coord}).
The new rod-diagram for a static black Saturn is shown in Figure 10c.
One can notice the existence of two horizons, at $r=r_0$ and at $\theta=\pi/2$, $r_d\leq r\leq r_s$, respectively.

Finally, we give also the expression of the metric functions for a static di-ring in
$(r,\theta)$-isotropic coordinates (the corresponding diagram is shown in Figure 10d)
\begin{eqnarray} 
\nonumber
&&f_0(r,\theta)=(\frac{r^2+r_0^2}{r^2-r_0^2})^2\frac{R_1}{R_4 },~~~
f_2(r,\theta)= \frac{(r^2-r_0^2)^4}{4r^4}\frac{R_4}{R_1R_5 }\sin^2 2\theta,~~~
f_3(r,\theta)=R_5,
\\
\label{diring-rt}
&&f_1(r,\theta)=\frac{ R_5\sin^2\theta}{r^8}(r^2-r_0^2)^4(r^2+r_0^2)^2
\left(
\frac{R_1 +\frac{(r^2+r_0^2)^2}{r^2}\cos^2\theta}
{R_1+\frac{(r^2-r_0^2)^2}{r^2}\cos^2\theta}
\right)^2
\left(
\frac{R_4 +\frac{(r^2-r_0^2)^2}{r^2}\cos^2\theta}
{R_4+\frac{(r^2+r_0^2)^2}{r^2}\cos^2\theta}
\right)^2
\\
\nonumber
&&{~~~~}\times
\left(
\frac{R_5 +\frac{(r^2+r_0^2)^2}{r^2}\cos^2\theta}
{R_5^2+\rho^2(r,\theta)}
\right) 
\frac{(R_4R_5+\rho^2(r,\theta))(R_1R_4+\rho^2(r,\theta))^2}
{(R_1^2+\rho^2(r,\theta))(R_4^2 +\rho^2(r,\theta)(R_1R_5+\rho^2(r,\theta))
(R_3R_5+\rho^2(r,\theta))
},
\end{eqnarray}
The auxiliary functions $R_i$ have the following expression 
\begin{eqnarray}
\nonumber
&&R_1(r,\theta)= \frac{1}{2r^2r_s^2}
\bigg(
-r^2(r_0^4+r_s^4)-(r^4+r_0^4)r_s^2\cos 2\theta
\\
\nonumber
&&{~~~~~~~~~~~~~~~~~~~~~~~~~~}
+\sqrt{(r^4+r_s^4+2r^2r_s^2\cos 2\theta)(r_0^8+r^4 r_s^4+2r^2r_0^4r_s^2\cos 2\theta)}
\bigg),
\\
\nonumber
&&R_4(r,\theta)= \frac{1}{2r^2r_d^2}
\bigg(
r^2(r_0^4+r_d^4)-(r^4+r_0^4)r_d^2\cos 2\theta
\\
\nonumber
&&{~~~~~~~~~~~~~~~~~~~~~~~~~~}
+\sqrt{(r^4+r_d^4-2r^2r_d^2\cos 2\theta)(r_0^8+r^4 r_d^4-2r^2r_d^4r_0^2\cos 2\theta)}
\bigg),
\\
\nonumber
&&R_5(r,\theta)= \frac{1}{2r^2r_f^2}
\bigg(
r^2(r_0^4+r_f^4)-(r^4+r_0^4)r_f^2\cos 2\theta
\\
\nonumber
&&{~~~~~~~~~~~~~~~~~~~~~~~~~~}
+\sqrt{(r^4+r_f^4-2r^2r_f^2\cos 2\theta)(r_0^8+r^4 r_f^4-2r^2r_f^4r_0^2\cos 2\theta)}
\bigg),
\end{eqnarray}
Different from the previous cases, the line $r=r_0$
corresponds in this case to a finite $\psi$-rod with a conical excess.
The first event horizon is located at $\theta=0$, $r_0\leq r\leq r_s$, and
the second one at $\theta=\pi/2$, $r_0\leq r\leq r_d$.

These examples make clear that the functions $f_i$ have a manageable expression
also in $(r,\theta)$-coordinates.
In particular, the singularities in the expression of $f_1$
at $\rho=0$, $z=\pm u$ are eliminated by the coordinate transformation (\ref{new-coord}).
Also, it is straightforward to perform a systematic study of these solutions in these coordinates,
similar to that considered in Appendix A.

\subsection{Rotating solutions: balanced black ring and Myers-Perry black hole}

The coordinate system introduced above turns out to be very useful also in the
numerical construction of $d=5$ rotating solutions.
We illustrate that by exhibiting some results
for the balanced Emparan-Reall black rings and Myers-Perry black holes
with a single angular momentum.

The spinning solutions can be constructed within a simple generalization of (\ref{re1metric}), with
\begin{eqnarray}
\label{metric-rot5}
ds^2=-f_0(r,\theta) dt^2+ \frac{1}{f_1(r,\theta)}(dr^2+r^2 d\theta^2)
+ f_2(r,\theta) d\psi^2
+ f_3(r,\theta) (d\varphi+\frac{w(r,\theta) }{r}dt)^2,~~{~~~}
 \end{eqnarray}
 such that the horizon is located at a fixed value of $r=r_0$.
 Expanding the Einstein equations in the vicinity of the horizon in powers of $r-r_h$, one finds   
$f_i(r,\theta)=f_{i0}(\theta)+f_{i2}(\theta)(r-r_h)^2+O(r-r_h)^3$, $w(r,\theta)=w_h+w_{2}(\theta)(r-r_h)^2+O(r-r_h)^3$,
(where the functions $f_{ik}(\theta),w_{2}(\theta)$ are
solutions of a complicated set 
of nonlinear second order ordinary differential equations
and $f_{00}(\theta)=0$), which leads to an
  event horizon metric 
\begin{eqnarray}
\label{eh-m}
d\sigma^2=\frac{r_h^2 d\theta^2}{f_{10}( \theta)}
+f_{20}( \theta)d\psi^2
+f_{30}(\theta)d\varphi^2.
\end{eqnarray}
The Hawking temperature, entropy and the event horizon 
velocity\footnote{Note that the Killing vector
$\partial/\partial t+\Omega_H \partial/\partial \varphi$ is null at the horizon. Also, the Einstein equation $G_r^\theta=0$
implies that the Hawking temperature is constant.} of the solutions are given by
\begin{eqnarray}
\label{num-quant}
T_H= \frac{1}{2\pi}\sqrt{f_{02}(\theta)f_{10}(\theta)},~~S= {\pi^2 r_0} \int_{\theta=0}^{\pi/2}
d\theta
\sqrt{\frac{f_{20}(\theta)f_{30}(\theta)}{f_{10}(\theta)}},~~\Omega_H=\frac{w_h}{r_0}.
 \end{eqnarray}
  
 For any topology of the horizon, as $r\to \infty$, the Minkowski spacetime background is recovered,
 with $f_0=f_1= 1$, $f_2=r^2\cos^2 \theta$,  $f_3=r^2\sin^2 \theta$, 
 $w=0$.
  The mass $M$ and the angular momentum $J$ of the solutions are read from the asymptotic expansion 
   of the metric functions,
 $f_0=1- {8\pi  M}/{3 \pi r^2}+\dots$,  $w=  {4  J}/{\pi r^2}+\dots$.  
 
However, the expression of the metric functions $f_i,w$
are quite complicated for any topology of the horizon.
For example, a straightforward but cumbersome  computation based on the Weyl-coordinate expressions in \cite{Harmark:2004rm}
leads to the following metric functions of a balanced black ring in the ($r,\theta$)-coordinates proposed above: 
\begin{eqnarray}
\label{funct-balanced-rings}
&&f_0=\frac{1}{(r^2+r_0^2)^2}\frac{S_1^2S_3S_4S_5}{U_1Q},
~~
f_1 =2(r_b^4-r_0^4)^2r^6\frac{R_3}{S_1S_7},~~
f_2=-\frac{(r^2+r_0^2)^2}{2r^2r_b^2(r^2-r_0^2)^2}S_6,~~~{~~~~}
\\
\nonumber
&&f_3=\frac{(r^2-r_0^2)^2}{2r^2r_b^2(r^2+r_0^2)^2}\frac{Q}{S_1S_3},~~
w=-4\sqrt{2}\frac{r^3r_0^2r_b(r^2+r_0^2)^2(r_0^2+r_b^2)\sqrt{r_0^4+r_b^4}}{(r^2-r_0^2)^2(r_b^2-r_0^2)} 
\frac{S_2S_3}{Q},
\end{eqnarray}
where, in order to simplify the expresssions, we have defined
\begin{eqnarray}
\nonumber
&&S_1=(r^4+4r^2r_0^2+r_0^4)(r_0^4+r_b^4)+4 r^2r_0^4r_b^2\cos 2\theta -4r^2r_0^2r_b^2R_3,
\\
\nonumber
&&S_2=-(r^4+r_0^4)r_b^2+r^2(r_0^2+r_b^2)^2-2r^2r_0^2r_b^2\cos 2\theta+2r_b^2r^2R_3,
\\
\nonumber
&&S_3=(r^4-4r^2r_0^2+r_0^4)(r_0^4+r_b^4)+4 r^2r_0^4r_b^2\cos 2\theta -4r^2r_0^2r_b^2R_3,
\\
&&S_4= (r_b^4+r_0^4)r^2-r_b^2(r^4+r_0^4) \cos 2\theta+2r_b^2r^2R_3,
\\
\nonumber
&&S_5=(r^4+4r^2r_0^2+r_0^4)(r_0^4+r_b^4)+4 r^2r_0^4r_b^2\cos 2\theta +4r^2r_0^2r_b^2R_3,
\\
\nonumber
&&S_6=(r_b^4+r_0^4)r^2-r_b^2(r^4+r_0^4) \cos 2\theta-2r_b^2r^2R_3,
\\
\nonumber
&&S_7= (r^4+r_0^4)(r_0^4+r_b^4) -4r^2r_0^4r_b^2\cos 2\theta+4r_0^2r_b^2r^2R_3,
\end{eqnarray}
and
\begin{eqnarray}
\nonumber
&&U_1=(r^4+r_0^4)(r_0^4-r_b^4)^2 +2r^2r_0^2(3r_0^4+r_b^4)(r_0^4+3r_b^4)
+16r^2r_0^4r_b^2(r_0^4+r_b^4)\cos 2\theta,
\\
&&
Q=S_1^2S_4-\frac{16r^2r_0^4(r^2+r_0^2)^2(r_0^2+r_b^2)^2(r_0^4+r_b^4)}{(r^2-r_0^2)^2(r_0^2-r_b^2)^2}S_2^2,
\\
\nonumber
&&
R_3=\frac{1}{2r^2r_b^2}
\sqrt{(r^4+r_b^4-2r^2r_b^2\cos 2\theta)(r_0^8+r^4 r_b^4-2r^2r_0^4r_b^2\cos 2\theta)}~.
\end{eqnarray}
The physical quantities  are complicated functions of the 
input parameters $r_0,r_b$
\begin{eqnarray}
\label{th-value}
&&M= {3\pi } \frac{r_0^2(r_0^4+r_b^4)}{(r_0^2-r_b^2)^2},~~
J= {\sqrt{2}\pi } \frac{r_0^2(r_0^2+r_b^2)^3\sqrt{r_0^4+r_b^4}}{r_b(r_0^2-r_b^2)^3},~~
\Omega_H=\frac{r_b(r_0^2-r_b^2)}{\sqrt{2}(r_0^2+r_b^2)\sqrt{r_0^4+r_b^4}},~~~~{~~~~}
\\
\nonumber
&&
T_H=\frac{(r_0^2-r_b^2)^2}{8\sqrt{2}\pi r_0^2r_b \sqrt{r_0^4+r_b^4}},~~~
S= {8\sqrt{2}\pi^2} \frac{r_0^4r_b\sqrt{r_0^4+r_b^4}}{(r_0^2-r_b^2)^2}.
\end{eqnarray}

The existence of this exact solution  
allows us to test the  scheme developed in this work by recovering numerically
the balanced black ring starting with the static solution.
Then the Einstein equations are solved for the metric ansatz (\ref{metric-rot5})
and the rod structure in Figure 10b, looking for balanced solutions.
 Again, in practice we use a set of 
 background functions which takes automatically into account the  sets
 of conditions on the boundaries and determines the topology of the horizon.
Therefore one defines $f_i=F_i f_i^{(b)}$, where
  $f_i^{(b)}$ are the functions of the  static black ring as given by (\ref{ring-rt}).
  In our approach, the position of the horizon $r_0$ and the radius of the ring $r_b$
  are kept fixed and one varies the event horizon velocity $\Omega_H$.
  All other relevant quantities are evaluated from the numerical output.
When increasing the boundary parameter $w_h$, the absolute value of the 
angular deficit excess decreases, such that $\delta$ becomes zero for a critical value of
the event horizon velocity (afterwards the ring becomes over-rotating with $\delta>0$).
By varying the value of $r_b$ (or the position of the horizon), the full spectrum of
Emparan-Reall balanced black rings can be recovered numerically. 

This approach turns out to be consistent and usually provides very good accuracy results\footnote{Note, however,
that the black rings with a  large radius 
or those close to the naked singularity point are difficult to obtain with enough accuracy.}.
A crucial ingridient of our approach is that all numerical singularities are absorbed 
\newpage
\setlength{\unitlength}{1cm}
\begin{picture}(15,21)
\put(-1,0){\epsfig{file=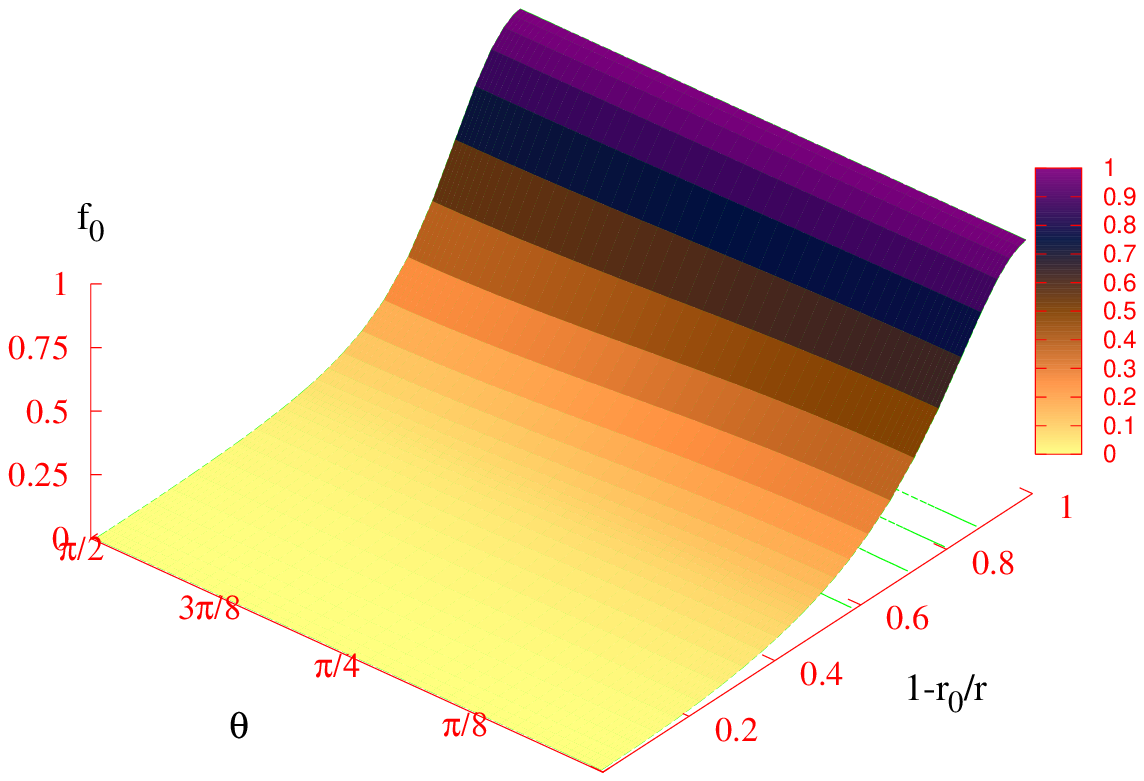,width=7.5cm}}
\put(7,0){\epsfig{file=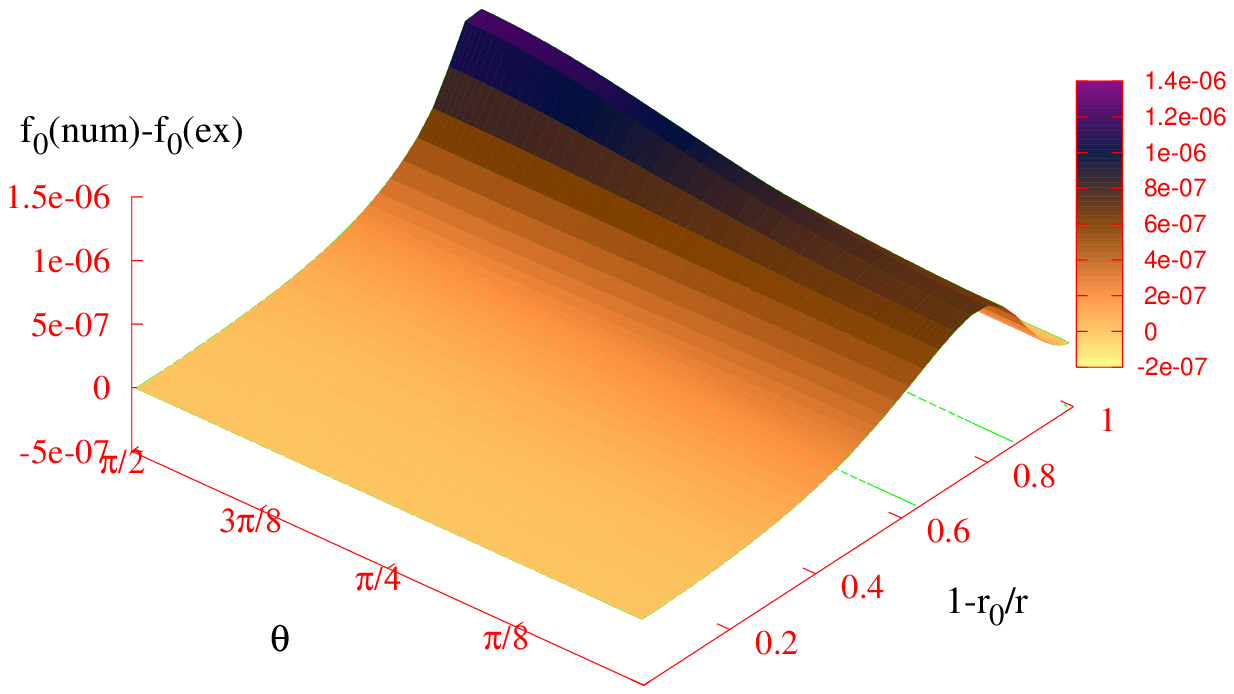,width=7.5cm}}
\put(-1,6){\epsfig{file=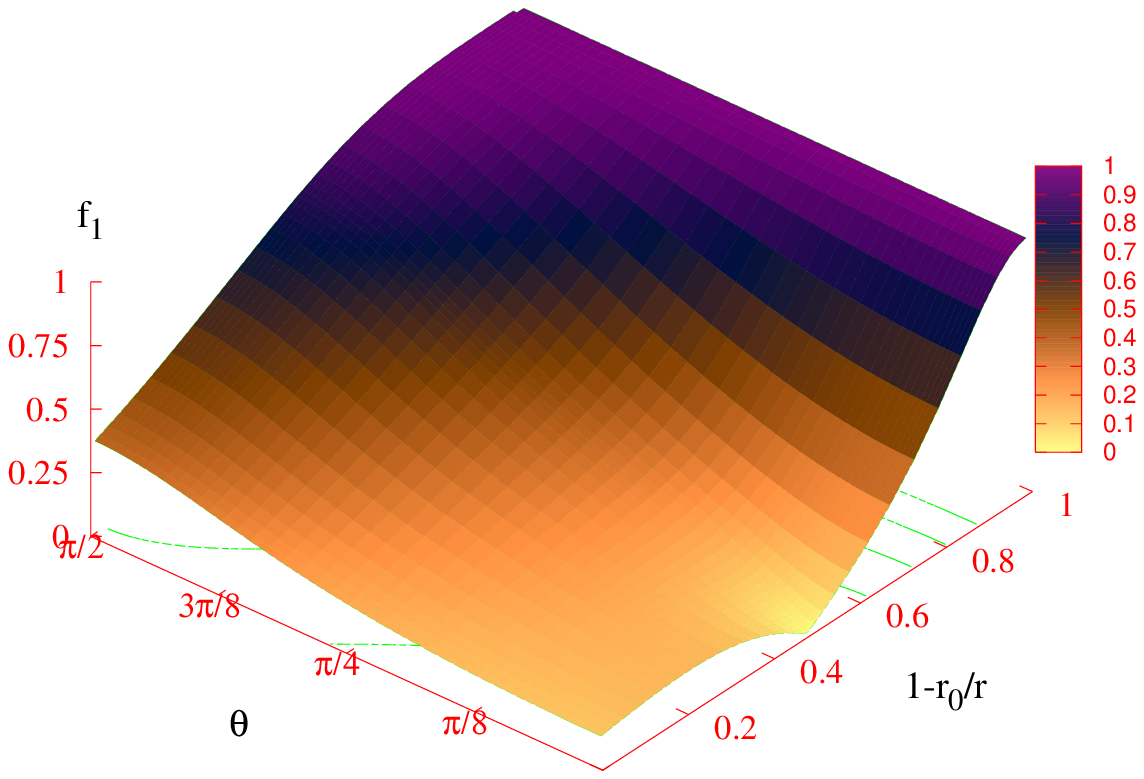,width=7.5cm}}
\put(7,6){\epsfig{file=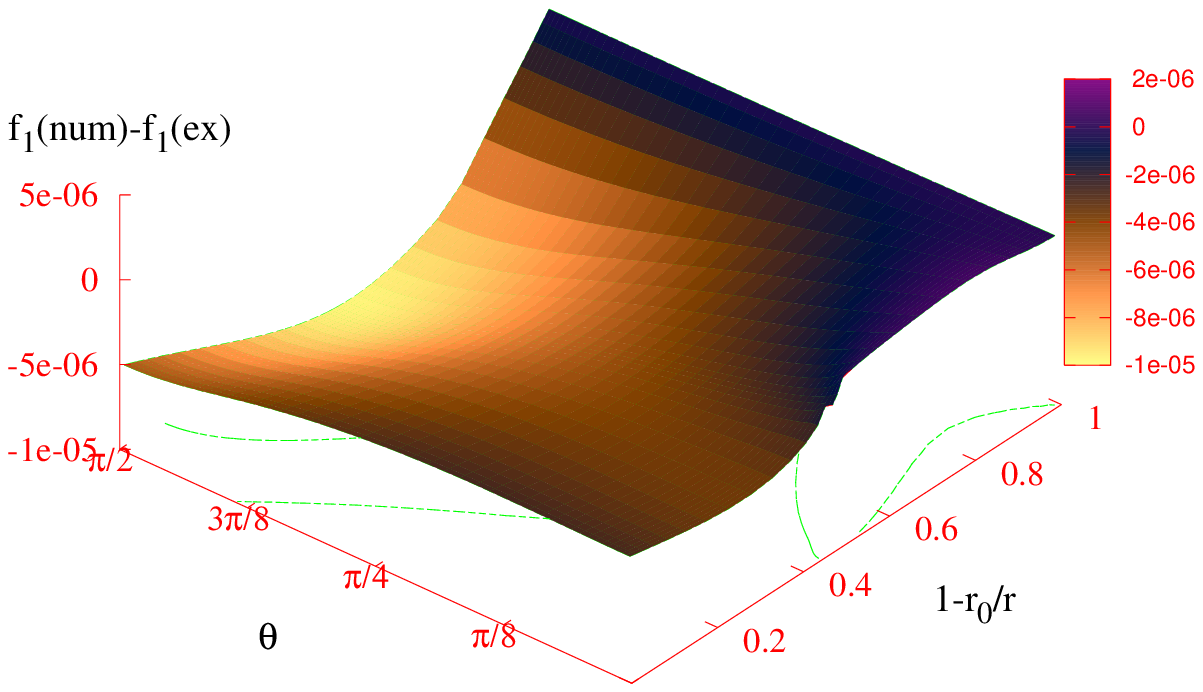,width=7.5cm}}
\put(-1,12){\epsfig{file=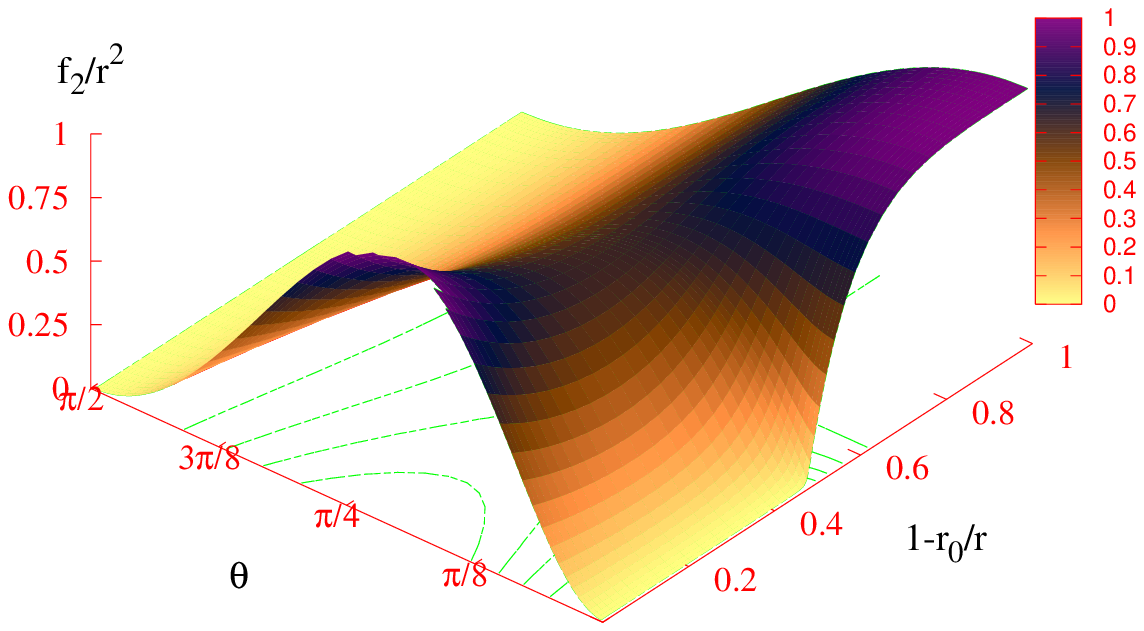,width=7.5cm}}
\put(7,12){\epsfig{file=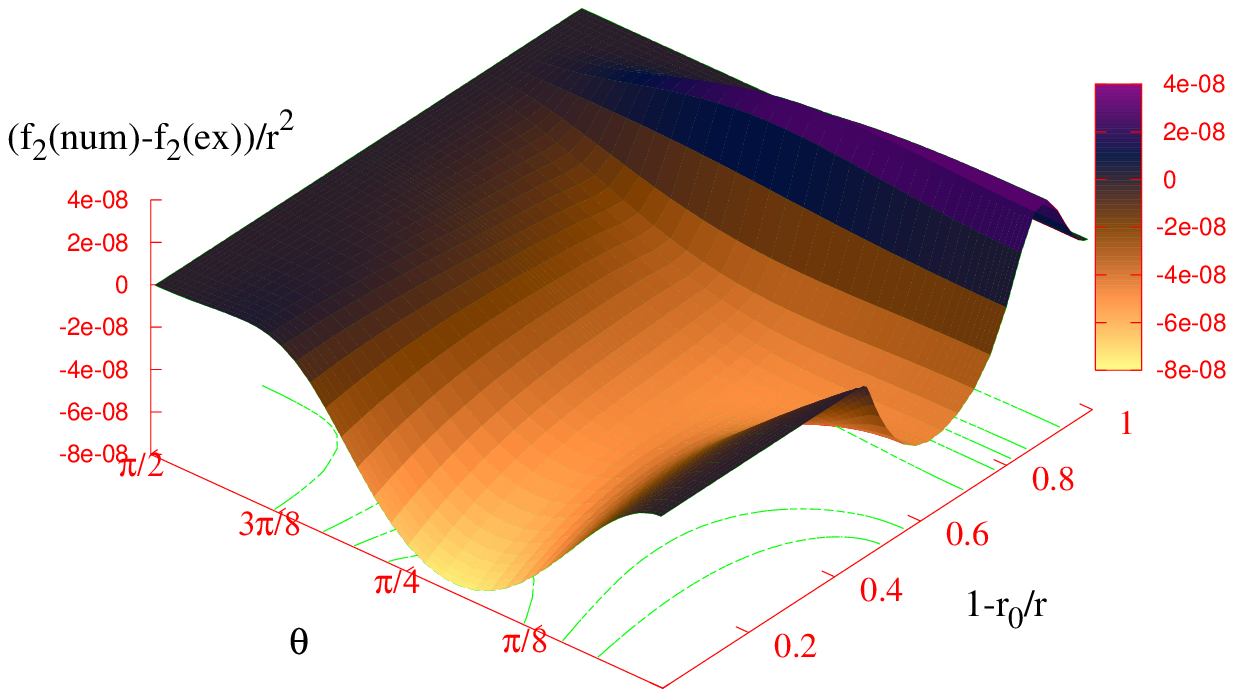,width=7.5cm}}
\put(-1,18){\epsfig{file=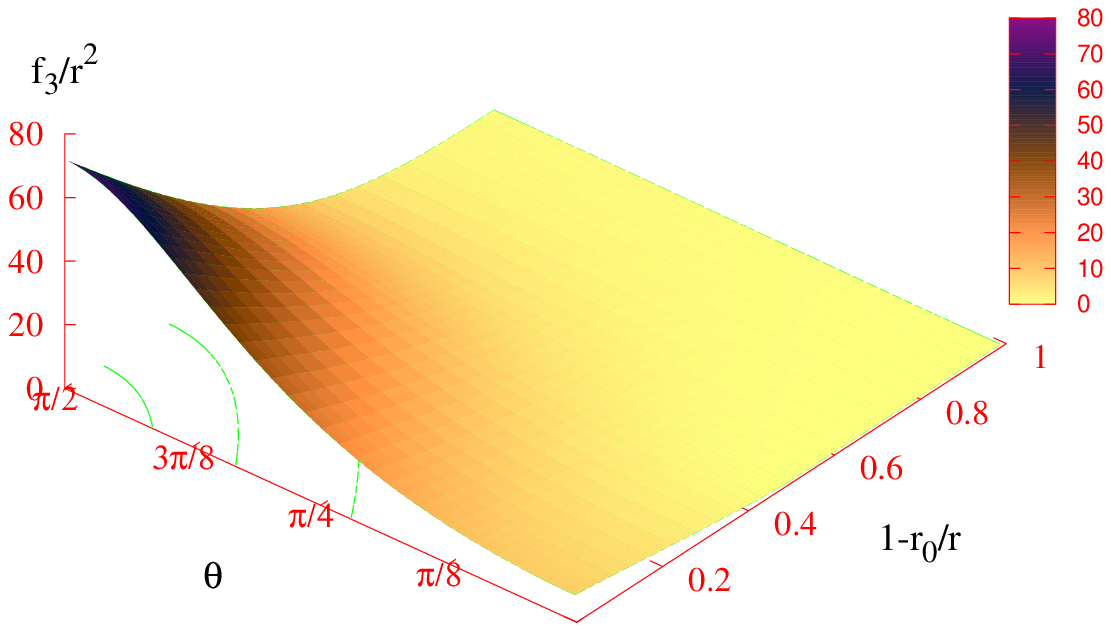,width=7.5cm}}
\put(7,18){\epsfig{file=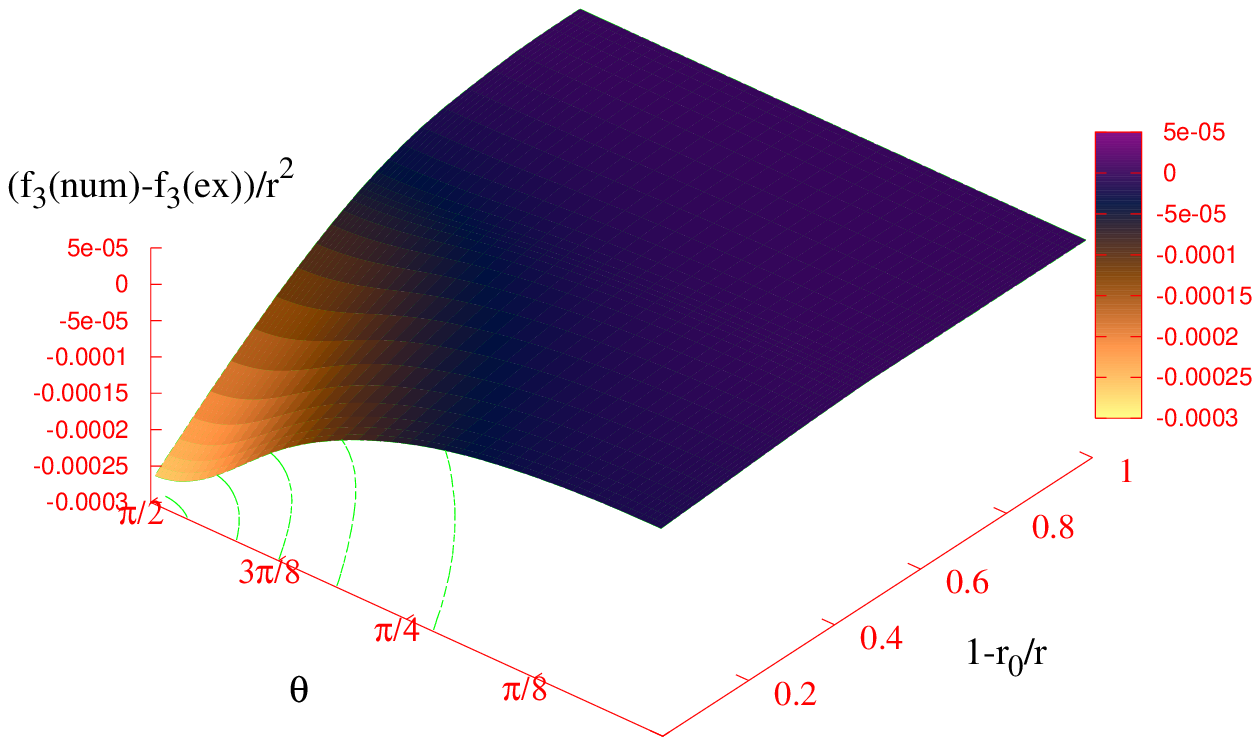,width=7.5cm}}
\end{picture} 
\\
{\small {\bf Figure 11.}
 The metric functions $f_i$ as given by (\ref{funct-balanced-rings})
 together with difference between the exact solution and the
 numerical solution
are shown for a typical
$d=5$ balanced black ring.
} 
  \newpage
\setlength{\unitlength}{1cm}
\begin{picture}(8,4)
\put(-1,0){\epsfig{file=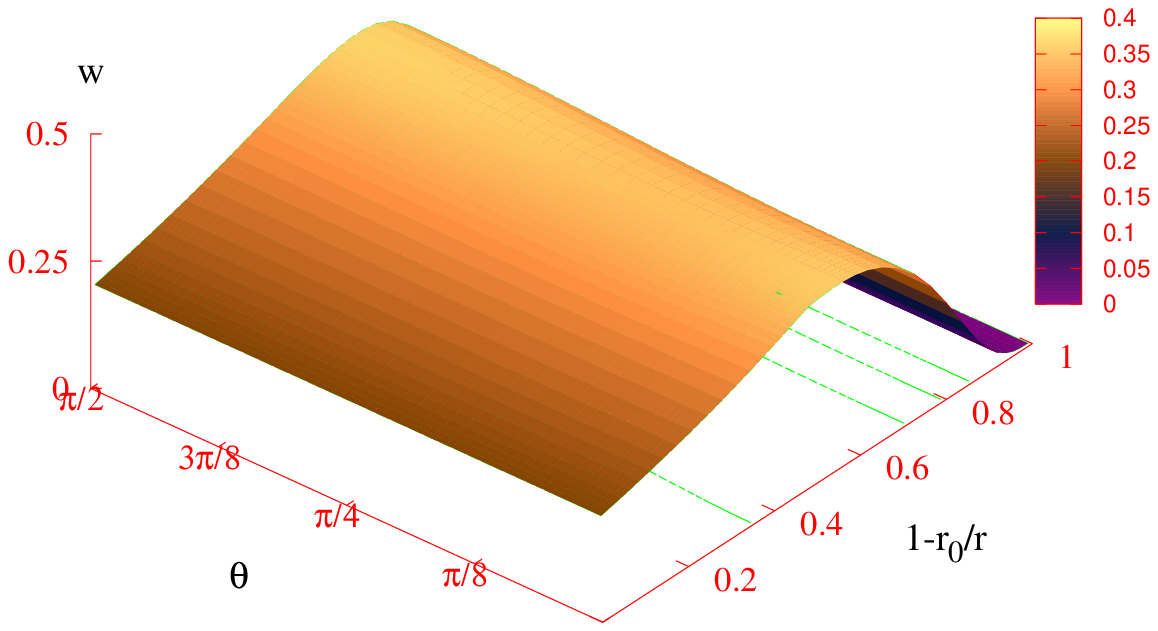,width=7.5cm}}
\put(7,0){\epsfig{file=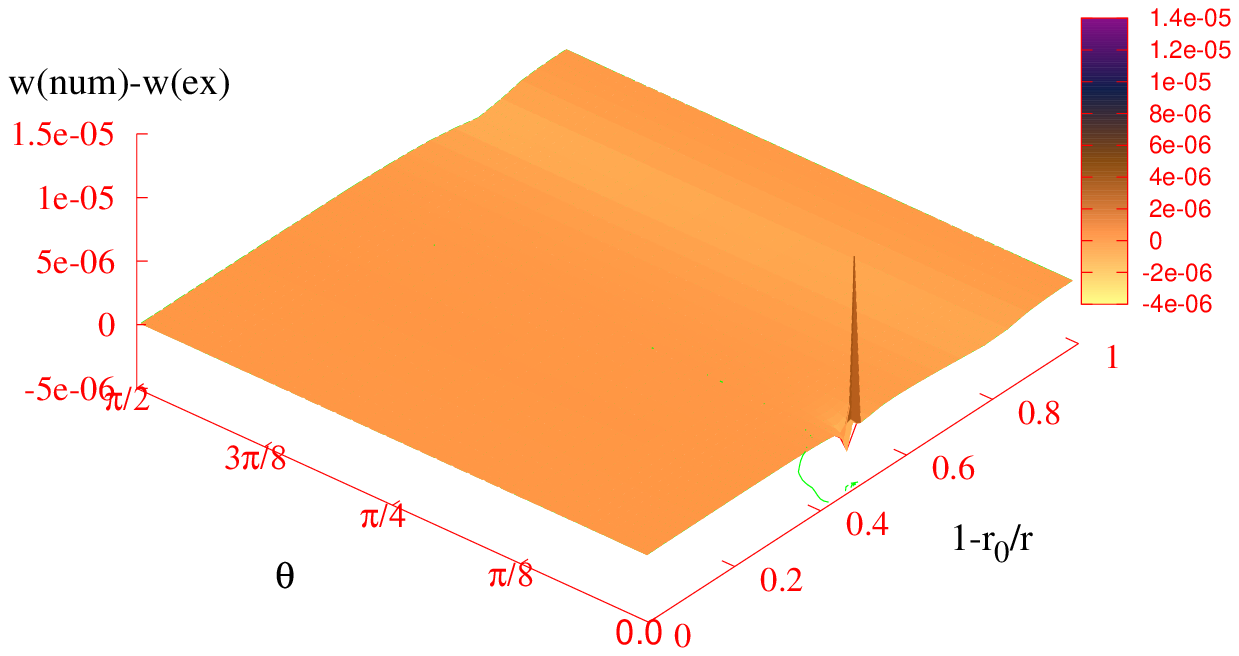,width=7.5cm}} 
\end{picture} 
\\
{\small {\bf Figure 11 (continued).} 
}   
\\
\\ 
already by the 
background functions of the static solution, such that the rotation leads to smooth functions $F_i,w$.
The description of the numerical method presented in Section 2 is also 
valid in this case.
In particular, we have used a compactified radial coordinate $x=1-r_0/r$ 
and a nonequidistant grid in $\theta$.

In Figure 11, we plot the metric functions of an exact solution with $r_0=1$, $r_b=1.93$,
 as well as the difference between the exact solution and the numerical result.
One can see that the differences are on the order of $10^{-6}$ everywhere.
As shown in Table 1, the global quantities computed numerically according to (\ref{num-quant}) are also in excellent agreement with 
the theory values (\ref{th-value}). 

For completeness, we give here the expression of the metric functions which enter the Myers-Perry
solution with a single angular momentum within the coordinate system (\ref{metric-rot5})
\begin{eqnarray}\nonumber
&&f_0= (1-(\frac{r_0}{r})^2)U ,~ f_1=((1+(\frac{r_0}{r})^2)^2+\frac{a^2}{r^2}\cos^2\theta)^{-1},~ 
f_2=r^2\cos^2\theta (1+(\frac{r_0}{r})^2)^2,~f_3=r^2\sin^2\theta U ,
\\
&&w=\frac{a}{r}(\frac{a^2}{r^2}+\frac{4r_0^2}{r^2})\frac{f_1}{U },
~{\rm and}~
U=(1+(\frac{r_0}{r})^2)^2+\frac{a^2}{r^2}+\frac{a^2}{r^2}\frac{(\frac{a^2}{r^2}+\frac{4r_0^2}{r^2})\sin^2\theta}
{(1+(\frac{r_0}{r})^2)^2+\frac{a^2}{r^2}\cos^2\theta},
\end{eqnarray}
the relevant quantities being given by
\begin{eqnarray}
&&M=\frac{ 3\pi }{8 }a(a^2+4r_0^2),~~
J=\frac{ \pi }{4 }a(a^2+4r_0^2),~~
\Omega_H=\frac{a}{a^2+4r_0^2},
\\
\nonumber
&&
T_H=\frac{r_0}{\pi(a^2+4r_0^2)},~~~
S= { \pi^2}  r_0(a^2+4r_0^2).
\end{eqnarray}
 
We have verified that  within the same numerical scheme as above (with the background functions $f_i^{(b)}$
given by the expressions (\ref{ST-rt}) of the Schwarzschild-Tangherlini metric in isotropic coordinates)
one  recovers the set of rotating black holes with an $S^3$ topology of
the horizon and a single angular momentum.
The numerical accuracy is even better in this case, since the expression of the background functions
contains no square roots.

It would be interesting to recover within the same approach the  balanced black Saturn and balanced
black diring solutions starting with the corresponding static configurations.

  To conclude, we have proposed a numerical scheme
which could reproduce physically interesting  $d=5$
spinning solutions starting with the corresponding static configurations. 
This opens the possibility to study generalizations
of the Emparan-Reall balanced black rings and Myers-Perry black holes in various theories
where closed form solutions are unlikely to exist ($e.g.$ in Einstein-Gauss-Bonnet  
or Einstein-Yang-Mills  theory).
We hope to report on that in future work.

\begin{table}[t]
\begin{center}
\begin{tabular}{|cr|c|c|c|c|c|c|c| }
 \hline
&  $r_b$  & $\Omega_H$  & $M(num)$  & $M(ex)$ & $|J(num)|$  & $|J(ex)|$ & $A_H(num)$  & $A_H(ex)$\\
\hline 
\hline          
 & $ 1.61803$   & $0.182574 $  & $24.0003$  & $24.0000$  & $109.547$  & $109.545$   & $773.616$   & $773.605$\\
 & $ 1.93186$   & $0.204124 $  & $16.0000$  & $16.0000$  & $58.7880$  & $58.7878$   & $446.647$   & $446.645 $\\
 & $ 2.18890$   & $0.207020 $ &  $13.3332$   & $13.3333$   & $45.0836$   & $45.0843$    & $ 332.911$    & $ 332.909$\\
 & $ 2.41421$   & $  0.204124$   & $  12.0001$   & $  12.0000$   & $ 39.1922$  & $  39.1918$    & $ 273.514$   & $  273.518 $\\
 & $   2.80588$   & $  0.193649$   & $  10.6671$   & $  10.6667$ & $   34.4289$ & $   34.4265$  & $   210.552$ & $    210.563 $\\
 & $  3.14626$   & $  0.182574 $& $    9.99982$  & $   10.0000$  & $  32.8624$ & $   32.8634$ & $    176.553$ & $    176.555 $\\
 & $   3.45197$  & $   0.172516$  & $   9.59981$  & $   9.60000$  & $  32.4596$ & $   32.4607$ & $    154.723$ & $    154.726 $\\
 & $   3.99215$ & $   0.155902$  & $   9.14274$  & $   9.14286$  & $  32.9869$ & $   32.9877$ & $    127.614$ & $    127.616 $\\
 & $   4.46653$  & $   0.143019$  & $   8.88879$  & $   8.88889$  & $  34.1828$ & $   34.1834$ & $    110.970$ & $    110.973 $\\
 \hline 
\end{tabular}
\end{center} 
\vspace{0.5cm} 
{\small
{\bf Table 1.} 
The   values of the event horizon velocity $\Omega_H$,  mass parameter $M$,
angular momentum $J$ and
of the event horizon area $A_H$  are shown for rotating balanced black ring solutions
with $r_0=1$ and several values of $r_b$.
For comparison, both the numerical and exact values are given here. 
}
\end{table}
  
 \begin{small}
 
 \end{small}

\end{document}